\documentclass[floats]{revtex4}
\usepackage{amsmath}
\usepackage{graphicx}
\usepackage{dcolumn}
\usepackage{xcolor}
\usepackage[caption=false]{subfig}
\usepackage[final,author={Alfredo},markup=Highlight,color=yellow,open=false]{pdfcomment}
\usepackage{hyperref}
\usepackage[final,color]{showkeys}

\hypersetup{
	pdfstartview={XYZ null null 1.00}, pdfauthor={Alfredo Correa},
	pdftitle={Typeset a document in Sans Serif fonts (sansserif.pdf)},
	pdfsubject={subject}, pdfkeywords={},  pdfcreator={\pdftexbanner}
}

\begin{document}

\preprint{}

\title{A first-principles global multiphase equation of state for hydrogen}


\author{Alfredo A. Correa}
\email[]{correaa@llnl.gov}
\affiliation{Lawrence Livermore National Laboratory, Livermore CA 94550}

\author{Lorin X. Benedict}
\affiliation{Lawrence Livermore National Laboratory, Livermore CA 94550}


\author{Miguel A. Morales}
\affiliation{Lawrence Livermore National Laboratory, Livermore CA 94550}

\author{Philip A. Sterne}
\affiliation{Lawrence Livermore National Laboratory, Livermore CA 94550}

\author{John I. Castor}

\affiliation{Lawrence Livermore National Laboratory, Livermore CA 94550}

\author{Eric Schwegler}

\affiliation{Lawrence Livermore National Laboratory, Livermore CA 94550}


\date{\today}

\begin{abstract}
				
	We present and discuss a wide-range hydrogen equation of state model based on a
	consistent set of ab initio simulations including quantum protons and electrons.
	Both the process of constructing this model and its predictions are discussed in
	detail. The cornerstones of this work are the specification of simple physically
	motivated free energy models, a general multiparameter/multiderivative fitting
	method, and the use of the most accurate simulation methods to date. The
	resulting equation of state aims for a global range of validity ($T =
	1-10^9~\mathrm{K}$ and $V_\mathrm{m} = 10^{-9}-1~\mathrm{m^3/mol}$), as the
	models are specifically constructed to reproduce exact thermodynamic and
	mechanical limits. Our model is for the most part analytic or semianalytic and
	is thermodynamically consistent by construction; the problem of interpolating
	between distinctly different models --often a cause for thermodynamic
	inconsistencies and spurious discontinuities-- is avoided entirely.
	
	LLNL-JRNL-745184-DRAFT
\end{abstract}

\pacs{PACS}


\maketitle

\tableofcontents

\section{Introduction\label{sec:intro}}

The equation of state (EOS) of elemental hydrogen has been of singular interest
to the scientific community for many years. The reasons for this are: 1.
Hydrogen is the simplest element, from an atomic physics point of view, and it
has therefore long been a subject of discussion regarding what its properties
might be at extreme compressions \cite{WignerHuntington}. 2. Being the most
abundant element in the universe, it is thought to be a major component of
stars and giant planets, and the understanding of its compressive properties
are key to the modeling of these objects \cite{astro}. 3. Isotopes of hydrogen,
deuterium and tritium, not only form the components of stars, but are also the
primary constituents in fuel capsules used in current designs to achieve
inertial confinement fusion (ICF) \cite{ICF}.

Taken together, these and related studies require a knowledge of hydrogen EOS
(as well as the EOS of deuterium-tritium --DT-- mixtures) across a range that
spans from densities of $\mathrm{mg/cc}$ to $\mathrm{kg/cc}$, and temperatures
from a few $\mathrm{K}$ to $10^8~\mathrm{K}$. This includes multiple solid
phases (high-$\rho$, low-$T$), the diatomic molecular gas (low-$\rho$, low-$T$),
and the atomic liquid/gas (high-$T$). This last category includes plasma states
in which the electrons are not particularly associated to individual protons
(ultra-high-$T$). Such a diversity of states makes a uniform theoretical
treatment extremely difficult. For instance, it is widely believed that
mean-field electronic structure theory approaches, such as approximations (e.g.
LDA or GGA) to Density Functional Theory (DFT), should work well for highly
compressed states, while they are known to fail in many respects for the
description of very low-density matter. Indeed, this is not specific to
hydrogen. However, hydrogen in particular poses additional challenges which
arise from its small ionic mass: The quantum mechanical nature of the protons
must be taken into account for an accurate description of the EOS, \emph{even
for temperatures at or exceeding the melt temperature}, a fact which is made
obvious by noting that the fundamental vibrational frequency of the gas-phase
$\mathrm{H-H}$ bond in $\mathrm{H_2}$ corresponds to a temperature of $\sim
10,000~\mathrm{K}$, while the melting temperature is far below this for all pressures at
which hydrogen has been experimentally interrogated thus far\cite{RMP}.

The essential physics governing hydrogen's thermal and compressive properties
--molecular dissociation due to temperature and pressure, ionization, melting,
etc.-- have been included in highly detailed EOS models, some old enough to
predate the current spate of ab initio electronic structure calculations
\cite{Kerley72, Kerley72pub, Kerley03, SandC91, SandC92, SCvH, Saumon07, Young, Caillabet}. The EOS models of Kerley\cite{Kerley72, Kerley72pub, Kerley03} 
and Saumon et al.\cite{SandC91,SandC92,SCvH,Saumon07} in particular have found wide
use in ICF and astrophysical applications, respectively. In each family of
models, experimental data available at the time of model construction was used
for comparisons, and in some cases was used to constrain the EOS models
themselves. Such data includes cryogenic temperature EOS information
\cite{Silvera, Souers}, principal shock Hugoniot\cite{Nellis, DaSilva, Collins,
Hicks, Knudson01, Knudson04, HicksQuartz} and reverberating shock wave
measurements\cite{Knudson04}, and diamond anvil cell (DAC) studies in which
melting was inferred \cite{Deemyad,Eremets,Sano}. In some recent H EOS
models\cite{Caillabet,HEDP}, heavy use has been made of electronic structure
calculations of the DFT \cite{Galli, Lenosky, Desjarlais, BonevMilitzer,
BonevSchwegler, TamblynPRL, TamblynPRB, Vorberger, Holst} and quantum Monte
Carlo (QMC)\cite{Militzer, Levashov, HuPRL, HuPRB, MoralesPierleoni,
MoralesPNAS, Geng} varieties to provide constraints in regimes where no
experimental data was available.

Well into compression ($\rho \sim$ a few g/cc and above), the EOS of hydrogen
can be treated much like that of other materials \cite{Wallace98,Wallace2002},
in which first-order phase transitions separate distinct solid phases, and solid
from liquid. Though the known solid phases of hydrogen are many and various
\cite{RMP}, it is currently thought that an adequate description of the EOS (for
many applications) can be achieved by averaging these many individual allotropes
into a single effective solid phase, and this is indeed the choice that was made
in the aforementioned models \cite{Kerley72} \cite{Kerley72pub} \cite{Kerley03,
SandC91, SandC92, SCvH, Saumon07, Young, Caillabet}. At lower densities, the
coexistence of molecular ($\mathrm{H_2}$) and atomic ($\mathrm{H}$) units in the
fluid and gas force the EOS modeling to be quite subtle. Here, it has proved
necessary to invoke notions of {\it chemical equilibrium} \cite{FGvH}
\pdfcomment{Why are we citing this? Why in the context of "chemical equilibrim"}
, in which the molar fractions of $\mathrm{H}$ and $\mathrm{H_2}$ are determined
at a given density and temperature (or, alternatively, at a given pressure and
temperature) by minimizing the free energy subject to the constraint of a fixed
number of particles. This has been accomplished in existing H EOS models in two
distinct ways: 1.~By constructing independent, though somewhat artificial, free
energies for pure-$\mathrm{H}$ and pure-$\mathrm{H_2}$ fluids, and then mixing
them together while applying appropriate constraints \cite{Kerley72,
Kerley72pub, Kerley03}, and 2.~By constructing 2-body potentials for the various
constituents ($\mathrm{H-e^-}$, $\mathrm{H-H^+}$, etc.) and then determining the
free energy of the heterogeneous mixture of particles interacting via these
potentials using \pdfmarkupcomment{various means}{What means???}
\cite{SandC91, SandC92, SCvH, Saumon07}. Such chemical equilibrium models are
perfectly suitable at low, gas-phase densities; at higher-$\rho$, where distinct
species are less well-defined, they are much harder to justify. Still, the fact
that such constructs produce wide-ranged EOS models which respect known limits
has made them an attractive starting-point for the construction of hydrogen EOSs
which use ab initio quantum molecular dynamics (QMD) data as input, even though
the QMD itself invokes no assumptions of chemical equilibrium mixing of
individual $\mathrm{H}$ and $\mathrm{H_2}$ fluids \cite{Caillabet, HEDP}.
\pdfcomment{Why are we citing this? Why in the context of "QMD itself invokes no
assumptions"?}

In this work, we present a multiphase EOS for hydrogen which is based on:
1.~Legacy thermodynamic data \cite{Silvera}\cite{Souers} \pdfcomment{What
data from Souers are we using??? I don't have that book, what page???.} to
constrain the behavior in the neighborhood of the initial conditions for ICF
capsules ($\rho_{\rm DT}\sim 0.25~\mathrm{g/cc}$, $T\sim 10\mathrm{K}$),
2.~Known properties of the $\mathrm{H_2}$ molecule in its gas phase
\cite{Souers} \pdfcomment{again, what from Souers are we using???} , 
and 3.~A host of ab initio simulation data on individual-phase EOSs at elevated density and
temperature (pressure and internal energy as functions of $\rho$ and $T$); this
also includes phase lines, as well as high-$(\rho,T)$ limits of EOS determined
from average-atom calculations \cite{INFERNO, Purgatorio}. When possible, the
individual-phase free energy models are built from the assumed decomposition:
$\text{cold}+\text{ion-thermal} + \text {electron-thermal}$, like the models of
Kerley and its derivatives\cite{Kerley72, Kerley72pub, Kerley03,HEDP}.
for liquid hydrogen is constructed with a chemical equilibrium mixing procedure
in which individual $\mathrm{H}$ and $\mathrm{H_2}$ liquids are defined and
combined. The various parameters of the EOS model, to be enumerated and
discussed in detail below, are fit to the types of data numbered 1 - 3 above
with a nonlinear optimization scheme (NLOpt) specifically designed for this
application. Throughout the construction of our EOS, particular care is taken to
fit to the best available electronic structure theory data (item 3, above). Many
of these data have been produced specifically for this project and are reported
here for the first time; the bulk of this data is produced by a scheme in which
the electrons are treated within DFT, while quantum path integral is
performed on the ions. In this way, the all-important quantum nature of the ions
is taken into account in an essential way.

We describe the ab initio simulations used to produce the bulk of our EOS data
in Section~\ref{sec:simulations}. We discuss the models and assumptions used for
the various components (cold, ion-thermal, electron-thermal), and of the free
energy of each phase in Section~\ref{sec:models}; some technical details are
left to the Appendices. In Section~\ref{sec:fitting} we present the EOS data (in
a multitude of plots) together with the precise mathematical expressions used to
fit these data and the fitting strategy. Section~\ref{sec:results} contains a
discussion of the resulting EOS table and the prediction of important derived
thermodynamic information (such as the principal Hugoniot), and their comparison
with experimental data. In the final Section~\ref{sec:comparison}, we discuss
the resulting similarities and differences with the other available wide-range
hydrogen EOS models \cite{Kerley72, Kerley72pub, Kerley03, SandC91, SandC92,
SCvH, Saumon07, Young}, and comment on the likely implications for applications
such as ICF.

To approach a wide audience and for consistency we decided to use SI units in
the plots, but formulas and parameters (equations and tables in
Section~\ref{sec:fitting}) are casted such that can be used in a variety of unit
systems. In the following we use molar (per mole of atomic nuclei) quantities,
since number-normalized (as opposed to mass normalized) quantities make easier
to compare isotopic effects. The model developed and data presented in plots
referes to Hydrogen unless stated otherwise. To transform molar volumes to
density use the inverse relation for H, D and T, $1~\mathrm{m^3/mol^{-1}} \to
1.00794/2.014/3.016049\times10^{-6}~\mathrm{g/cm^3}$ and $1~\mathrm{g/cm^3} \to
1.00794/2.014/3.31559\times10^{-6}~\mathrm{mol/m^3}$ respectively. (This simple
density scaling does not introduce quantum isotopic effects which may be
important at low temperatures.)

\section{Simulations}\label{sec:simulations} 

\begin{figure}
	\includegraphics[page=1,height=5.5cm]{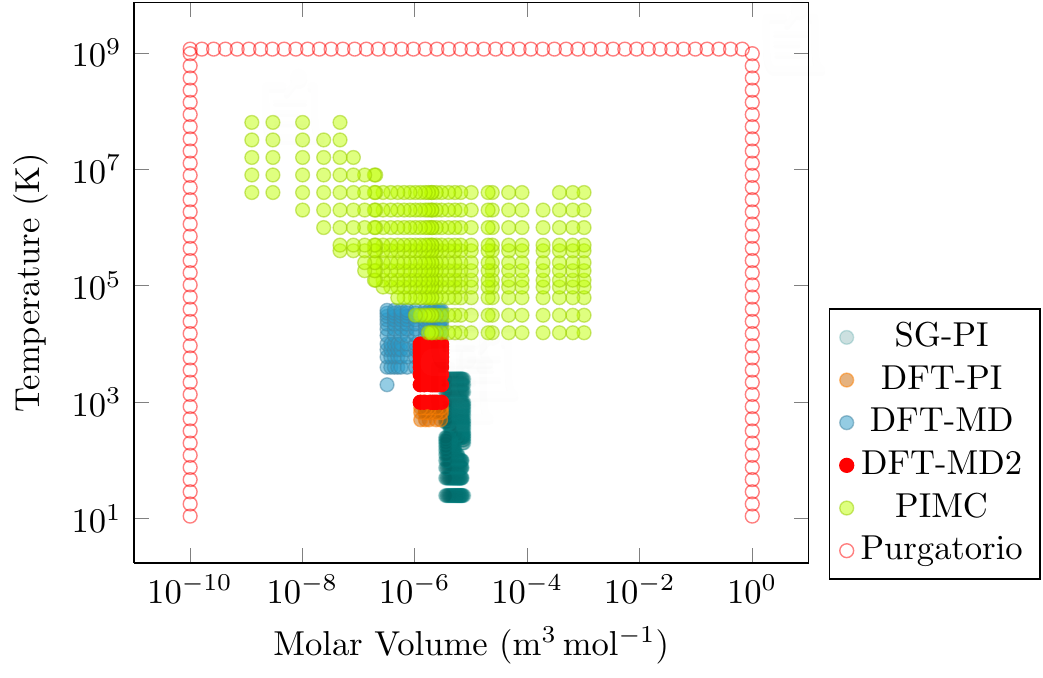}
	\includegraphics[page=2,height=5.5cm]{fig/ranges.pdf}
	\caption{ 
		(color online) Simulation data used in this work, (i) SG-PI (nuclei path integral Monte
		Carlo with the empirical Silvera-Goldman intermolecular potential), (ii) DFT-PI
		(nuclei path integral molecular dynamics with ab initio DFT inter-ionic potential),
		(iii) DFT-MD (classical nuclei molecular dynamics with ab initio DFT inter-ionic
		potential), (iv) PIMC (nuclei + electron path integral quantum Monte Carlo; not
		used for fitting in this work, only for validation), (v) \sc{Purgatorio} (relativistic
		electrons treated within DFT in an average atom embedded in jellium; the red
		boundary indicates the range {\it within} which the
		\sc{Purgatorio} calculations are performed (so, they envelop the all the other
		calculations).
	}
	\label{fig:ranges}
\end{figure}

In this work, simulation results are used to construct and to fit the free
parameters of the EOS models. Due to the extensive range of the EOS, extending
over many orders of magnitude in density and temperature\ref{fig:ranges}, and
the existence of very different regimes within, we used various simulation
techniques to produce reliable thermodynamic data across the phase diagram.

In general, we use path integral methods to treat the ions at lower temperatures
in order to properly account for nuclear quantum effects. At higher
temperatures, typically above 10,000 K, we can safely use classical simulation
methods for the ions. As described below in more detail, we use effective
interactions between protons in the molecular phases at lower densities, where
due to the dilute nature of the system this level of description is accurate
enough for the present purposes. At higher densities, we must resort to an
ab-initio description of the electronic degrees of freedom due to the lack of
experimental results or appropriate models for the dense phases.

Underlying methods utilized are different depending on the density regime (e.g.
below or above $4\times 10^{-6}~\mathrm{m^3/mol}$) and temperature (e.g. below
or above $10000~\mathrm{K}$ )  as depicted in Fig.~\ref{fig:ranges}. Detailed simulation results are shown together compared with the resulting model.

\subsection{Low Density Molecular Phases} 

At low density, particularly in the molecular phases, the interaction between
ions can be accurately modeled using empirical pair potentials. 
These potentials do not require an explicit treatment of the
electronic degrees of freedom, making them particularly efficient and
computationally inexpensive. They are typically obtained from a combination of
experimental information and accurate calculations on clusters of atoms. There
are several well known empirical potentials designed for condensed phases of
hydrogen \cite{Schaefer79,Silvera78,Silvera80,Ross83,Hemley90}. In this work we
use the well-known Silvera-Goldman (SG) potential \cite{Silvera78}, including
the modifications proposed by Hemley, \emph{et. al} \cite{Hemley90} to improve
its agreement with experiment at higher densities. The SG potential describes
only the interaction between hydrogen molecules, it contains no intramolecular
terms. In order to describe the intramolecular properties correctly, which
strongly influence thermodynamical properties at finite temperature, we use the
ground state potential energy surface for the isolated molecule as calculated by
W. Kolos and L. Wolniewicz, \cite{Kolos69} (KW). This combined description (SG +
KW) is accurate as long as molecules don't dissociate and temperature is low
enough to justify the neglect of electronic excitations. Both conditions are
well satisfied in this regime.

While empirical pair potentials produce sufficiently accurate results at low
densities and temperatures, their use in this work is further motivated by the
following reasons. As the density is decreased DFT calculations become more
computationally expensive, which makes a full ab-initio description in this
regime less attractive. This is particularly important when a full path integral
description for the ions is intended. In addition, typically
exchange-correlation potentials used in DFT do not properly account for
dispersion interactions, which are dominant in low density hydrogen. While the
use of dispersion-corrected functionals in hydrogen has been explored at higher
densities, close to molecular dissociation \cite{Morales13a,Morales13b}, their
use at low density has not been explored yet.

We performed quantum Path Integral Molecular Dynamics (PIMD) simulations with
this empirical (SG + KW) potential, in both liquid and solid molecular hydrogen
phases, for densities in the range of $4-7\times10^{-6}~\mathrm{m^3/mol}$.
(Fig.~\ref{fig:solidpidftU}) were performed assuming phase I of
hydrogen~\pdfcomment{cite Miguel, for example who discovered Phase I ???}, where the molecules reside
on an `hcp' (hexagonal close-packed) lattice and there is no orientational order.
We used 360 atoms and a time step of $0.2~\mathrm{fs}$ ($8~\mathrm{Ha^{-1}}$) in
all simulations. We discretized the path integrals with a time step of
$1.5~\mathrm{fs}$ ($0.000208~\mathrm{K^{-1}}$), which corresponds to 48 beads at
a temperature of $100~\mathrm{K}$. Simulations in the molecular liquid
(Fig.~\ref{fig:mliq1pisgU}) were performed up to a temperature of $2500~\mathrm{K}$, since above this temperature we expect dissociation to occur .

\subsection{High Density Molecular and Atomic Phases}

At higher densities the interaction between nuclei becomes considerably more
complicated, making the use of empirical potentials unreliable. This is
particularly true close to molecular dissociation in the warm dense matter
regime. In this case, we use an ab-initio description of the electronic degrees
of freedom based DFT. DFT calculations were performed with various simulation
packages including: CPMD \cite{CPMD}, \textsc{Qbox} \cite{qbox} and
\textsc{Quantum Espresso}\cite{QE}. The use of multiple simulation software
packages is due to the large set of simulations used in this work, which were
produced over an extended period of time. Some of these simulations have been
reported in previous publications
\cite{Morales09,MoralesPierleoni,MoralesPNAS,Morales13a}. 
All simulations in the high density molecular solid were done with PIMD (Figs.~\ref{fig:solidpidftP} and \ref{fig:solidpidftU}), 
(Figs.~\ref{fig:mliqpidftP} and \ref{fig:mliqpidftU}), 
above this temperature we used a classical description of the ions based on QMD. 
In all cases, we used Troullier-Martins
\cite{Troullier91} norm-conserving pseudopotentials with a cutoff radius of 0.5
Ha \pdfcomment{What is the cutoff radius in length units???}. Simulation
sizes ranged from 128-432 atoms, and we used a time step of $0.2~\mathrm{fs}$
($8~\mathrm{Ha^{-1}}$). In this case, the path integrals were discretized with a
time-step $0.95~\mathrm{fs}$ ($0.000125~\mathrm{K^{-1}}$), which corresponds to
8 beads at 1000 K. All simulations were performed at the Gamma point with a
plane wave cutoff of 90 Ry. We added corrections to the equation of state to
account for the finite cutoff and the limited k-point sampling in the
simulations. To do this, we used 15-20 snapshots from simulations at each
density and performed well converged calculations with a plane-wave cutoff of
$300~\mathrm{Ry}$ and k-point sampling with a $3\times 3\times 3$ Monkhorst-Pack
grid. 
\pdfcomment{Miguel, any quantification of the correction???}

We used Coupling Constant Integration \pdfcomment{\cite{Miguel, citation needed}} to calculate the Helmholtz free energy on
both solid and liquid phases. This gives us access to the
entropy (Fig.~\ref{fig:solidpidftS}), which can not be calculated from single-point
equilibrium simulations, but nonetheless is a crucial to constrain the EOS model. For more
details on these calculations, see references
\cite{MoralesPierleoni,MoralesPNAS}.

\subsection{Additional Simulations}

\pdfcomment{Miguel, THIS SECTION IS CONFUSING ARE WE COMPARING WITH CEIMC OR WITH RPIMC??? THE DISTINCTION CEIMC, QMC and RPIMC IS NOT CLEAR}

In this subsection, we briefly describe simulation sets that were used to
compare against the resulting EOS (Section~\ref{sec:comparison}), but were not
used in the fitting process. Accurate calculations have been published in the
atomic liquid regime below 10,000 K using the Coupled Electron-Ion Monte Carlo
(CEIMC) method \cite{Dewing02, Pierleoni06, MoralesPierleoni}. CEIMC is a method
based on an accurate description of the electronic degrees of freedom using
quantum Monte Carlo (QMC) methods combined with a classical description of the
ions. QMC is an accurate many-body method that does not suffer from many of the
theoretic deficiencies of DFT approximate methods, and is particularly accurate
in hydrogen. While these calculations were not used directly in the fitting
process, as mentioned above, they have been consistently used to test the
accuracy of EOS models for hydrogen \cite{Caillabet, Vorberger13}. At very high
temperatures, the restricted Path Integral Monte Carlo (RPIMC) method provides
very accurate results for the equation of state of light elements, particularly
hydrogen \cite{Ceperley96,Militzer,Militzer01}. This method is based on a path
integral description for both electrons and ions simultaneously. The only
approximation of the method comes from the use of an approximate nodal surface
for the thermal density matrix, which is needed due to the fermion sign problem
that appears as a consequence of the fermion symmetry of the problem, see Ref.
\cite{Ceperley96} for a detailed description of the method. This approximation
is exact in the limit of infinite temperature, and remains very accurate at
temperatures above $\approx$ 0.5 $T_\mathrm{F}$, where $T_\mathrm{F}$ is the Fermi temperature of the electrons. 
This makes RPIMC calculations a very accurate benchmark of the
thermodynamic properties of hydrogen at high temperatures. Recently, RPIMC
calculations were reported on an extended regime of the hydrogen phase diagram
\cite{HuPRB}, which we use in this work to test the accuracy of our EOS model at
high temperatures.

\section{Models}\label{sec:models}

In our multiphase equation of state, each thermodynamic phase has its own model
for the Helmholtz free energy, $F$ ($= E - T S$), from which all thermodynamic
quantities, such as energy $E= -T(\partial F/\partial T)_V + F$, pressure $P=
(\partial F/\partial V)_T$, entropy $S = -(\partial F/\partial T)_V$, and other
thermodynamic potentials are derived. Transitions between phases at constant $P$
and $T$ are computed by equating Gibbs free energies $G= F + PV$, or
equivalently by performing the common-tangent construction for $F$, resulting in
a single multiphase free energy if needed. The choice of the free energy as the
generating function for the multiphase EOS is a straight-forward way to ensure
thermodynamic consistency, as embodied for instance by the Maxwell relations
(equality of mixed partial derivatives of the free energy- e.g.,
$\partial^{2}F/\partial V \partial T = (\partial S/\partial V)_{T} =
\partial^{2}F/\partial T \partial V = (\partial P/\partial T)_{V}$). The local
stability of the individual phases is represented by convexity requirements on
the individual-phase free energies; stability of the multiphase system is then
ensured by the convexity brought about by the common-tangent construction
\cite{Wallace98, Wallace2002}. Operationally, all our free energy models are
constructed with $V$ and $T$ as the independent variables, $F(V, T)$; this
facilitates connection to simple statistical mechanical models with $V$- and
$T$-dependent partition functions, $Z(V,T)$, through the relation $F=-k_{B}{\rm
	ln}Z$ \cite{Wallace98,Wallace2002}.


Broadly speaking, the basic free energy of each phase is constructed from
 component free energy models. An important example of these
components which we will use is the decomposition of a single-phase free energy
into `cold', ion-thermal (IT), and electron-thermal (ET) pieces
\cite{Wallace2002}, \begin{equation}\label{eq:coldITET} F(V, T) =
\phi_\mathrm{cold}(V) + f_\mathrm{IT}(V,T) + f_\mathrm{ET}(V,T). \end{equation}
Here, the first term (cold) represents the energy of classical ions in the
idealized absence of excitations (i.e. without zero point energy) for a given
atomic structure; the second term (IT) contains information about the phonons or
collective modes in that given structure and under some average electronic
state, while the third term (ET) includes thermal electronic excitations. These
component free energies, individually, need not satisfy convexity stability
requirements everywhere. It is crucial to note that in many instances this
separation is only nominal, for some collection of these terms must be
determined {\it together, as a whole}. In this sense, while these separations
allow us to associate the label of each term with different aspects of the
physical problem. It is only the concrete identification of these labels (such
as those given above) with simple idealized models what makes the model less
general.

Another example of a different type of free energy we will use, is the chemical
equilibrium mixing of liquid $\mathrm{H}$ and liquid $\mathrm{H_2}$ components.
In this case, the identification of individual (additive) free energy terms is
not physically possible, although the separation is still used at a more basic
level, to model the component free energies. \cite{Kerley72, Kerley72pub,
Kerley03, FGvH}, as we discuss below.

The EOS data culled from the simulations we describe below certainly need not
respect the above decompositions, such as that of Eq.\ref{eq:coldITET} and those
arising from chemical equilibrium mixing. Nevertheless, such decompositions
still will prove fruitful for constructing our EOS model, as they have in the
past for other models\cite{Kerley72, Kerley72pub, Kerley03, SandC91, SandC92, SCvH, Saumon07, Young, Caillabet, Wallace2002}.

In this Section, the models and equations are discussed and outlined; for the
concrete mathematical representation of the models, see the equations in
Section~\ref{sec:fitting}.

\subsection{Molecular Solid}\label{sec:modelsmsolid}

There are many known molecular solid phases of hydrogen \cite{RMP}. We choose to
lump all of these into a {\it single} representative molecular solid phase
instead and avoid the distraction of multiple distinct solid phases. A more
detailed multi-solid phase description can be added in future work in a
straightforward way. For our representative solid allotrope, we concentrate on
the molecular `hcp' phase, both in the simulations and in the modeling. This
assumption has also been made, to varying degrees, in previous work
\cite{Kerley72, Kerley72pub, Kerley03, SandC91, SandC92, SCvH, Saumon07, Young,
Caillabet}. Because the nuclei of this representative phase are light,
consisting of individual protons, the delocalized quantum nature of these nuclei
render the construct of \emph{individual} `cold' and `IT' pieces
(Eq.~\ref{eq:coldITET}) fundamentally ill-posed. We will, however, define these
separate terms for convenience. 
Though we consider the solid
at elevated temperatures, and though the higher-pressure (atomic, which we do
not consider) solid is predicted to be a metal \cite{RMP}, we refrain from
adding the comparatively small electron-thermal term, $f_{\rm ET}$, for the
molecular solid phase. This term will be considered explicitly in the section on
the atomic liquid (Section~\ref{sec:modelsatomicfluid}), where its inclusion is essential.

\subsubsection{Cold Curve}

The common procedure for determining a cold-curve,  $\phi_{\rm cold}(V)$, from
electronic structure calculations for normal solid (higher mass number)
materials involves considering a certain fixed (and mechanically stable) crystal
structure, in which the total energy at $T=0$ is calculated (e.g., using DFT and
its associated approximations) on a fine grid of volumes for ions fixed in these
crystalline positions. This yields the energy as a function of volume $V$,
assuming the electrons to be in their ground state, and the ions to have
\emph{infinite mass} (classical positions). The first term `cold' in
Eq.\ref{eq:coldITET} can be therefore completely defined from static
calculations alone. Additional contributions from ionic and electronic
excitations are then added by computing phonons \cite{caveat1} and electronic
excitations in this particular crystal structure, which yield $f_\mathrm{IT}(V,
T)$ and $f_\mathrm{ET}(V, T)$ \cite{Correa}.


The hydrogen case, however, is completely different from normal materials
\emph{at solid densities} due to its molecular nature and its exceedingly light
nuclei; the $\mathrm{H}_2$ molecules are freely rotating even though the lattice
of such molecules is well- defined\cite{RMP}. Furthermore, no
\emph{fixed}-nuclei crystalline solid structure is known to be (classically)
mechanically stable from modern theories (e.g. DFT). Thus, it is not possible to
define the second term in Eq.\ref{eq:coldITET} from perturbations about a stable
minimum where the ions are bolted in place. Although the data from our
simulations need not be consistent with the separation in Eq.\ref{eq:coldITET},
it is at least operationally possible to assume it. The strategy we employ in
this case is then to define the first and second terms in Eq.\ref{eq:coldITET}
as a single unified object in which they are determined together.

These complexities notwithstanding, we take $\phi_\text{cold}(V)$ to have the
Vinet form \cite{Vinet}

\begin{subequations}
	\begin{equation}\label{Vinet}  
		\phi_{\rm cold}(V)=
		\phi_{0} + \frac{4V_{0}B_{0}}{(B_{1} - 1)^{2}}[1 - (1 + X)\exp(-X)] + \cdots
	\end{equation}
	\[
		X= \frac{3}{2}(B_{1} -1)[(V/V_{0})^{1/3} - 1], 
	\]
						
	where $V_0$ is the molar volume at which $\phi_\text{cold}(V)$ is minimum and
	equal to $\phi_0$, $B_{0}$ is the (cold) isothermal bulk modulus, and $B_1$ is
	the pressure-derivative of the isothermal bulk modulus at $V_0$. We also add
	corrections of the form
	\begin{equation}\label{ThomasFermi}
		\cdots + E_\mathrm{TF} {\mathopen{}\left(\frac{V}{V_\mathrm{TF}}\right)\mathclose{}}^{-2/3} \exp
		\mathopen{}\left[-{\mathopen{}\left(\frac{V}{V_\mathrm{TF}}\right)\mathclose{}}^{2/3}\right],
	\end{equation}
\end{subequations}

that are important for $ V < V_\text{TF} $, in order to alter the otherwise
incorrect high-compression behavior of the Vinet form in a way which respects
the bounds imposed by the high compression Thomas-Fermi limit \cite{Holzapfel}.
$V_\mathrm{TF}$ and $E_\mathrm{TF}$ are volume and energy scale parameters that
for the most part ensure the relative stability of the atomic phase model (that
respects the Thomas-Fermi limit by construction, see below).

From the above discussion, the parameters of this cold curve model cannot be
determined independently of the choice of the second term in
Eq.\ref{eq:coldITET}, the ion-thermal part, as defined next.

\subsubsection{Ion Thermal}

Though, at this point, the separation of `cold' and `IT' is merely nominal due
to the peculiarities of the molecular solid, we assume that the ion-thermal free
energy is based in the first place by the quasiharmonic
expression\cite{Wallace98,Wallace2002} (plus certain high energy corrections
explained later), 

\begin{equation} \label{QH}
	f_\text{harm}(V,T)=
	\frac{3}{\text{atom}}\int_{0}^{\infty}\mathrm{d}\omega
	D_V(\omega)\left[\frac{1}{2}\hbar\omega + k_\mathrm B T\log[1 -
	\mathrm{e}^{-\frac{\hbar\omega}{k_\mathrm B T}}]\right], 
\end{equation} 

where $D_V(\omega)$ is a normalized volume-dependent effective phonon density of
states. Furthermore, we take $D_V(\omega)$ to have the double-Debye
form\cite{Correa} for each $V$, in which two separate Debye-like peaks exist in
$D_V(\omega)$, here denoted `A' and `B', each with its own $V$-dependent Debye
temperature, $\theta_\mathrm A(V)$ and $\theta_\mathrm B(V)$. The larger of
these two, $\theta_\mathrm B$, is meant to embody the intramolecular vibrations
and librations of the $\mathrm{H}_2$ units, while the smaller, $\theta_\mathrm
A$, is meant to represent the vibrations of the significantly softer
intermolecular bonds. \pdfmarkupcomment{
	At high compressions these modes should hybridize, and this tendency can be
	captured naturally in the model by allowing $\theta_\mathrm A(V) \to
	\theta_\mathrm B(V)$, eventually describing a situation in which a single Debye
	temperature will suffice}{Lorin: I agree that this is the ideal case, but
	I don't think this happens, the functional forms used for theta(V) don't permit
	them to have the same limit, besides the fact that acoustic and optical modes
	merges is taken into account by the fact that one transitions to an atomic
	phase.
	}

The free energy of the double-Debye model can be derived simply from
Eq.~\ref{QH} once normalization of $D_V$ is enforced,

\begin{equation}\label {double-Debye} 
	f_\text{harm}(V,T)=
	\xi_\mathrm{A} f_\mathrm A(V,T) + \xi_\mathrm{B} f_{\rm B}(V,T), 
\end{equation} 

where $\xi_\mathrm{A} + \xi_\mathrm{B} = 1$ and
 
\begin{equation}\label{f}
	f_{\rm A,B}(V,T)= \frac{k_\mathrm{B}}{\text{atom}}\left[\frac{9}{8}\theta_{\rm A,B}(V) + 3T\log(1 -
		\mathrm e^{-\theta_{\rm A,B}(V)/T}) - T\mathcal
	D_3(\theta_\mathrm{A,B}(V)/T)\right] 
\end{equation} 

are the familiar single-Debye free energies with 

\begin{equation}\label{D3} \mathcal D_3(x)=
	\frac{3}{x^{3}}\int_{0}^{x}\frac{y^{3}}{\exp(y)-1}\mathrm dy. 
\end{equation} 

The $\theta_{0}$ appearing in Eq.\ref{double-Debye} denotes the logarithmic
moment \cite{Wallace2002} of $D_{V}(E)$ and is given by

\begin{equation}\label{theta0} 
	\theta_{0}(V)= \mathrm
	e^{1/3}\exp\left[\int_0^\infty\log(\omega) D_V(\omega)\mathrm d\omega\right].
\end{equation} 

This double-Debye model, introduced first in Ref.\cite{Correa}
for a different material, was applied recently to hydrogen \cite{Caillabet}.
									
Since it is not possible to compute vibrations about fixed ionic configurations
for solid hydrogen, owing to the freely rotating (and delocalized, in the
quantum sense) $\mathrm{H}_2$ units, we have no way to determine the $D_V(\omega)$ that
enters Eqs.\ref{QH} and \ref{theta0}. Instead, we choose to work with Eqs.\ref
{double-Debye},\ref{f} and \ref{D3} {\it directly}, and use the Debye
temperatures, $\theta_{\rm A}(V)$, $\theta_{\rm B}(V)$, and $\theta_{0}(V)$, as
parameters with which the resulting EOS is fit. To facilitate this fitting by
reducing the number of free parameters, each Debye temperature is assumed to
have a specific $V$-dependence. In particular for the molecular solid phase, we choose that the higher Debye temperature and factors $\xi_\mathrm{A, B}$ be constant and that the lower Debye temperature has a constant Gr\"uneisen parameter $\gamma$:
\begin{equation}
	\theta_\mathrm{A}(V)  = \theta^0_\mathrm{A} \left(\frac{V}{V^0_\mathrm{A}}\right)^{-\gamma_\mathrm{A}}
\end{equation}
									
									
									
									
It is important to note that the true excitations of the molecular solid include
$\mathrm{H_2}$ rotations in addition to the intra- and intermolecular
vibrations. Since, however, these rotations are hindered at most densities and
morph into optical branch phonons as density increases, we allow them to be
lumped into the lower of the two Debye peaks in our double-Debye description. As
long as the simulations discussed in Section~\ref{sec:simulations} provide an adequate description
of molecular rotations, the values that they generate for our fitting will suffice
for our modeling. In the low-density molecular gas, where we do not fit to ab
initio MD data, other constructs are used which take the effects of rotations
into account (see below).
									
Because the free energy of our ion-thermal model for the solid is based on the
notion of a harmonic vibrational spectrum at each $V$ (the so-called
quasiharmonic assumption \cite{Wallace98}), as represented in Eq.~\ref{QH}, we
do not include any detailed effects of anharmonicity at the lower temperatures
where the solid is stable. In this model, as for molecular rotations, any
effects of anharmonicity present at low temperature in the simulation data serve
merely to renormalize the Debye temperatures. What this simple picture lacks so far is
the means to describe deviations at high-$T$ from the Dulong-Petit limit of the
ionic specific heat at constant-$V$, $C_V^{\rm ion}$ \cite{Wallace2002, Correa}.
In comparing our resulting solid-H EOS model to data from our simulations (which
includes anharmonicity), we will see that this is not a major problem. The only
high temperature anharmonic effects in the solid phase we include are of a very specific type, as described immediately below.
									
We now digress briefly to address a fundamental issue specific to the multiphase nature of
the EOS model: The high-$T$ limit of the solid-phase free energy. The solid is
unfavored with respect to the liquid above $T_\text{melt}$, and is certainly not
thermodynamically stable at, say, a few times the Debye temperature. Thus, it
would be tempting to remove the solid phase from the picture
altogether at such high temperatures. As we discuss below, however, our fitting
procedure for the multiphase hydrogen EOS benefits from each of our phases being
defined everywhere, since we allow the phase lines to move in the course of the
fitting procedure as their phase-dependent free energy parameters are optimized.
Furthermore, since our solid model, so far described, has $C_V^{\rm harm} \to
3k_\mathrm{B}/\text{atom}$ at high $T$, the resulting Helmholtz free energy of the solid at
extreme temperatures would necessarily be \emph{lower} than that of
the liquid/gas (which must have $C_V^{\rm IT} \to \frac{3}{2}k_\mathrm{B}/\text{atom}$ at
high-$T$), in contradiction to reality. To prevent such pathological behavior,
we add a term to $f_\mathrm{IT}$ for the solid that forces the high-$T$ limit of
its $C_V^\mathrm{IT} \to 0$. Though this may seem somewhat arbitrary, this
limiting value for $C_V^\mathrm{IT}$ can be seen as a manifestation of the
limits imposed in configuration space (phase space) for a system with long-range
order \cite{caveat3}\cite{Carpenter}. This behavior can be obtained with an additive term in the ion thermal free energy:
\begin{equation}
	f_\mathrm{IT} = f_\text{harm} + f_\text{cell}
\end{equation}
									
The term added has the form,
\begin{equation}\label{cellsolid} f_\text{cell} \propto -\frac{k_{\rm B}}{\text{atom}} T \log\left[{\rm
			erf}\left(\sqrt{\frac{T^{*}}{T}}\right) -
\frac{2}{\sqrt{\pi}}\sqrt{\frac{T^{*}}{T}}\mathrm e^{-T^{*}/T}\right],
\end{equation} where the proportionality constant is the appropriate one which gives the desired
high-$T$ limit for $C_{V}^{\rm ion}$, and $T^*$ (generally larger than $T_\text{melt}$)
is a $V$-dependent temperature scale. The motivation for this particular
type of expression and the proportionality contant, as well as its label `cell', will be explained in detail
below when we use a similar a scheme for the fluid phase. (Precise expressions are given in Section~\ref{sec:fittingmsolid}.) For
the moment, we are content that our solid-phase free energy is defined
everywhere, even in regions where other phases are bound to be more stable, both
physically and by construction. 
				
\subsection{Fluid}
				
Fluid phase referres here to a set of (somewhat loosely defined) thermodynamic states. Namely, it emcompases the molecular liquid/gas, the dense atomic liquid, the atomic gas and the plasma state. 
				
We mentioned in the previous section that the decomposition of
Eq.~\ref{eq:coldITET} is somewhat problematic, particularly for molecular solid hydrogen.
For the a liquid phase, it is even less meaningful to univocally define individual terms such as
$f_{\rm cold}$, $f_{\rm IT}$, etc., since a liquid cannot be obviously described by
perturbing about a configuration in which the ions are held fixed in position.
Nevertheless, it has been shown that the EOS of the liquid, as computed by
state-of-the-art DFT-based MD simulations, can indeed be represented quite well
by Eq.\ref{eq:coldITET} with the appropriate choices of cold, ion-thermal, and
electron-thermal pieces\cite{HEDP}. We thus adopt this approach here, taking
for instance the $T$-independent (i.e., cold) piece of the molecular liquid free
energy to have the form assumed in Eq.\ref{Vinet} (but with different parameter values than for the solid).
				
In simple dense monoatomic liquids, it has been argued that the low-$T$ thermodynamics
is rather solid-like, as evidenced by $C_V^{\rm ion}$ having values close to
those of the high-$T$ solid. The attempt to understand and describe this has led
to the Chisolm-Wallace model (CW) \cite{Chisolm-Wallace,Wallace2002}, which has now
been applied to a number of monoatomic liquids\cite{Correa,ChisolmAl,LXBBe}. In
this approach, a simple Mie-Gr\"uneisen term is used for $f_{\rm IT}$; its
characteristic temperature, $\tilde{\theta}(V)$, reflects both the average
curvature of a local energy well in configuration space, and the multiplicity of
such wells, per particle, such that for a classical fluid:
\begin{equation}\label{CW} 
	f_\text{CW} \propto - \frac{3k_\mathrm{B}}{\text{atom}} T \log\left(\frac{\tilde{\theta}}{T}\right). 
\end{equation} 
This Mie-Gr\"uneisen form is equivalent to the high-$T$ limit of the Debye model, as
used for the solid (with a Debye temperature equal to $\tilde\theta$).
We use this as one contribution to $f_\mathrm{IT}$ for
liquid hydrogen, the other contribution being $f_\text{cell}$, mentioned above
and described in detail below. A normalization factor is also needed, as determined by the number of degrees of freedom of the species (molecular, atomic) in question (for molecular phases, intramolecular degrees of freedom give yet another contribution to the free energy.)
				
To complete the picture for fluid hydrogen, we must address two additional
issues: 1. Electronic excitations must be included in a manner which takes into
account the increased propensity for ionization at elevated temperatures and
pressures. 2. As mentioned in the Introduction, the fluid at any given
$(\rho,T)$ is a mixture of H and H$_{2}$ units, and these units are particularly
distinct at lower gas-like densities. To this end, we now discuss our models for
individual {\it primitive} atomic and molecular liquid free energies, and then
the means by which they are combined to produce a single-phase free energy,
having the property that along {\it certain} paths in $(\rho,T)$, the transition
between molecular and atomic extremes is completely continuous \cite{RMP}.
				
\subsubsection{Atomic fluid}\label{sec:modelsatomicfluid}
				
We envision a liquid consisting solely of an unstructured assembly of hydrogen atoms (or an interleaving gas of dissociated protons and electrons, in the plasma). Though this phase presents similar challenges as does the solid, in that the individual terms of Eq.\ref{eq:coldITET} are difficult to define in general, at the extremes of $(\rho,T)$ the IT term becomes well-defined. This term is determined by the following constraints: Low-$\rho$, high-$T$: $f_{\rm IT}$ should be the free energy of a non-relativistic gas of classical protons \cite{caveat4}. High-$\rho$, low-$T$: $f_{\rm IT}$ should be described by the Chisolm-Wallace model \cite{Chisolm-Wallace} (Eq.\ref{CW}), where the excitations of the disordered nuclei system are phonon-like in nature.
				
\begin{figure}
	\includegraphics[scale=2]{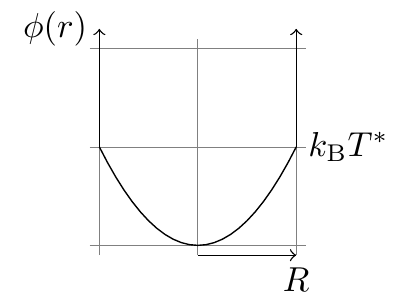}
	\caption{
		Schematic representation of the potential energy and position-space of an 
		ion in the cell model. 
		An ensemble of such one-particle systems is coupled to a heat bath at temperature $T$. 
		$R$ is the radius of a sphere in which the ion resides. For $r < R$, the potential is quadratic. 
		The ion is not permitted to visit $r > R$. 
		When $T \ll T^*$, the ion behaves as an Einstein oscillator. 
		When $T \gg T^*$, the ion undergoes free-particle motion, 
		save the inconsequential reflections off the cell boundary.
	}
	\label{cellmodel}
\end{figure}
				
Based on these constraints, we motivate a form for
$f_\mathrm{IT}$ that respects these limits; details are to be found in the Appendix. We call it the {\it cell model},
because one way to interpret it is that each nucleus is confined to its own
individual volume element, or cell. The derivation of this model begins with the
classical partition function of a particle in a harmonic potential. If evaluated
normally by integrating over position and momentum degrees of freedom, this
partition function (through the relation $f_\mathrm{IT}= -k_{B}T\log Z_\mathrm{IT}$) yields a
Mie-Gr\"uneisen free energy, like the one of Eq.\ref{CW}. But instead, we
perform the integral over the positional degrees of freedom only over a sphere
of radius $R$, thereby confining the representative particle to its cell. This
results in 
\begin{equation}\label{cell1} f_{\rm IT}=
	-3k_{\rm B} T\log\left[\frac{T}{\theta}\right] - k_{\rm B} T\log\left[{\rm
			erf}\left(\sqrt{\frac{T^{*}}{T}}\right) -
\frac{2}{\sqrt{\pi}}\sqrt{\frac{T^*}{T}}\mathrm e^{-T^*/T}\right], \end{equation} with
\begin{equation}\label{Tstar} k_\mathrm BT^*= \frac{mk_\mathrm B^{2}\theta^2
		R^2}{2\hbar^2}, \end{equation} where $\theta$ in these equations represents the
volume-dependent Mie-Gr\"uneisen characteristic temperature arising from the potential well
curvature and $k_\mathrm{B}T^{*}$ is roughly the energy at which the parabolic potential
meets the cell wall (see Fig.\ref{cellmodel}). This form for $f_{\rm IT}$ has
the property that $C_V^\text{IT}\to 3k_{\rm B}$ (classical harmonic value)
for $T \ll T^*$, and $C_V^\text{ion}\to\frac{3}{2}k_{\rm B}$ for $T \gg T^*$
(ideal gas value). Judicious choice of the $V$-dependence of $R$, and hence
$T^{*}$, then guarantees that the ideal gas pressure, $ k_{\rm B}T/V$, is
reached as $T\to\infty$. Similar choices can also be made which guarantee that
the ideal gas entropy is reached. These details can be found in the Appendix.
The final step is to introduce the quantum behavior ($C_{V}^{\rm ion}= 0$ at $T=
0$) by simply replacing the Mie-Gr\"uneisen term in Eq.\ref{cell1} by the Debye
model free energy. In this sense, the second term of Eq.\ref{cell1}, responsible
for the high-$T$ limiting behavior of the thermodynamics, appears as an addition
to the Debye model free energy, \begin{equation}\label{cell2} f_{\rm IT}= f_{\rm
	Debye} + f_{\rm cell}. \end{equation} Note that this model, although reasonable
across a wide range of conditions, does not pretend to describe all the details
of condensed phases at elevated temperatures; we expect it to apply best when
the nuclei are in essentially unstructured configurations, as is assumed for our
idealized atomic-fluid phase. Note also that the detailed dependence of
$C_{V}^{\rm ion}$ on $T$ can, in principle, be altered to match ab initio
simulation or experimental data by further controlling the manner in which the
quadratic piece of the potential meets the hard walls of Fig.\ref{cellmodel}.
Such alterations would likely prevent a model to have an analytical expression, however,
and we will see below that the simplest approach we have outlined here suffices
for our purposes \cite{caveat5}.

\begin{figure}
	\includegraphics[scale=0.66]{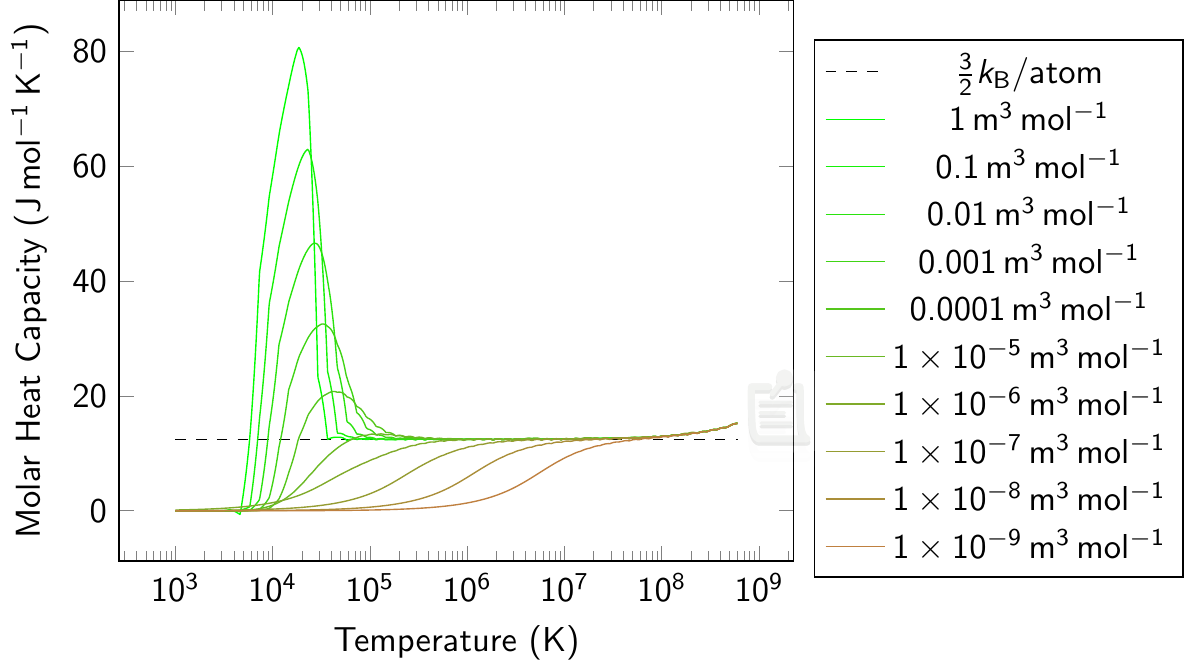}
	\caption{Electronic heat capacity in the atomic fluid, as obtained from the \textsc{Purgatorio} method. Positive pressure values (left) and negative (binding) pressures (right).
		This partial heat capacity is defined as $C_V^\text{Purga} = -T (\partial^2 F^\text{Purga}/\partial V^2)_T$ and does not contain the ion thermal part. Electronic shell structure is evidenced by the peak in the heat capacity related to atomic ionization. High temperature deviations from the ideal gas specific heat ($3/2 k_\text{B}$ per electron or per atom) is a consequence of the treatment of relativistic effects within \textsc{Purgatorio}.  
	}
	\label{fig:purgacv}
\end{figure}

\begin{figure}
	\includegraphics[scale=0.66]{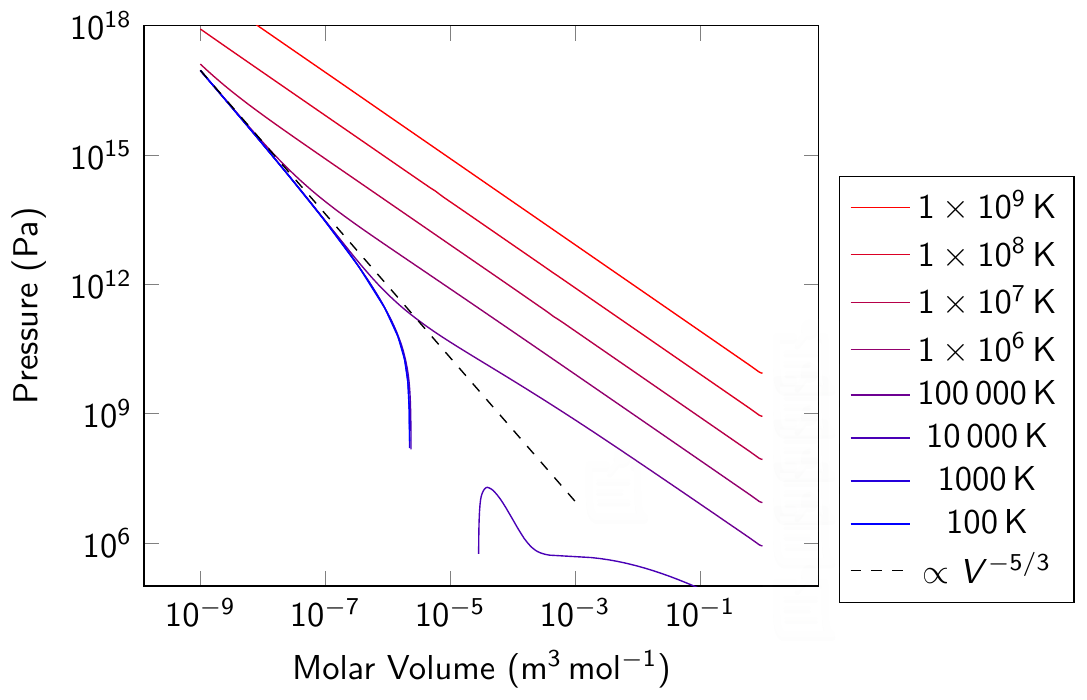}
	\includegraphics[scale=0.66]{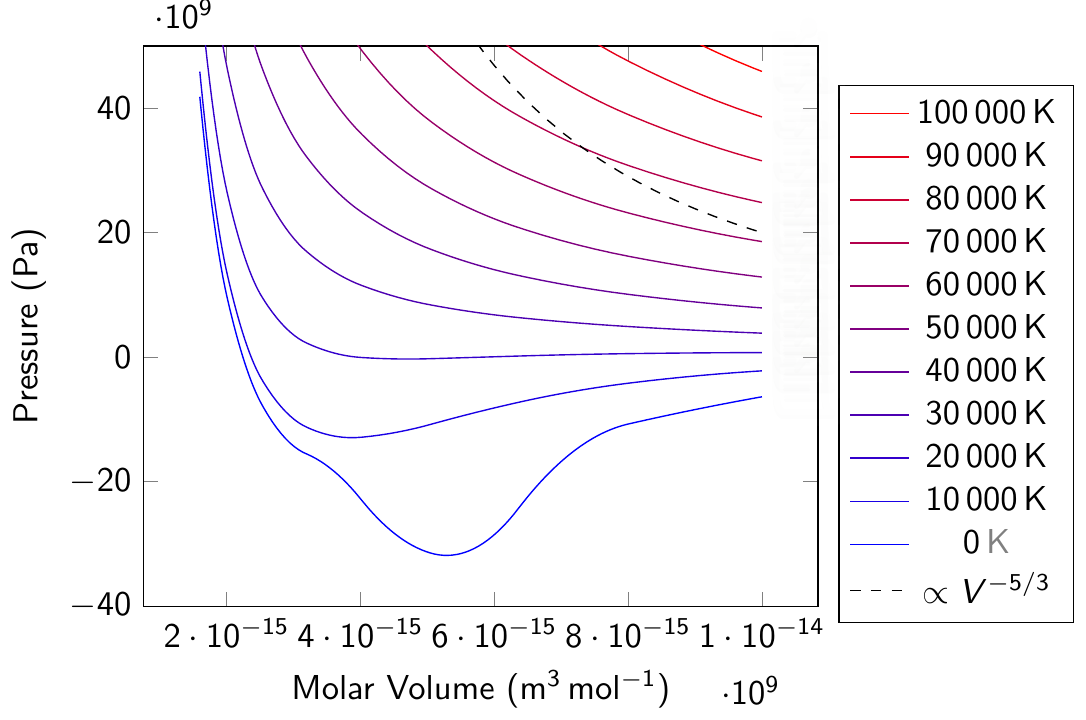}
	\caption{Electronic pressure isotherms in the atomic fluid, as obtained from the \textsc{Purgatorio} method. Positive pressure values (left) and negative (binding) pressures (right).
		This partial pressure is defined as $P^\text{Purga} = -\partial F^\text{Purga}/\partial V|_T$ and does not contain the ion thermal part. Electronic shell structure considered in \textsc{Purgatorio} provides the binding mechanism characteristic of condensed matter and it enters into the model as a negative contribution to the total pressure and compressibility. Most of the negative pressure region however is (in the final equation of state) dominated by the molecular fluid phase (which does not contain important electronic contributions). 
		The Thomas-Fermi ($T=0$) pressure for free electron gas is shown for comparison (dashed line), note in particular the high density limit.
	}
	\label{fig:purgaptv}
\end{figure}

For the rest of the free energy of the atomic liquid, i.e. the combined `cold' and electron-thermal (ET) terms, we employ a DFT atom-in-jellium model known as
\textsc{Purgatorio}\cite{Purgatorio}, which is an updated implementation of the
\textsc{Inferno} model pioneered by Liberman \cite{INFERNO}. In these approaches
\cite{caveat6}, a single atom is placed in a neutral spherical cell, the volume
of which is determined by the density. Outside the spherical cell, the density
is taken to be strictly uniform 
\pdfcomment{(THE DENSITY OF WHAT???)}. 
The finite temperature relativistic Kohn-Sham
\cite{KohnSham} (KS) equations for electrons within a Local
Density Approximation (Hedin-Lundqvist form of the exchange-correlation --XC-functional-- potential \cite{HL}) are solved self-consistently, and
the resulting electronic free energy is computed. (Where the electronic entropy is obtained from single particle occupations.) Since the method exploits
spherical symmetry, it is possible to perform the computations even at extreme
temperatures (in excess of billions of Kelvin), all while avoiding the prohibitively large number of explicit KS states needed in a more conventional condensed matter calculations. \textsc{Purgatorio} is well-suited to the
monoatomic phase for several reasons: i)~The average-atom representation is
appropriate for the random unstructured configurations of the atomic phase at high-$T$.
ii)~The low-temperature Thomas-Fermi high density limit \cite{Hora} is
reproduced exactly. iii)~Contrary to Thomas-Fermi, the explicit consideration of single-particle orbitals in the theory allow for atomic shell structure, which in turn
creates binding, a characteristic of condensed matter (Fig.~\ref{fig:purgaptv}). 
iv)~The proper high-$T$
electron ideal gas limit is recovered (e.g. $C_V^\text{ET} \to \frac{3}{2} \frac{k_\mathrm
	B}{\text{electron}}$ for nonrelativistic; and $C_V^\text{ET} \to 3\frac{k_{B}}{\text{electron}}$ for ultrarelativistic electrons\cite{Hora}) (Fig.~\ref{fig:purgacv}). 

A possible deficiency of \textsc{Purgatorio} is the electron self-interaction problem that ultimately affects the ionization in the atomic (low density) regime, plus the fact that an inherently $T=0$ XC-functional is utilized at all temperatures.

The \textsc{Purgatorio} electronic free energy does not have an analytic expression,
but gives a deterministic and smooth result in a virtually unbounded range of densities and temperatures and without introducing extra parameters. The complexity of the electronic problem amerits the use of this semianalytic portion of the model.
Thus, we interpolate a high resolution table of
\textsc{Purgatorio} free energies for hydrogen ($Z=1$), and the resulting contribution is used as
a combined $f_\text{cold + ET} = f^\text{Purga}$ term for the atomic liquid.

\subsubsection{Molecular liquid}

The free energy for the molecular liquid is once again constructed from
individual pieces, as per Eq.~\ref{eq:coldITET}, however here we discard the
electron-thermal term (ET) altogether. This is partly because an assembly of
$\mathrm H_2$ molecules possesses only rather high-lying electronic excitations,
and more importantly, even the dense $\mathrm H_2$ fluid is known to be
insulating\cite{RMP}. Mostly, however, we set the molecular ET term to zero
because electronic excitations are already included in the atomic liquid, and
the full free energy of liquid hydrogen will be assembled by mixing atomic and
molecular models together (see below)\cite{caveat7}. This same choice regarding
the ET term of the molecular liquid was made in the most recent hydrogen EOS model of Kerley\cite{Kerley03}. Unlike for the atomic liquid (where the cold piece was subsumed into the
electron-thermal term with the \textsc{Purgatorio} model, see later), we include here an explicit
cold curve, again in the form of Eq.~\ref{Vinet}. 

For the ion-thermal contribution, we include both intermolecular and intramolecular terms. 
Our intermolecular free energy is the sum of a CW liquid term, and a cell model that is scaled from that
written in Eq.\ref{cell1} and \ref{Tstar} by identifying the center of mass of
the $\mathrm{H}_2$ molecule as the unit that constitutes the fluid state.
The intramolecular partition function has two pieces: vibrational and rotational. In this work we simply assume decoupled vibrations and rotations, though in reality, vibrations and rotations are coupled; sophisticated theories can be employed which take into account coupled roton-vibron states\cite{Kerley03}. However, even this more sophisticated description is rendered inaccurate at high densities, due to strong inter-molecular coupling. In our simplified description, we do account for the propensity for a molecule to dissociate when it is highly excited, albeit in an approximate way. For the vibrational contribution to the molecular liquid free energy, we use
\begin{equation}\label{vibrational}
	f^{\rm vib} = \frac{1}{4}\frac{\hbar \omega}{\text{atom}} + \frac{1}{2}\frac{k_{\rm B}T}{\text{atom}}\log\left[1 - \exp\left(\frac{-\hbar\omega}{k_{\rm B}T}\right)\right] - 2\frac{k_{\rm B}T}{\text{atom}}\log\left[{\rm erf}\left(\sqrt{\frac{T^{\rm v}}{T}}\right)\right],
\end{equation}
where $\omega= 7.94\times 10^{14}$ Hz, and $T^{\rm v}$= 51100 K. The first two terms are the result of summing over an infinite number of 1D harmonic oscillator states, for an oscillator with angular frequency $\omega$. The third term is an {\it approximate} correction to this ideal harmonic oscillator contribution, which takes into account the fact that above some temperature, the real molecules will be sufficiently excited to exhibit behavior which deviates from the purely harmonic (ultimately leading to molecular dissociation, considered by the transition to the atomic model, see below). This term is derived from a generalization to the 1D version of the cell model mentioned above, as presented in Sections~\ref{sec:cell1d} and \ref{sec:cellmoment} of the Appendix. The values of $\omega$ and $T^{\rm v}$ were chosen by considering the effective inter-H potential derived from detailed coupled-cluster calculations for an isolated H$_{2}$ molecule \cite{Kolos69}. For our molecular rotation free energy, we use 
\begin{equation}\label{rotational}
	f^{\rm rot}= -\frac{k_{\rm B}T}{2\text{atom}}\log\left[\sum_{\ell= 0}^{\ell_{\rm max}}(2\ell + 1)\exp\left(\frac{-\ell(\ell+1)B}{k_{\rm B}T}\right)\right],
\end{equation}
where $B= \frac{\hbar^{2}}{2I}$, with $I= 4.61 \times 10^{-48}$ kg m$^2$. Here, the maximum rotational quantum number allowed, $\ell_{\rm max}$, is taken to be a free parameter to be optimized in the course of fitting to data (see below). Its value is expected to depend on density \cite{Kerley03}, due to the increased propensity for H$_{2}$ to dissociate in a dense environment \cite{RMP}; we allow for only a single density-independent value which necessarily averages over the behavior throughout a range of densities. The value of the molecular moment-of-inertia, $I$, is taken from the molecular distance calculated by W.~Kolos and L.~Wolniewicz\cite{Kolos69}. 


As is the case for the other phases, it is useful to have the molecular liquid defined for all thermodynamic conditions. Again, phase space arguments show that the IT heat
capacity has a limiting value of $\frac{3}{2} k_\text{B}/\text{molecule}$. (The
contribution from internal degrees of freedom is absent in the high
temperature limit as for both we have defined a cutoff). This ensures that it will be more stable than the
solid (with $C_V\to 0$, using the cell model correction we imposed for the solid; see above) but less stable than a
monoatomic gas at some high temperature ($C_V\to \frac{3}{2} k_\text{B}/\text{atom}$).
This further clarifies the reasons for choosing 
the high-$T$ limits of the specific heats for each phase as we have \pdfcomment{(Fig.~\ref{fig:limitcv})}.

\subsubsection{Fluid Mixing Model}

In our picture, the atomic and molecular liquid free energies can be regarded as
representing individual phases; indeed, in the appropriate limits, they each
provide a reasonable description of liquid hydrogen. A robust description of liquid
hydrogen throughout a wide range of density and temperature must, however,
involves a dynamic mixture of atomic and molecular states. Therefore these primitive models presented so far must be combined somehow to provide a description of the mixed regime. 

\pdfmarkupcomment{Following earlier
	work}{This downplays our contribution, we have yet to find if indeed Kerley's
	mixing is in some degree similar to this, certainly not in Kerley72, I'll keep
	looking} \cite{Kerley72,Kerley72pub,Kerley03}, we construct a free energy of the
mixture as a suitable combination of atomic liquid and molecular liquid free
energies. The justification for this on statistical mechanical grounds parallels
the development of the Saha model\cite{Hora}, which treats ideal gases of
mixtures of atoms in different states of excitation. The details of our mixing
model are found in the Appendix; here we review its most important features and
assumptions.

We begin by proposing that the partition function for the mixture is a simple
product of individual atomic and molecular partition functions for $M$ {\it
	independent} molecules, and $A$ {\it independent} atoms.
\begin{equation}\label{Zmix} Z_{\rm mix}(A,M,V)= \frac{z_{\rm
		M}(V)^{M}}{M!}\frac{z_{\rm A}(V)^{A}}{A!}, \end{equation} where $z_{\rm M}(V)$
and $z_{\rm A}(V)$ are given molecular and atomic partition functions,
respectively. The numbers of molecules and atoms can vary subject to the
constraint, \begin{equation}\label{constraint} 2M + A= 2M_{0}= A_{0}.
\end{equation}
The minimum of the mixture free energy (which corresponds to the maximum in $Z_{\rm mix}$) is attained for specific values of $A$ and $M$ subject to this constraint. This optimal free energy for the mixture is given by (see the Appendix for details): 
\begin{equation}\label{mix1}
	f_{\rm mix}= (1 - x)f_{\rm M} + xf_{\rm A} + \frac{k_{B}T}{\text{atom}}\left[\frac{1}{2}(1 - x)\log\left[(1 - x)\right] + x\log(x)\right],
\end{equation}
where
\begin{equation}
	x= \frac{1}{2}\exp\left[\frac{-2(f_{\rm A} - f_{\rm M})}{\text{atom} k_\mathrm{B}T}\right] \left(\sqrt{1 + 4\exp\left[\frac{2(f_{\rm A} - f_{\rm M})}{\text{atom} k_\mathrm{B}T}+1\right]} - 1\right).
\end{equation}
The variable $0 \leq x \leq 1$ can be viewed as a variational parameter which controls the admixture of
atomic liquid and molecular liquid pieces in the total fluid free energy. This
parameter depends on both $V$ and $T$ and is a function of the difference
between atomic and molecular liquid free energies (per atom) relative to the temperature scale. We take these
primitive free energies, $f_{\rm M}(V,T)$ and $f_{\rm A}(V,T)$, to be those
determined by the models described in the previous two subsections. Three things
are worth noting: 1) The mixing parameter, $x$, depends on both $V$ and $T$, so
thermodynamic functions derived from $f_{\rm mix}$, such as $P_{\rm mix}=
-(\partial f_{\rm mix}/\partial V)_{T}$, have contributions resulting from this
dependence. This means, for instance, that $P_{\rm mix}$ is {\it not} in general
equal to $(1 - x)P_{\rm M} + xP_{\rm A}$, etc. 2) Though the individual atomic
and molecular liquid models have been constructed using the paradigm of
Eq.~\ref{eq:coldITET}, the $V$ and $T$ dependence of $x$ makes such a
decomposition invalid for $f_{\rm mix}$. 
3) No additional free parameters have been added at this stage of the construction.

It bears repeating that this mixing prescription is necessarily suspect for
dense systems, because the major assumption embodied in the above model is that,
for instance, the presence of atoms does not affect the statistical properties
of the molecules. If the system described is a low-density gas, this is
justified and the mixing prescription reduces precisely to that of the Saha
model\cite{Hora}. Much of our interest in hydrogen EOS is in regimes where this
is clearly not the case. Nevertheless, we use this model as a base for the mixing at any rate; at
low-$\rho$ (where we have little or no ab initio simulation data) we are
confident of its validity, and at high-$\rho$ we force the EOS to be
essentially determined by the results of simulations which do \emph{not} invoke
these chemical equilibrium mixing assumptions. Though the use of this Saha-like picture is a gross simplification for dense systems, it has the advantage of affecting a continuous transition between the atomic and molecular liquid \cite{Kerley72,Kerley72pub,Kerley03} (which is the case at high temperatures). In the next
section, we relax the assumption that the molecules and atoms are statistically independent, thereby admitting a description in which the transition is not everywhere continuous (which is the case at low temperatures).



\subsubsection{Critical Fluid Model}

The use of the aforementioned mixing model gives us a practical way to describe the
continuous transition between the atomic and molecular liquids, preserving
not only the correct limits, but also the expected molecular and
monoatomic behaviors in conditions where each one is expected to be the stable
state. However, in light of recent theoretical results, it fails to describe
the liquid in conditions where the ideal molecular {\it and} atomic phases are
{\it both} competitive (thermodynamically stable) at low temperature. These recent simulation results 
\cite{MoralesPNAS,Scandolo2003} show with great confidence that there is a
liquid-liquid phase transition in the region defined by $T < 2000~\mathrm K$ and
and $V \sim 1.3\times 10^{-6}\mathrm{m^3/mol}$ or $P\sim 100-150~\mathrm{GPa}$), ending in a
critical point. The presence of this remarkable feature in the 
liquid indicates that there is a cooperative phenomenon which is ignored in the
treatment of ideal mixing, as we have presented above.
Qualitatively, this cooperative nature necessarily results from some
molecular-atomic interaction (coupling) that locally favors the occurrence of like-species (molecule-molecule and atom-atom) in the mixed liquid. The cooperative
tendency works to stabilize molecular-rich and atomic-rich liquids, respectively,  on either 
sides of the transition line. It is in this sense that the assumption of the statistical independence of molecules and atoms is necessarely violated. As with all cooperative phenomena, these additions give
rise to special types of thermodynamic critical points, with associated characteristic features in the
specific heat and other susceptibilities.

In the Appendix we give the details of the derivation of a mean field (MF)
version of this cooperative mixing model and its free energy, that allows us to
constrain the location of a critical point in agreement with simulations \cite{MoralesPNAS}\cite{Scandolo2003}. We approximate the physics in the
neighborhood of the critical point by introducing a coupling parameter, $J$, 
which gives rise to the following {\it non}-ideal mixing free energy (to be compared with Eq.\ref{mix1}):
\begin{equation}\label{solfmix}
	f^\mathrm{MF}_\mathrm{mix} = \min_x \left\{(1-x) [f_\mathrm M + J x] + x [f_\mathrm A + J(1-x)] + \frac{k_\mathrm B T}{\text{atom}} \left[(1-x) \log(1-x)/2 + x \log(x)\right]\right\},
\end{equation}
where again, $x$ is obtained by minimizing the free energy. 
This variational parameter now fullfills a {\it self-consistent} equation:
\begin{equation}\label{solxscf}
	x = \mathrm e^{-2 \beta[\Delta f + 2J(1-2x)] - 1} \left(\sqrt{1 + 4 \mathrm
		e^{2\beta [\Delta f + 2J(1-2x)]+1}} - 1\right). \end{equation} 
Where $\Delta f = f_\mathrm{M} - f_\mathrm{A}$.
The parameter $J$ can be interpreted as a pairwise coupling related to the \emph{average}
energy cost of having a molecule surrounded by all neighboring atoms, 
relative to the pure molecular (atomic)
configuration (and viseversa). A vanishing $J$ reduces the model to that of the ideal
case of the previous section.

$J$ is a single parameter (possibly dependent on density and temperature) introduced to
model the interactions between species of the mixture. Its value could, in principle, be determined
from simulations by computing the free energy cost of replacing a molecule by two
un-bonded atoms in an otherwise pure molecular fluid (or vice versa). In our work, we choose instead to determine $J$ by
fitting to liquid hydrogen simulation results for EOS (in particular, $P$ vs. $V$ at fixed $T$) which show clear signs of the critical point \cite{MoralesPNAS,Scandolo2003}. It is the value of this coupling parameter that will determine the location of the critical point. In order to capture the fact that the coupling should vanish in the dilute limit, we choose the parameterization:
\begin{equation}
	J(V) = J_0 \mathrm e^{-V/V_0}.
\end{equation}
The precise form is not especially crucial as long as the value of $J$ is of the appropriate magnitude in the neighborhood of the critical line (see the discussion of this near the end of the Section~\ref{sec:fitting}).

\section{Fitting}\label{sec:fitting}

The parameters of the various free energy models outlined in Section~\ref{sec:models} are fit to the simulation data obtained by the methods described in Section II.  Most of these data consist of pressure and internal energy at various densities and temperatures. Entropy is also available for a small number of conditions, as obtained by thermodynamic integration using potential-switching techniques (so-called '$\lambda$-integration'). Our fitting procedure also uses, where it is deemed appropriate, phase transition lines (such as the melt line: $T_{\rm melt}$ vs. $P$). For this reason,
low temperature (solid) phases are fit first; the higher-$T$ phases
are then fit to both single-phase data and the constraints given by the transition lines separating the high-$T$ and low-$T$ phases. 

The fitting procedure consists of minimizing the residuals of the model as compared to the
data, relative to the error intrinsic to the data itself. This error can
represent either a systematic error of the theoretical method, a statistical (simulation) error, or
an experimental error, depending on the context. In this work, the goodness of a
given fit is represented mathematically by the following dimensionless residual quantity to be
minimized as a function of the parameters, $\{p\}$, of the model:
\begin{subequations}
	\begin{equation}\label{eq:minparam}
		\chi^2 = \min_{\{p\}}
		\tfrac{\sum_i \left| \tfrac{P_{\{p\}}(V_i, T_i) - P_i}{P^\text{err}_i} \right|^2}{N_\text{P}}
		+	\tfrac{\sum_i \left| \tfrac{U_{\{p\}}(V_i, T_i) - U_i}{U^\text{err}_i} \right|^2}{N_\text{U}}
		+	\tfrac{\sum_i \left| \tfrac{S_{\{p\}}(V_i, T_i) - S_i}{S^\text{err}_i} \right|^2}{N_\text{S}}
		+   \cdots,
	\end{equation}
	where $N_\text{P, U, S}$ is the number of data points available (e.g. from simulations) for that thermodynamic variable, 
	$P_i$, $U_i$, and $S_i$ are the values obtained from simulations at conditions 
	$(V_i, T_i)$, the denominators in Eq.~\ref{eq:minparam} are the errors or uncertainties assigned to the data.
	Note that the parameter-dependent \emph{functions} $P$, $U$, $S$ and $G$ are not independent of each other, as all are obtained from the same free energy function $F$ of a given phase (e.g. a particular model for $F$). 
								
	The above is the full expression used to fit the molecular solid phase. 
	For a high-$T$ phase, since the fitting involves phase lines connecting this phase with a lower-$T$ phase, we generalize Eq.~\ref{eq:minparam} to include information pertaining to other phases:
	\begin{equation}\label{eq:minparamg}
		\cdots + \tfrac{\sum_i \left| \tfrac{F_{\{p\}}(T^\circ_i, V^\circ_i) - F^\circ_i}{F^\text{err}_i} \right|^2}{N_\text{F}}
		+ \tfrac{\sum_i \left| \tfrac{G_{\{p\}}(P^\circ_i, T^\circ_i) - G^\circ_i}{G^\text{err}_i} \right|^2}{N_\text{G}}.
	\end{equation}
\end{subequations}

Here, $F^\circ_i$ is data obtained (or modeled) from other phases at transition conditions $V^\circ_i, P^\circ_i, T^\circ_i$.
These extra terms allow us to impose additional constraints on the relative stability of the phase in question with respect to other phases.
In this way, we are able to constrain a specific phase transition line at constant pressure. 

The full minimization of Eq.~\ref{eq:minparam} with respect to all 
parameters is a difficult numerical task. Each phase is characterized by at least
10 model parameters, and the various models are all non-linear functions of these parameters. 
Moreover, the optimization procedure not only involves the values of, say, the functions $F$, 
but also their derivatives as well: $\partial_T F$ and $\partial_V F$. We are thus faced with a multidimensional, multiderivative optimization problem. Myriad numerical techniques are available for problems of this type, however 
the optimal approach depends sensitively on the peculiarities of the
specific problem\cite{JohnsonSOFT}.
Due to the complexity arising in large part from the high dimensionality, we divide the optimization into more manageable chunks by
performing the minimization one phase at a time. 

\begin{figure}
	\includegraphics[scale=0.4]{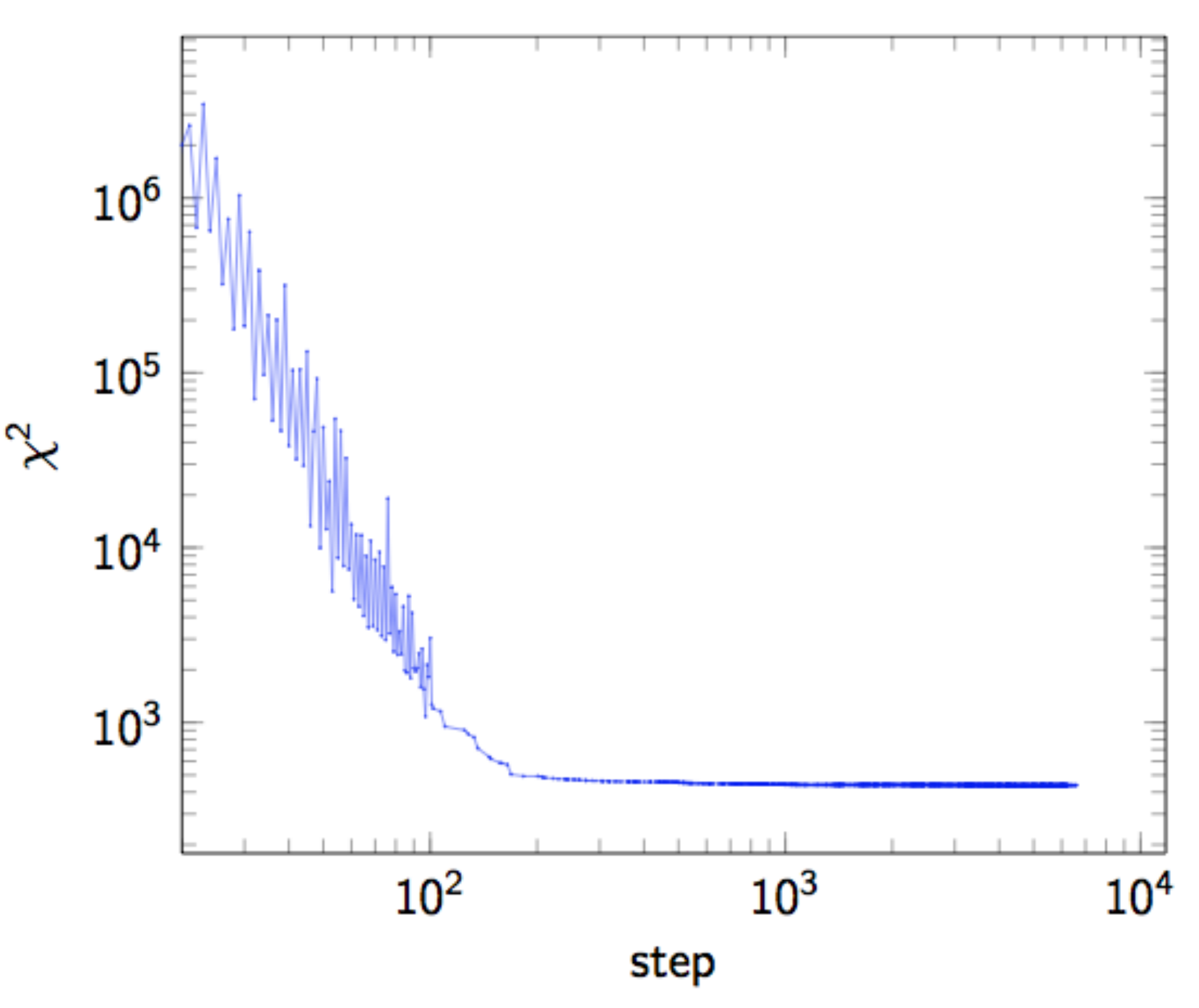}
	\caption{
		THIS FIGURE WILL SHOW A TYPICAL MINIMIZATION!!
		Residual for the optimization process for the determination of the 10 model parameters of the molecular solid phase.
		Typical values are $\chi^2 \sim 400$, therefore the average absolute error,
		relative to the error of the data, is $20$ times the error assigned to the data
		(See the figure captions for the typical error assigned to the data). This value depends
		on the number of parameters of the model; more parameters lower
		this ratio while complicating the model. Note that errors assigned to the data are primarily statistical, and are surely lower than the systematic errors (e.g., due to approximations within DFT).
	}
\end{figure}

\subsection{Molecular Solid}\label{sec:fittingmsolid}

\begin{figure}
	\includegraphics[scale=0.66]{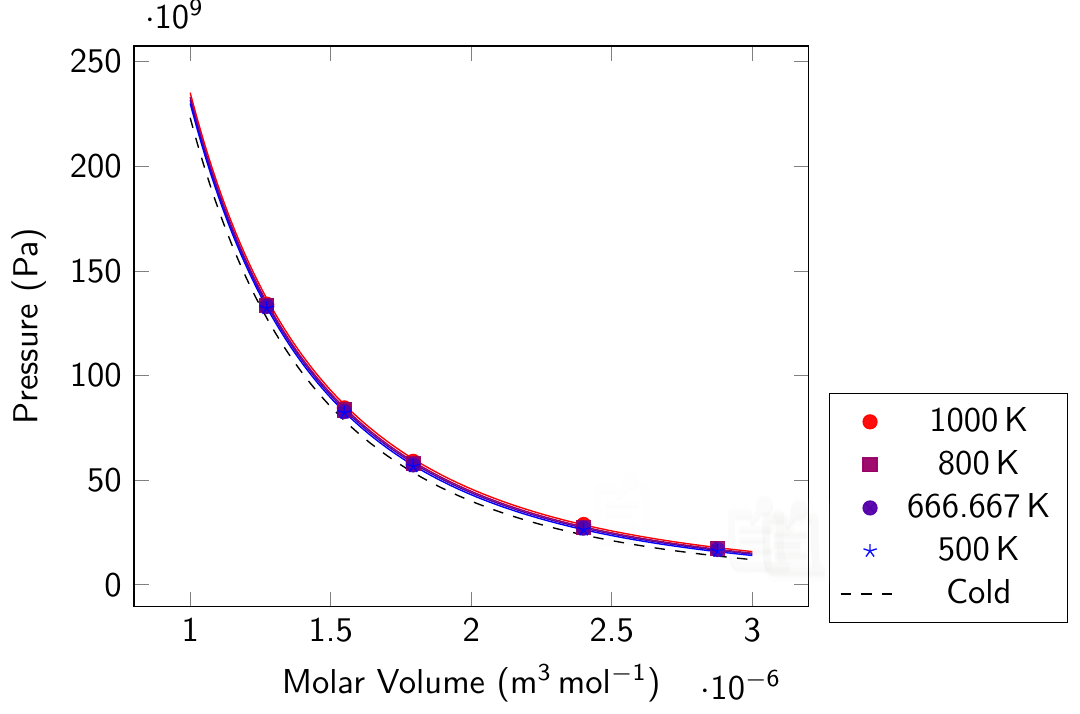}
	\includegraphics[scale=0.66]{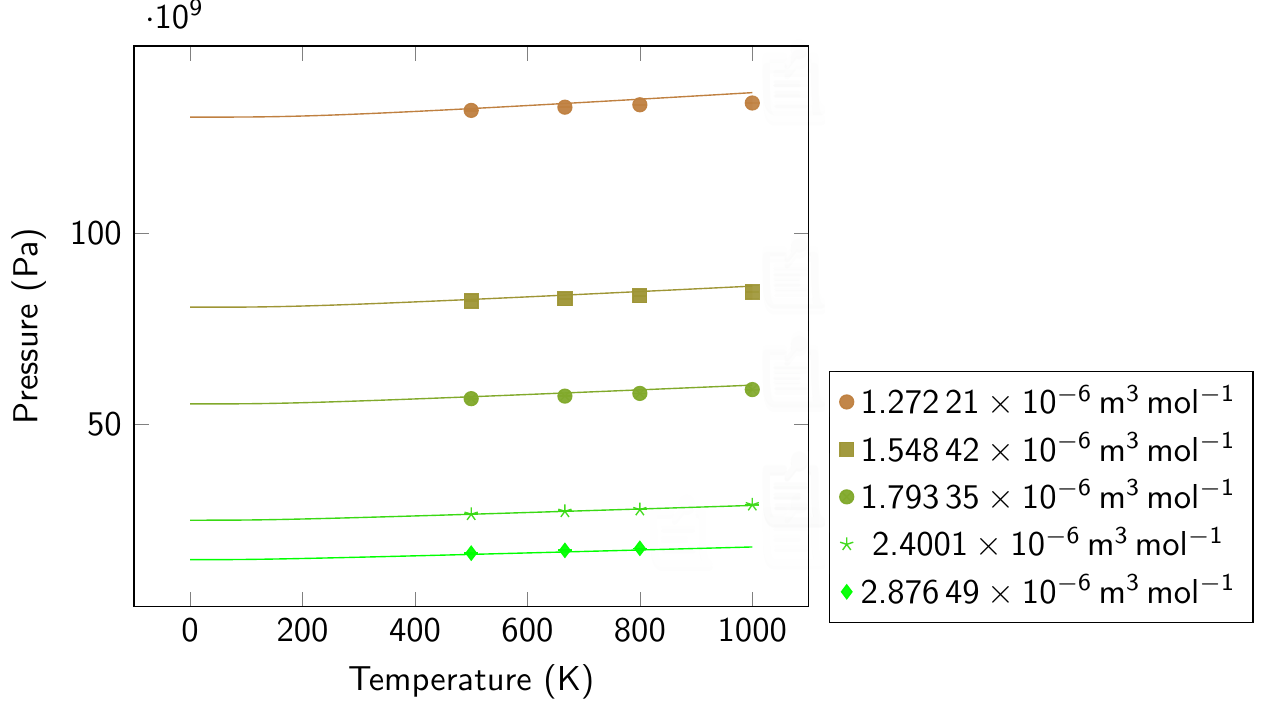}
	\caption{Molecular solid: Pressure isotherms (left) and isochores (right). 
		Symbols represent data obtained with PI-DFT simulations in the high-density solid in the HCP structure.
		Continuous lines represent the results of our EOS model (Eq.~\ref{eq:molsol} and Table~\ref{tab:msol}).
		Typical error bars (due to simulation statistics; not shown) are of the order of $5\times10^7~\mathrm{Pa}$.
		(The cold curve (dahed) in the left panel is the volume-derivative of the temperature-independent terms in Eqs.~\ref{eq:solidfvinet} and \ref{eq:solidftf}, and it does not include the zero-point motion.)
	}
	\label{fig:solidpidftP}
\end{figure}

\begin{figure}
	\includegraphics[scale=0.66]{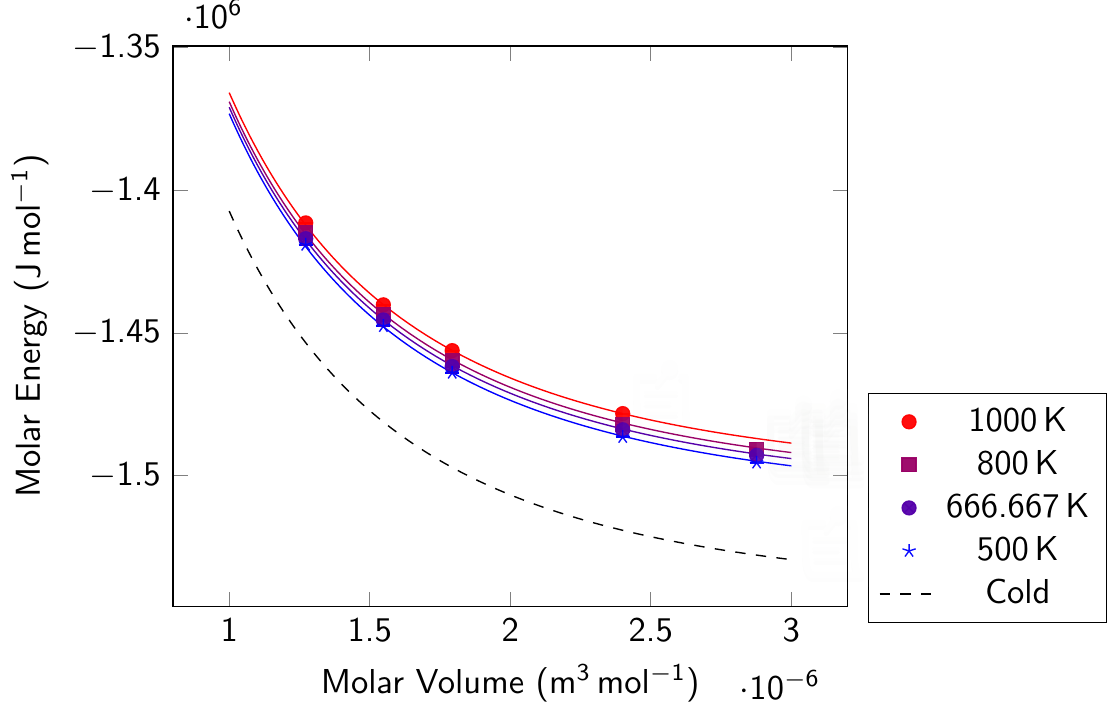}
	\includegraphics[scale=0.66]{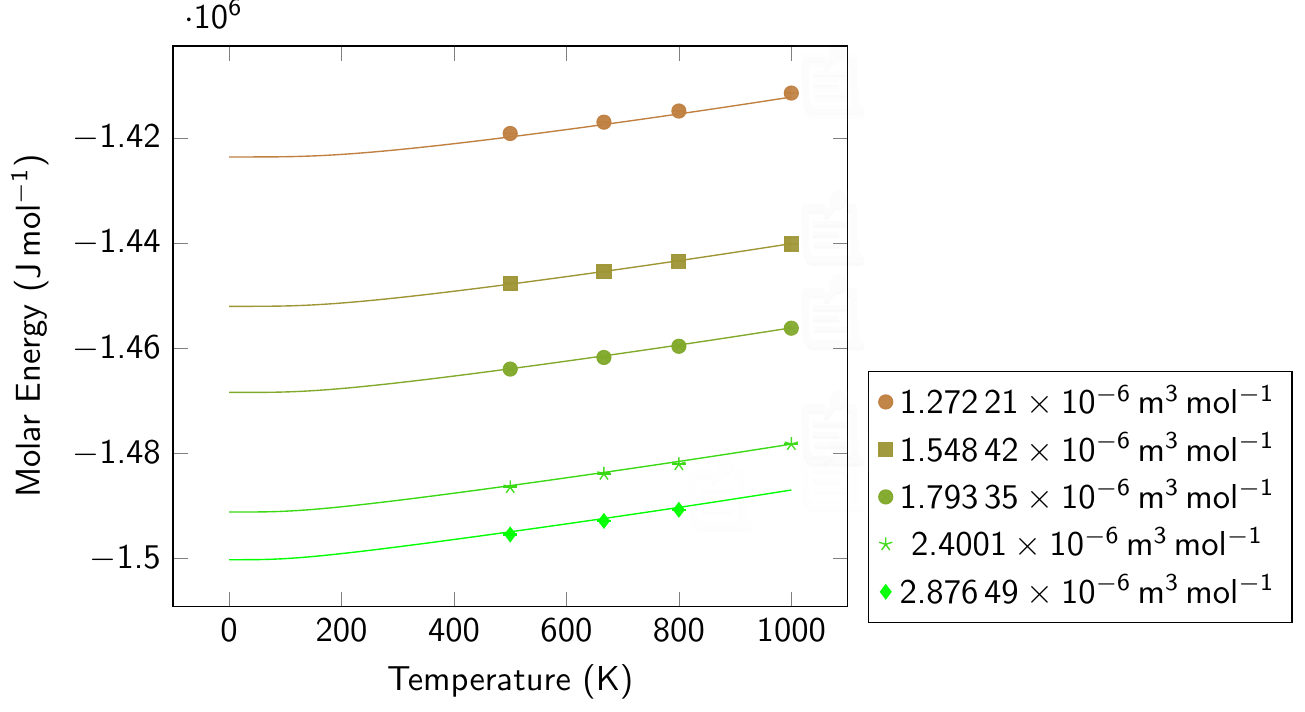}
	\caption{Molecular solid: Interal energy isotherms (left) and isochores (right). 
		Symbols represent data obtained with PI-DFT simulations in the high-density solid in the HCP structure.
		Continuous lines represent the results of our EOS model (Eq.~\ref{eq:molsol} and Table~\ref{tab:msol}).
		Typical error bars (due to simulation statistics; not shown) are of the order of $100~\mathrm{J/mol}$.
		(The cold curve in the left panel is the temperature-independent terms in Eqs.~\ref{eq:solidfvinet} and \ref{eq:solidftf}, and therefore does not include the zero-point energy.)
	}
	\label{fig:solidpidftU}
\end{figure}

The data used for the fitting of the molecular solid is based on PI-DFT
simulations (points in Figures \ref{fig:solidpidftP}, \ref{fig:solidpidftU},
\ref{fig:solidpidftS}), PI Silvera-Goldman potential\cite{Silvera} simulations
(Figures \ref{fig:solidpisgP}, \ref{fig:solidpisgU}) and experimental low
pressure data\cite{Silvera1978} (Figure~\ref{fig:solidexpsgP}). As discussed
in Section III, the solid free energy model is represented by the following detailed expression:

\begin{subequations}\label{eq:molsol}
	\begin{equation}\label{eq:solidfvinet}
		F\mathopen{}\left(V,\, T\right)\mathclose{}=\phi_{0}+\frac{4 V_{0} B_{0}}{\mathopen{}\left(B_{1}-1\right)\mathclose{}^2} \mathopen{}\left(1-\mathopen{}\left(1+X\right)\mathclose{} \exp\mathopen{}\left(-X\right)\mathclose{}\right)\mathclose{}+{} \cdots
	\end{equation}
	(where $X \to \frac{3}{2} \mathopen{}\left(B_{1}-1.\right)\mathclose{} \mathopen{}\left(\sqrt[3]{\frac{V}{V_{0}}}-1\right)\mathclose{}$)
	\begin{equation}\label{eq:solidftf}
		\cdots {}+E_\mathrm{TF} {\mathopen{}\left(\frac{V}{V_\mathrm{TF}}\right)\mathclose{}}^{-2/3} \exp\mathopen{}\left(-{\mathopen{}\left(\frac{V}{V_\mathrm{TF}}\right)\mathclose{}}^{2/3}\right)\mathclose{}+{} \cdots
	\end{equation} 
	\begin{equation}
		\cdots +\frac{\xi_\mathrm{A} k_\mathrm{B}}{\text{atom}} \mathopen{}\left(\frac{9}{8} \theta_\mathrm{A}+3 T \log\mathopen{}\left(1-\exp\mathopen{}\left(\frac{-\theta_\mathrm{A}}{T}\right)\mathclose{}\right)\mathclose{}-T \mathcal{D}_{3}\mathopen{}\left(\frac{\theta_\mathrm{A}}{T}\right)\mathclose{}\right)\mathclose{}+ \cdots
	\end{equation}
	\begin{equation}
		\cdots+\frac{\xi_\mathrm{B} k_\mathrm{B}}{\text{atom}} \mathopen{}\left(\frac{9}{8} \theta_\mathrm{B}+3 T \log\mathopen{}\left(1-\exp\mathopen{}\left(\frac{-\theta_\mathrm{B}}{T}\right)\mathclose{}\right)\mathclose{}-T \mathcal{D}_{3}\mathopen{}\left(\frac{\theta_\mathrm{B}}{T}\right)\mathclose{}\right)\mathclose{}+\cdots
	\end{equation}
	(where $\xi_\mathrm{B}=1-\xi_\mathrm{A}$, $\theta_\mathrm{A}=\theta_\mathrm{A}^0 \mathopen{}\left(\frac{V}{V^0}\right)\mathclose{}^{-\gamma_\mathrm{A}}$)
	\begin{equation}\label{eq:solidcell}
		\cdots -\frac{2 k_\mathrm{B} T}{\text{atom}} \log\mathopen{}\left(\mathrm{erf}\mathopen{}\left(\sqrt{\frac{T^{*}}{T}}\right)\mathclose{}-\frac{2}{\sqrt{\pi}} \sqrt{\frac{T^{*}}{T}} \exp\mathopen{}\left(-\frac{T^{*}}{T}\right)\mathclose{}\right)\mathclose{}
	\end{equation}
	(where $T^{*}=\frac{m_\text{H} k_\mathrm{B} \theta_{0}^{2} R^{2}}{2 \hbar}$, $R=\sqrt[3]{\frac{\frac{3}{4} V}{N_\mathrm{A} \pi}}$, $\log\mathopen{}\left(\theta_{0}\right)\mathclose{}=\xi_\mathrm{A} \log\mathopen{}\left(\theta_\mathrm{A}\right)\mathclose{}+\xi_\mathrm{B} \log\mathopen{}\left(\theta_\mathrm{B}\right)\mathclose{}$, $m_\text{H} = 1.67372\times 10^{-27}~\mathrm{kg}$ is the hydrogen atomic mass.)
								
\end{subequations}

\begin{figure}
	\includegraphics[scale=0.66]{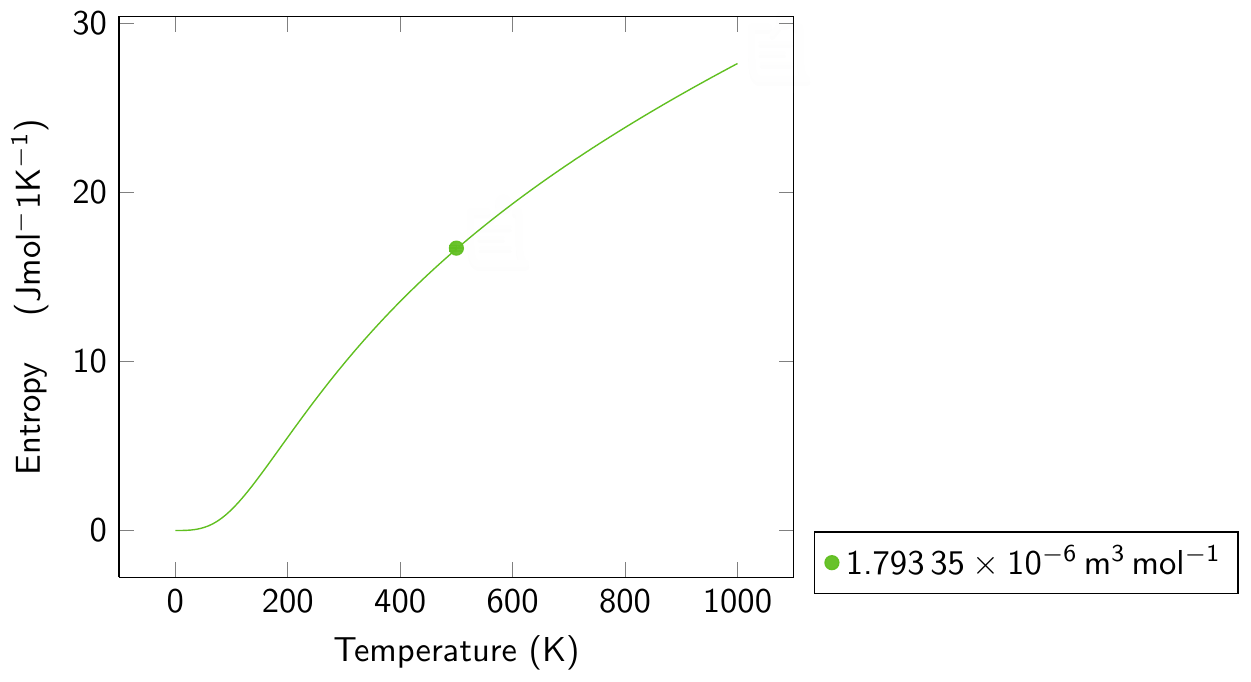}
	\caption{Molecular Solid: Entropy isochore. 
		The single symbol represents an entropy obtained with PI-DFT and a Coupling Constant Integration scheme.
		Continuous lines represent the result of our EOS model (Eq.~\ref{eq:molsol} and Table~\ref{tab:msol}).
		The error bar (due to simulation statistics; not shown) is estimated to be $0.2~\mathrm{J/mol/K}$.
	}
	\label{fig:solidpidftS}
\end{figure}
\begin{figure}
	\includegraphics[scale=0.66]{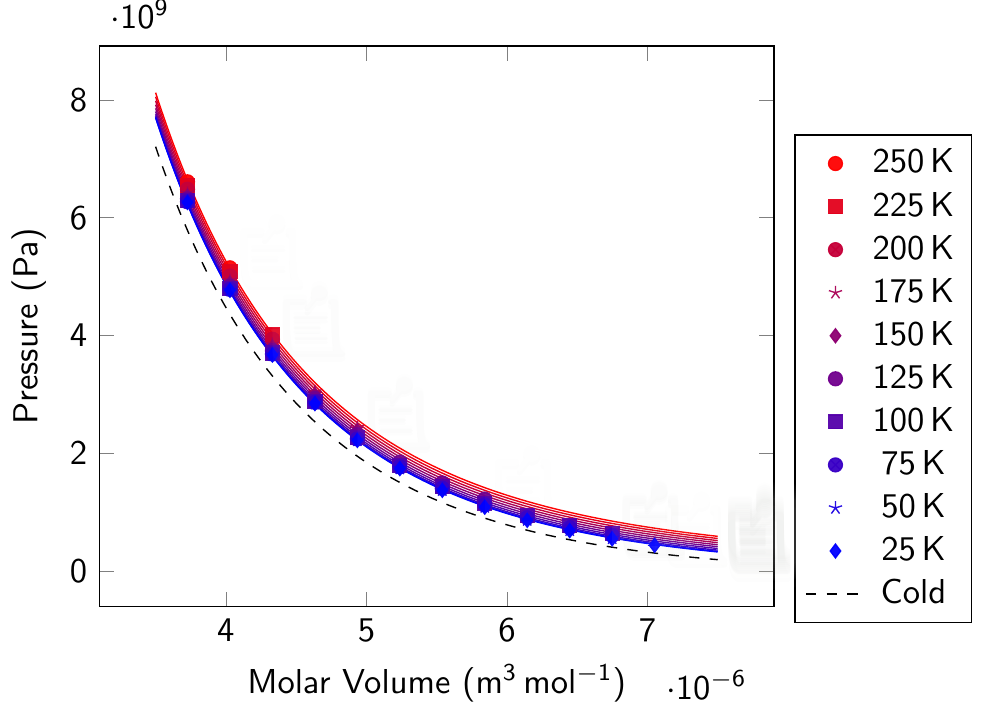}
	\includegraphics[scale=0.66]{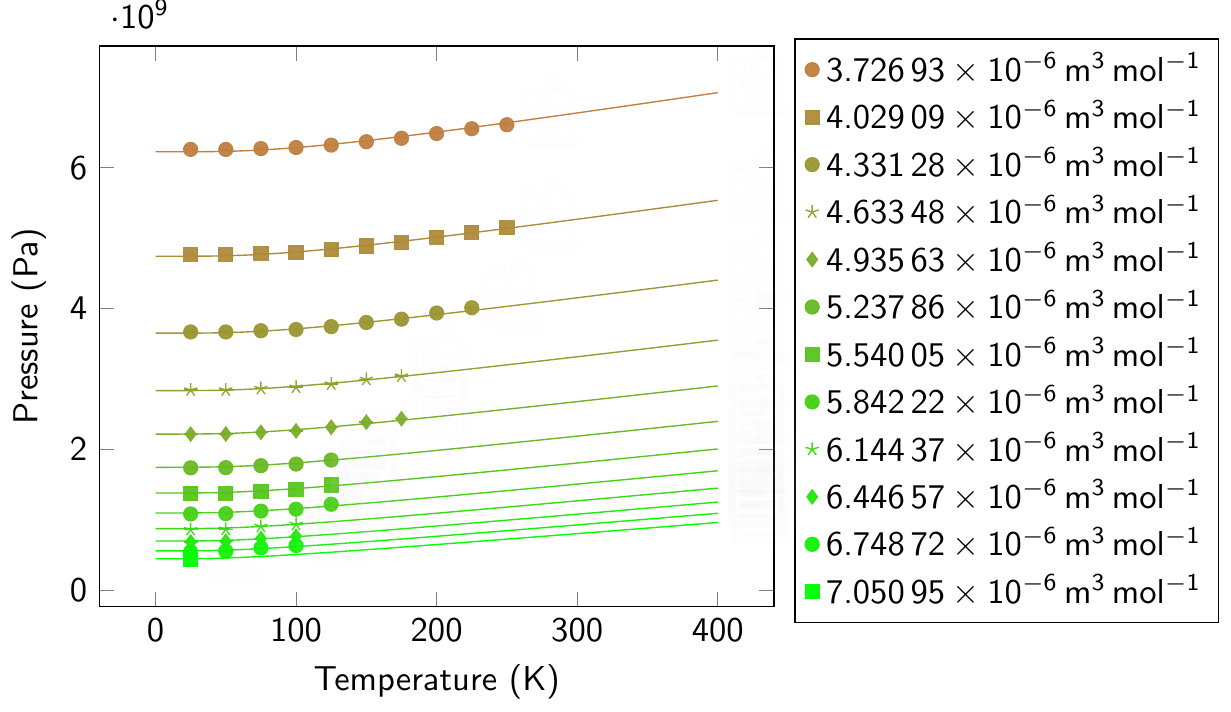}
	\caption{
		Molecular solid: Pressure isotherms (left) and isochores (right). 
		Symbols represent data obtained with PIMD using the Silvera-Goldman inter-molecular potential in the low-density solid HCP structure.
		Continuous lines represent the results from our EOS model (Eq.~\ref{eq:molsol} and Table~\ref{tab:msol}).
		Error bars (due to simulation statistics; not shown) are estimated to be $1\times 10^6~\mathrm{Pa}$.
		The `cold' curve is obtained by volume differentiation
		of the temperature-independent terms in Eqs.~\ref{eq:solidfvinet} and \ref{eq:solidftf}, and therefore does not include the zero-point energy.
	}
	\label{fig:solidpisgP}
\end{figure}
\begin{figure}
	\includegraphics[scale=0.66]{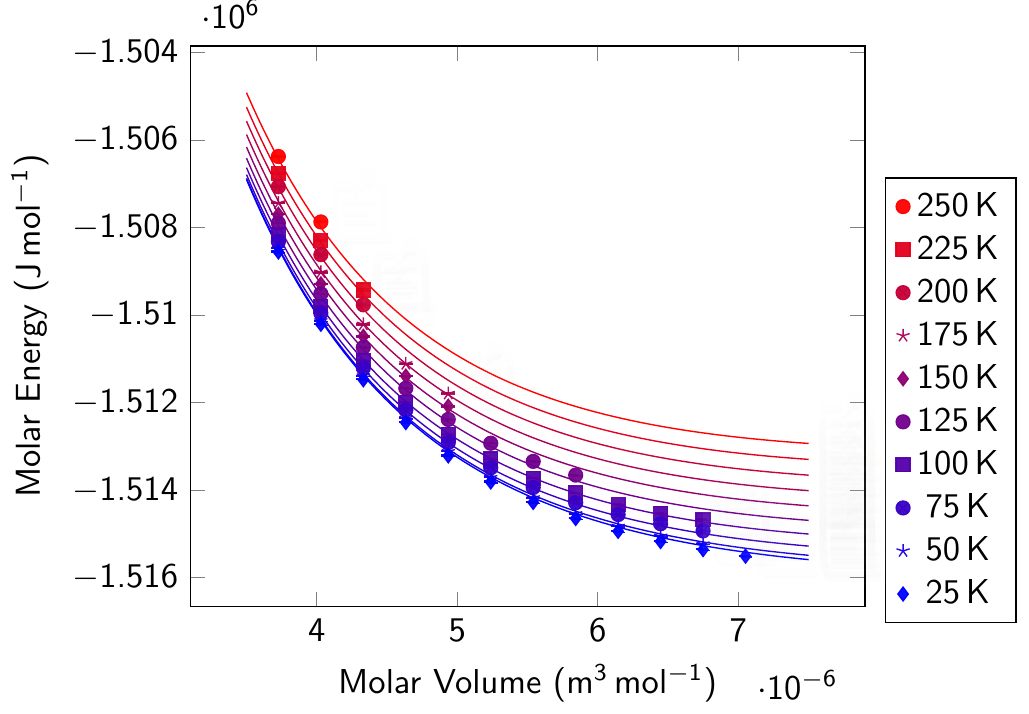}
	\includegraphics[scale=0.66]{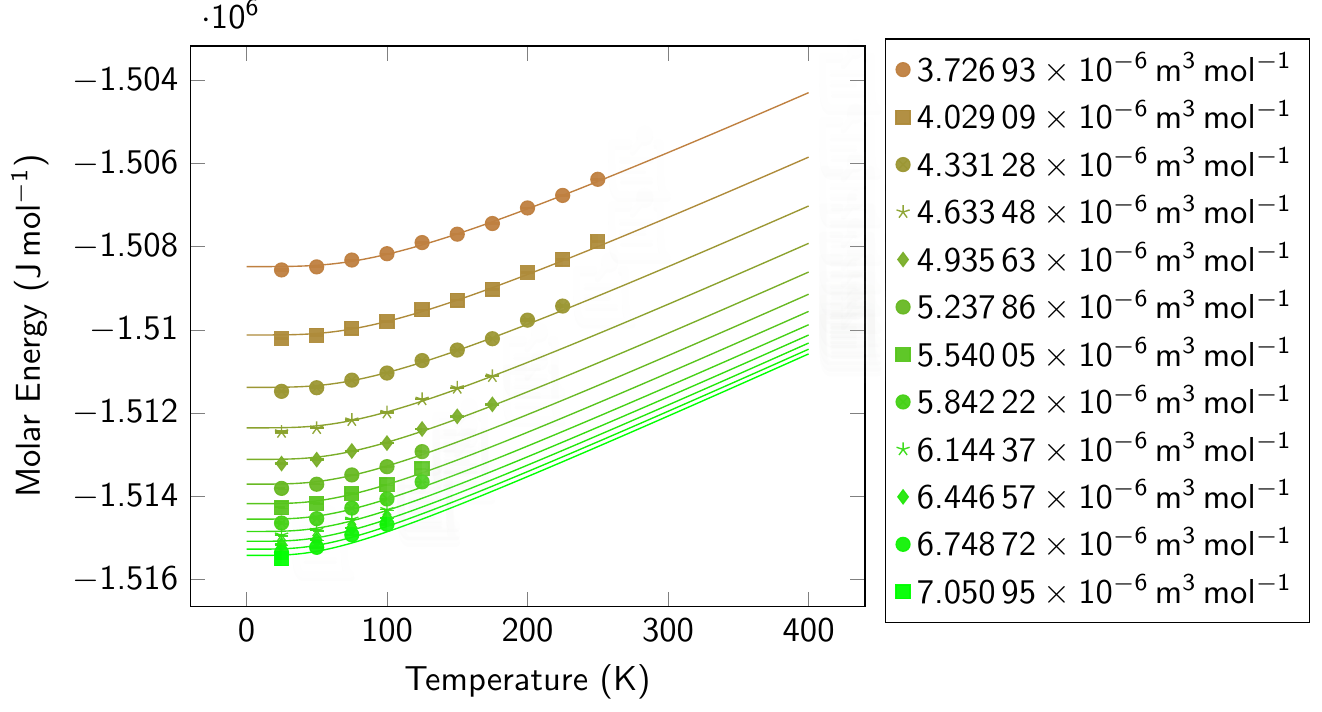}
	\caption{Molecular solid: Internal energy isotherms (left) and isochores (right). 
		Symbols represent data obtained with PIMD using the Silvera-Goldman inter-molecular potential in the low-density solid HCP structure.
		Continuous lines represent the results of our EOS model (Eq.~\ref{eq:molsol} and Table~\ref{tab:msol}).
		Error bars (due to simulation statistics; not shown) are estimated to be $\sim 20~\mathrm{J/mol}$.
		Here, the original simulation energy has been shifted up by $0.014\times 10^6~\mathrm{J/mol}$ to match the reference value of the PI-DFT simulations shown in Fig.~\ref{fig:solidpidftU}.
	}
	\label{fig:solidpisgU}
\end{figure}
\begin{figure}
	\includegraphics[scale=0.66]{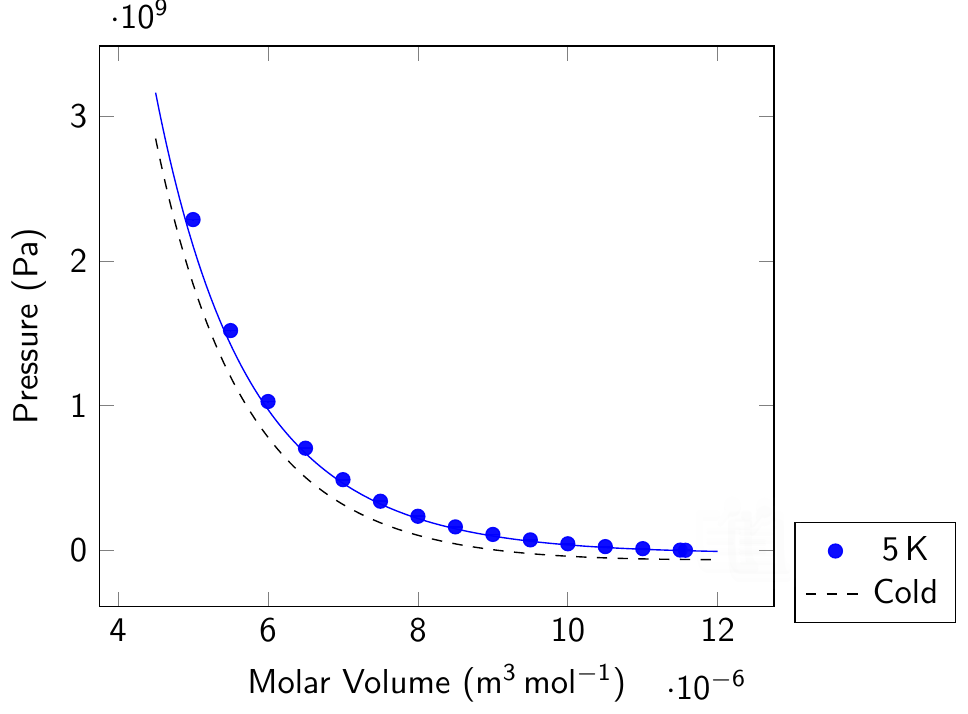}
	\caption{
		Molecular solid: $T$= 5 K pressure isotherms as measured in a high-pressure cell experiment by
		Silvera\cite{Silvera1978}. Symbols represent experimental data. Continuous
		lines represent the results of our EOS model. Experimental error bars (not shown)
		are $\sim 1.5\times 10^6~\mathrm{Pa}$. This is the only experimental data used
		in the fitting; it serves to ensure that the experimental equilibrium conditions
		are well-reproduced. The `cold' curve is obtained by volume differentiation
		of the temperature-independent terms in Eqs.~\ref{eq:solidfvinet} and \ref{eq:solidftf}, and therefore does not include the zero-point energy.}
	\label{fig:solidexpsgP}
\end{figure}

Figures show a comparison between the data and the model.
Once the molecular solid model and its parameters (Table~\ref{tab:msol}) are specified, we proceed to fitting the fluid phases.

\begin{table}
	\begin{tabular}{l|l|l|l}
		\hline\hline
		Parameter             & Value (SI units)                          & (cgs units)                                & (mixed units)                        \\ \hline
		$\phi_0$              & $ -1.53536\times 10^6~\mathrm{J/mol}$     & $-1.53536 \times 10^{13}~\mathrm{erg/mol}$ & $-15.9~\mathrm{eV/atom}$             \\
		$V_0$                 & $ 8.73389\times 10^{-6} \mathrm{m^3/mol}$ & $8.73389~\mathrm{cm^3/mol}$                & $14.503~\mathrm{\AA^3/\text{atom}}$  \\
		$B_0$                 & $ 6.6815\times 10^8 \mathrm{Pa}$          & $6.6815\times 10^9~\mathrm{barye}$         & $0.0041703~\mathrm{eV/\AA^3}$        \\
		$B_1$                 & $ 6.04994$                                & $\to$                                      & $\to$                                \\
		$E_\text{TF}$         & $ 1.08175 \times 10^8 \mathrm{J/mol}$     & $1.08175\times 10^{15}~\mathrm{erg/mol}$   & $1121.2~\mathrm{eV/\text{atom}}$     \\
		$V_\text{TF}$         & $ 3.78086\times 10^{-8} \mathrm{m^3/mol}$ & $0.0378086~\mathrm{cm^3/mol}$              & $0.06278~\mathrm{\AA^3/\text{atom}}$ \\
		$\theta_\mathrm{A}^0$ & $ 688.248~\mathrm{K}$                     & $\to$                                      & $0.0593~\mathrm{eV}/k_\mathrm{B}$    \\ 
		$(V_\theta)$          & $(2.\times 10^{-6}~\mathrm{m^3/mol})$     & ($2.~\mathrm{cm^3/mol}$)                   & ($3.321~\mathrm{\AA^3/\text{atom}}$) \\
		$\gamma_\mathrm A$    & $ 0.746467 $                              & $\to$                                      & $\to$                                \\
		$\theta_\mathrm B$    & $ 5813.38~\mathrm{K}$                     & $\to$                                      & $0.5009~\mathrm{eV}/k_\mathrm B$     \\
		$\xi_\mathrm A$       & $ 0.673068 $                              & $\to$                                      & $\to$                                \\
		\hline\hline
	\end{tabular}
	\caption{Molecular Solid: Optimal choice of parameters obtained from our fitting procedure. 
		These parameters can be directly plugged into Eq.~\ref{eq:solidfvinet}.
		$V_\theta$ is shown in parenthesis because it is {\it not} a fitting parameter, but simply the volume at which $\theta_{A}$ and $\theta_{B}$ take on the values $\theta_{A}^{0}$ and $\theta_{B}^{0}$. Note that $B_0$ does not represent the actual bulk modulus of the molecular solid at low-$T$; rather, it is the bulk modulus of the cold curve alone, which is quite different from the physical low-$T$ modulus due to the sizable zero-point energy. This applies to other quantities as well.}
	\label{tab:msol}
\end{table}

\subsection{Fluids}
The fitting of the fluid phase is much more complicated than the fitting of the solid. 
There are several reasons for this: 
(i) The fluid phase exists in a range which includes the extremes of density and temperature. 
Thus, dilute gas, ultra-dense fluid, and atomic ideal gas limits must be simultaneously respected. 
(ii) In total, the number of parameters ($\sim 20$) is much larger than for the solid.
(iii) The fitting depends on the fit for the molecular solid, for we must obtain a melt line in good agreement with prevailing data.
(iv) The extremes of density and temperature make it necessary to exercise components of the model which are not subject to variation resulting from the tuning of free parameters.
Point (ii) can be mitigated by breaking up the problem into two steps: First, atomic and molecular liquid parameters are fit separately using simulation data pertaining to each of them in turn. Second, all the data is reused to give a global fluid fit after the mixing has been invoked (see Section III B 3). 

\subsection{Molecular Fluid}

The molecular liquid free energy model is completely specified by the following expression:

\begin{subequations}\label{eq:mliq_model}
	\begin{equation}\label{eq:mliq_model_alpha}
		F\mathopen{}\left(V,\, T\right)\mathclose{}=\phi_{0}+\frac{4 V_{0} B_{0}}{\mathopen{}\left(B_{1}-1\right)\mathclose{}^2} \mathopen{}\left(1-\mathopen{}\left(1+X\right)\mathclose{} \exp\mathopen{}\left(-X\right)\mathclose{}\right)\mathclose{}+\cdots
	\end{equation}
	(where $X \to \frac{3}{2} \mathopen{}\left(B_{1}-1.\right)\mathclose{} \mathopen{}\left(\sqrt[3]{\frac{V}{V_{0}}}-1\right)\mathclose{}$)
	\begin{equation}\label{eq2TF}
		\cdots+E_\mathrm{TF} {\mathopen{}\left(\frac{V}{V_\mathrm{TF}}\right)\mathclose{}}^{-2/3} \exp\mathopen{}\left(-{\mathopen{}\left(\frac{V}{V_\mathrm{TF}}\right)\mathclose{}}^{2/3}\right)\mathclose{}+\cdots
	\end{equation}
	\begin{equation}
		\cdots+\frac{1}{2} \frac{k_\mathrm{B}}{\text{atom}} \mathopen{}\left(\frac{9}{8} \bar{\theta}+T 3 \log\mathopen{}\left(1-\exp\mathopen{}\left(\frac{-\bar{\theta}}{T}\right)\mathclose{}\right)\mathclose{}-T \mathcal{D}_{3}\mathopen{}\left(\frac{\bar{\theta}}{T}\right)\mathclose{}\right)\mathclose{}-T \frac{k_\mathrm{B}}{\text{atom}} \log\mathopen{}\left(w\right)\mathclose{}+\cdots
	\end{equation}
	(where $\bar{\theta}=\bar{\theta}^0 \mathopen{}\left(\frac{V}{V^0}\right)\mathclose{}^{-\gamma}$, $\log\mathopen{}\left(w\right)\mathclose{}=0.8$)
	\begin{equation}
		\cdots-\frac{1}{2} \frac{k_\mathrm{B}}{\text{atom}} \log\mathopen{}\left(\mathrm{erf}\mathopen{}\left(\sqrt{\frac{T^{*}}{T}}\right)\mathclose{}-\frac{2}{\sqrt{\pi}} \sqrt{\frac{T^{*}}{T}} \exp\mathopen{}\left(-\frac{T^{*}}{T}\right)\mathclose{}\right)\mathclose{}
	\end{equation}
	(where $T^{*}=\frac{2 m_\mathrm{H} k_\mathrm{B} \tilde{\theta}^{2} R^{2}}{2 \hbar}$, $R=\sqrt[3]{\frac{\frac{3}{4} 2 V}{N_\mathrm{A} \pi}}$, $\tilde{\theta}=\frac{\bar{\theta}}{w^{\frac{1}{3}}}$)
	\begin{equation}
		\cdots+\frac{\frac{1}{2} \omega \hbar}{2 \text{atom}}+\frac{k_\mathrm{B}}{2 \text{atom}} T \log\mathopen{}\left(1-\exp\mathopen{}\left(\frac{-\omega \hbar}{k_\mathrm{B} T}\right)\mathclose{}\right)\mathclose{}+\cdots
	\end{equation}
	(where $\omega=7.94\times 10^{14}~\mathrm{Hz}$)
	\begin{equation}
		\cdots-2 \frac{k_\mathrm{B} T}{2 \text{atom}} \log\mathopen{}\left(\mathrm{erf}\mathopen{}\left(\sqrt{\frac{T^\mathrm{v}}{T}}\right)\mathclose{}\right)\mathclose{}+\cdots
	\end{equation}
	(where $T^\mathrm{v}=51100~\mathrm{K}$)
	\begin{equation}\label{eq:mliq_model_omega}
		\cdots-k_\mathrm{B} T \log\mathopen{}\left(\sum_{\ell=0}^{\ell_\mathrm{max}} \mathopen{}\left(2 \ell+1\right)\mathclose{} \exp\mathopen{}\left(\frac{-\ell \mathopen{}\left(\ell+1\right)\mathclose{} B}{k_\mathrm{B} T}\right)\mathclose{}\right)\mathclose{}+{}
	\end{equation}
	(where $B=\frac{\hbar}{2 I}$, $I=4.61\times 10^{-48}~\mathrm{kg m^2}$)
\end{subequations}

\begin{figure}
	\includegraphics[scale=0.66]{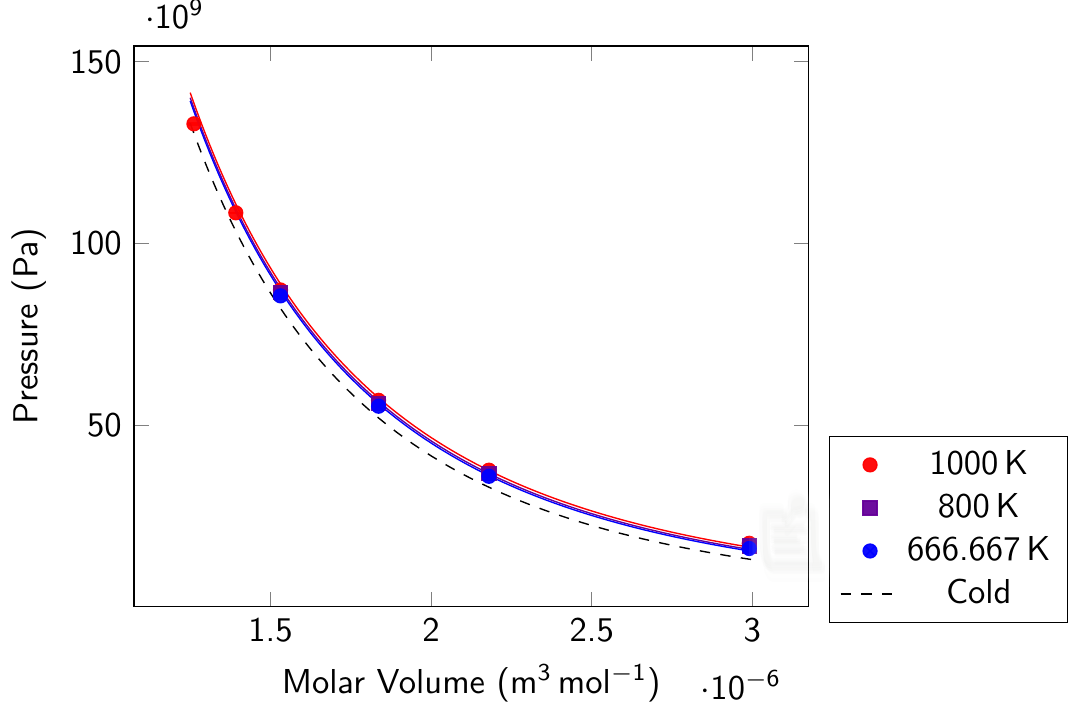}
	\includegraphics[scale=0.66]{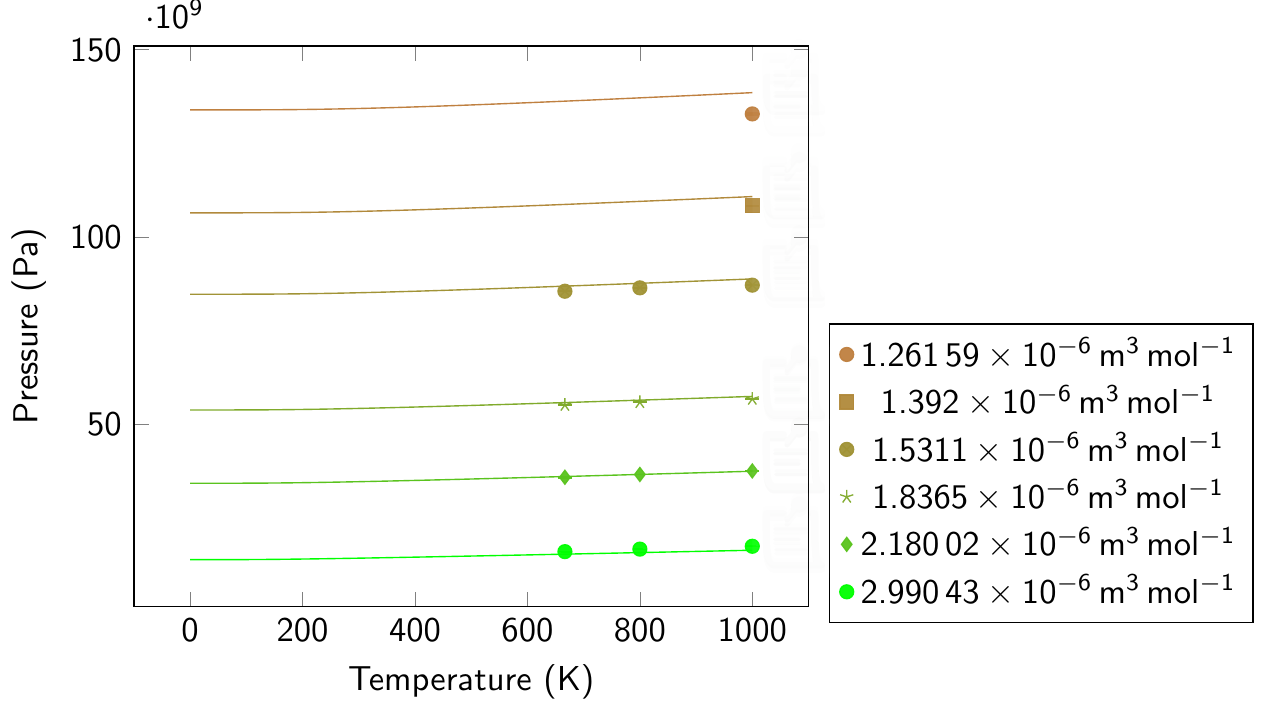}
	\caption{Molecular liquid: Pressure isotherms (left) and isochores (right). 
		Symbols represent data obtained with PI-DFT simulations in the liquid.
		Continuous lines represent the results of our EOS model.
		Typical error bars (due to simulation statistics; not shown) are of the order of $5\times10^7~\mathrm{Pa}$.
		The cold curve is obtained from the volume derivative of the temperature-independent first terms in Eqs.~\ref{eq:mliq_model_alpha} and \ref{eq2TF} and therefore does not include zero-point energy.
	}
	\label{fig:mliqpidftP}
\end{figure}

\begin{figure}
	\includegraphics[scale=0.66]{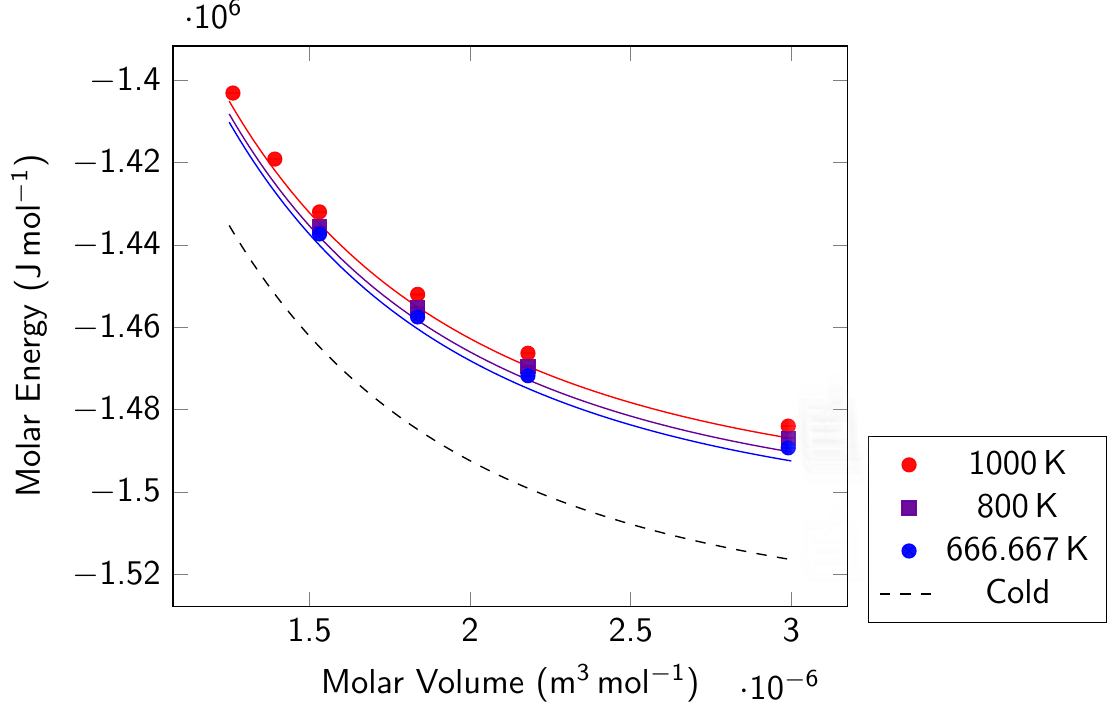}
	\includegraphics[scale=0.66]{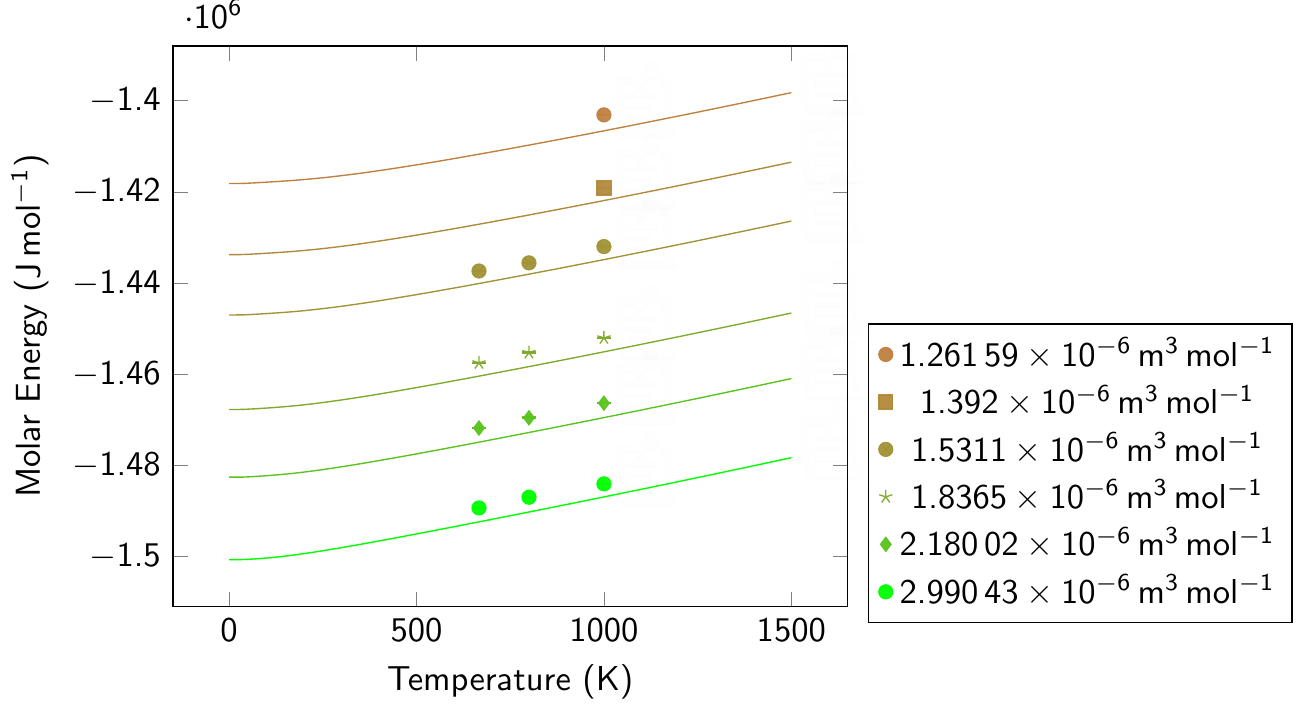}
	\caption{Molecular liquid: Internal energy isotherms (left) and isochores (right). 
		Symbols represent data obtained with PI-DFT simulations in the liquid structure.
		Continuous lines represent the results of our EOS model.
		Typical error bars (due to simulation statistics; not shown) are of the order of $100~\mathrm{J/mol}$.
		The cold curve is obtained from the volume derivative of the temperature-independent terms in Eqs.~\ref{eq:mliq_model_alpha} and \ref{eq2TF} and therefore does not include zero-point energy.
	}
	\label{fig:mliqpidftU}
\end{figure}

\begin{figure}
	\includegraphics[scale=0.66]{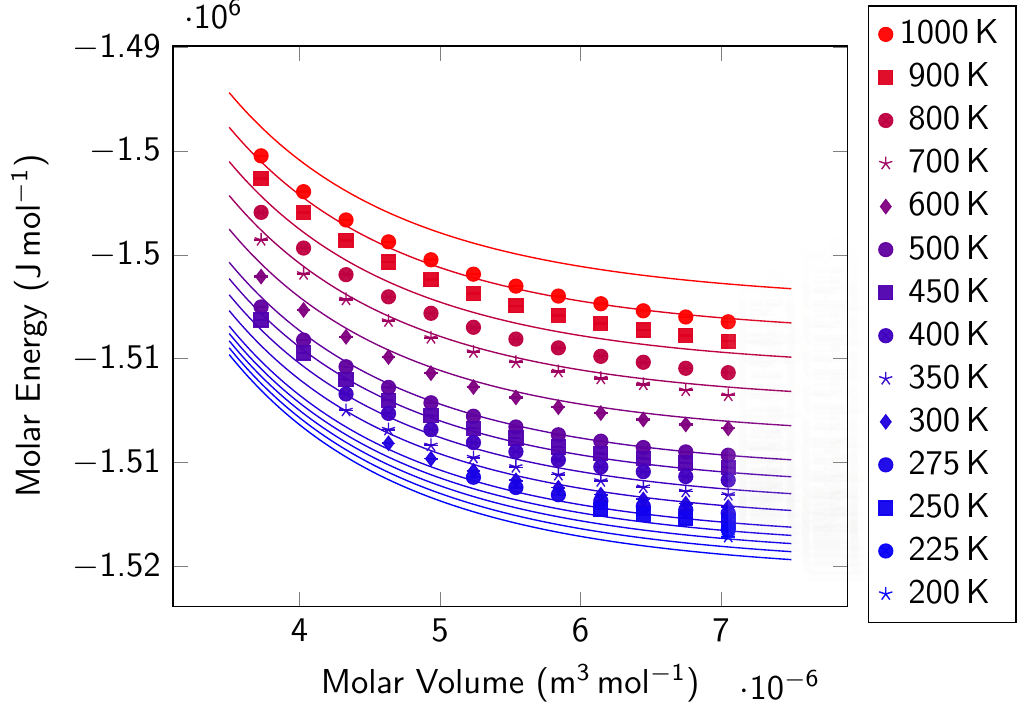}
	\includegraphics[scale=0.66]{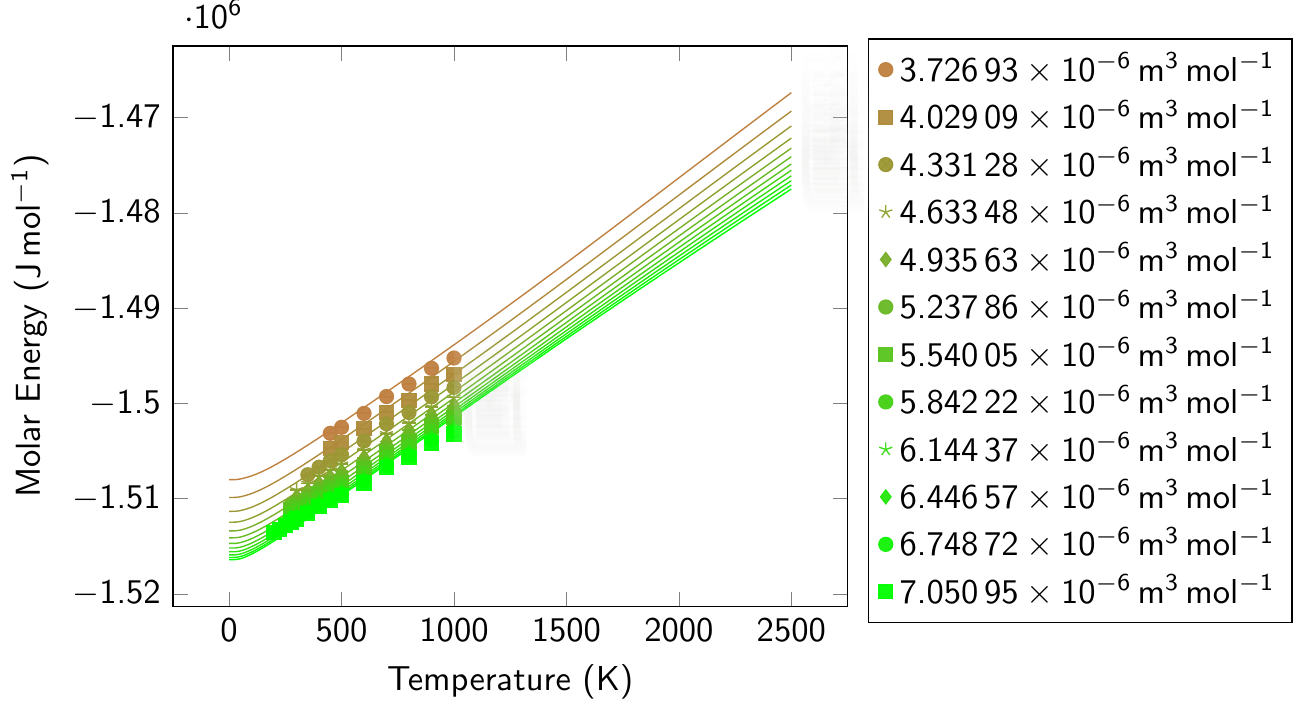}
	\caption{Molecular liquid: Internal energy isotherms (left) and isochores (right). 
		Symbols represent data obtained with PIMD using the Silvera-Goldman inter-molecular potential in the liquid structure.
		Continuous lines represent the results from our EOS model.
		Error bars (due to simulation statistics; not shown) are estimated to be $\sim 20~\mathrm{J/mol}$.
		The original simulation energy is shifted up by $0.014\times 10^6~\mathrm{J/mol}$ to match the reference value of the PI-DFT simulations shown in Fig.~\ref{fig:mliqpidftU}.
	}
	\label{fig:mliq1pisgU}
\end{figure}

\begin{figure}
	\includegraphics[scale=0.66]{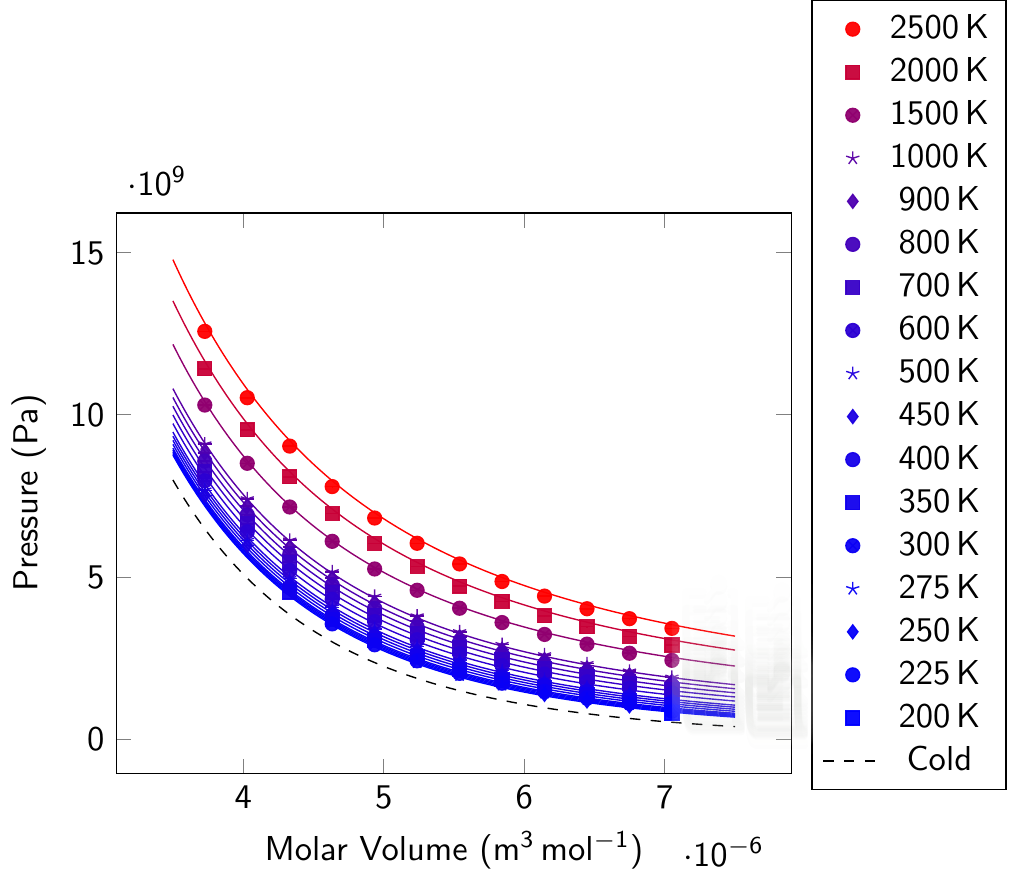}
	\includegraphics[scale=0.66]{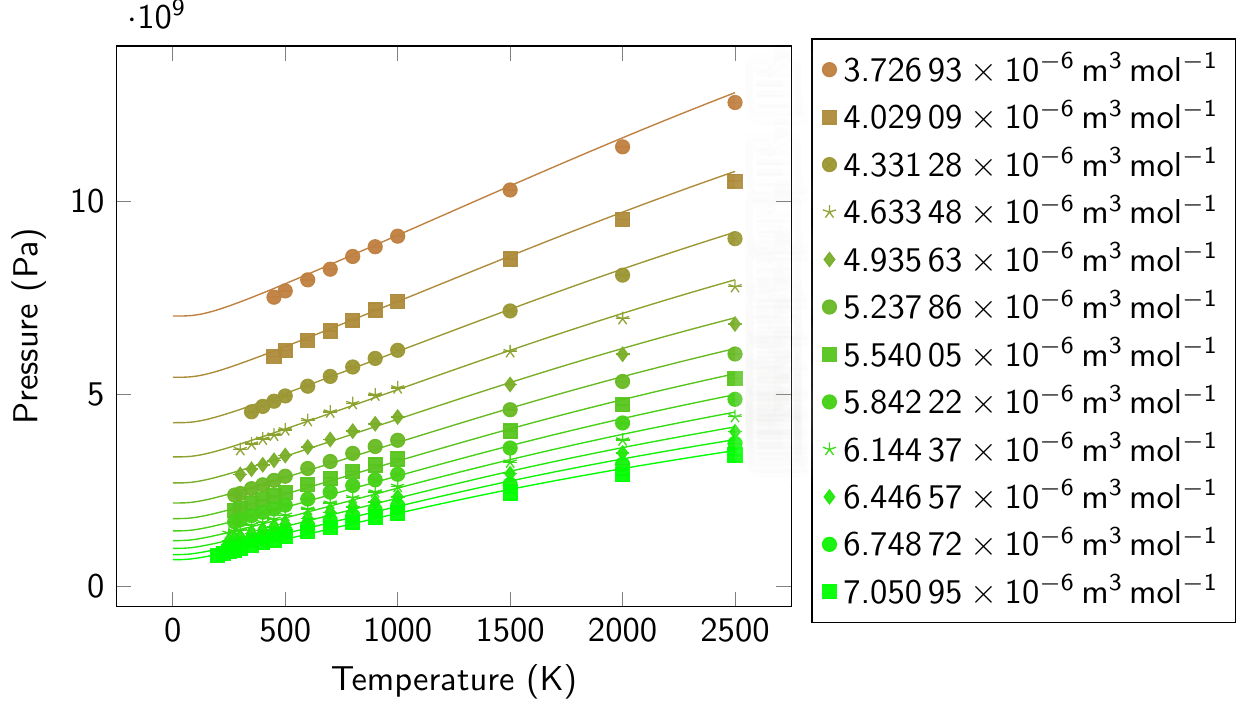}
	\caption{Molecular liquid: Pressure isotherms (left) and isochores (right). 
		Symbols represent data obtained with PIMD using the Silvera-Goldman inter-molecular potential in the liquid structure.
		Continuous lines represent the results of our EOS model.
		Error bars (due to simulation statistics; not shown) are estimated to be $1\times 10^6~\mathrm{Pa}$.
	}
	\label{fig:mliq1pisgP}
\end{figure}

The fitted parameters are presented in Table~\ref{tab:mliq}, the data used for fitting and the comparison with the model are shown Figures~\ref{fig:mliqpidftP} through~\ref{fig:melt}. Such data includes information provided by the melting of the molecular solid into the molecular liquid, as discussed next.

\begin{table}
	\begin{tabular}{l|l|l|l}
		\hline\hline
		Parameter         & Value (SIs units)                          & (cgs units)                           & (mixed units)                               \\ \hline
		$\phi_0$          & $ -1.52996 \times 10^6~\mathrm{J/mol}$    & $-1.53\times 10^{13}~\mathrm{erg/mol}$ & $-15.857~\mathrm{eV/atom}$                  \\
		$V_0$             & $ 4.49273\times 10^{-6} \mathrm{m^3/mol}$ & $4.49273~\mathrm{cm^3/mol}$               & $7.4604~\mathrm{\AA^3/\text{atom}}$      \\
		$B_0$             & $ 1.45338\times 10^8 \mathrm{Pa}$         & $1.45338\times 10^9~\mathrm{barye}$       & $9.0713\times 10^{-4}~\mathrm{eV/\AA^3}$ \\
		$B_1$             & $ 5.23459 $                               & $\to$                                 & $\to$                                \\
		$E_\text{TF}$     & $ 749369.~\mathrm{J/mol}$                 & $7.49369\times 10^{12}~\mathrm{erg/mol}$  & $7.7667~\mathrm{eV/\text{atom}}$        \\
		$V_\text{TF}$     & $ 6.14271\times 10^{-7} \mathrm{m^3/mol}$ & $0.614271~\mathrm{cm^3/mol}$               & $1.02002~\mathrm{\AA^3/\text{atom}}$     \\
		$\theta^0$        & $873.141~\mathrm{K}$                      & $\to$                                 & $0.075241~\mathrm{eV}/k_\mathrm{B}$       \\ 
		$(V_\theta)$      & $(2.\times 10^{-6}~\mathrm{m^3/mol})$     & ($2.~\mathrm{cm^3/mol}$)              & ($3.321~\mathrm{\AA^3/\text{atom}}$) \\
		$\gamma$          & $0.828897 $                               & $\to$                                 & $\to$                                \\
		$\ell_\text{max}$ & $40$                                      & $\to$                                 & $\to$                                \\
		\hline\hline
	\end{tabular}
	\caption{Molecular Liquid: Optimal choice of parameters obtained from our fitting procedure. 
		These parameters can be directly plugged into Eq.~\ref{eq:mliq_model} 
		$V_\theta$ is shown in parenthesis because it is {\it not} a fitting parameter, but simply the volume at which $\theta_{A}$ and $\theta_{B}$ take on the values $\theta_{A}^{0}$ and $\theta_{B}^{0}$. Note that $B_0$ does not represent the actual bulk modulus of the molecular solid at low-$T$; rather, it is the bulk modulus of the cold curve alone, which is quite different from the physical low-$T$ modulus due to the sizable zero-point energy. This applies to other quantities as well.}
	\label{tab:mliq}
\end{table}

\begin{figure}
	\includegraphics[scale=0.66]{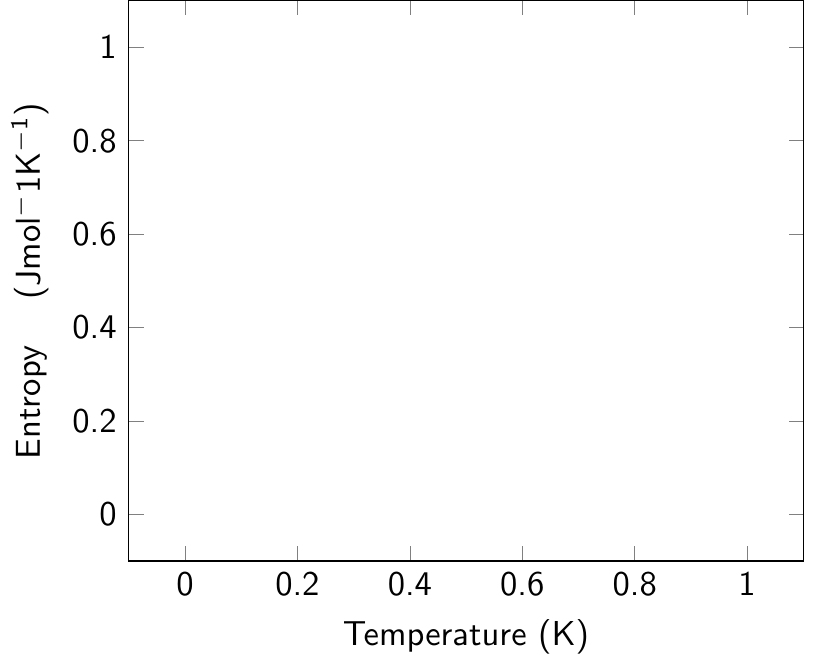}
	\caption{Molecular Liquid: Entropy isochore. The single symbol represents a datum obtained with PI-DFT and a $\lambda$-integration scheme.
		Continuous lines represent the results of our EOS model.
		The error bar (due to simulation statistics; not shown) is estimated to be $0.2~\mathrm{J/mol/K}$.
	}
	\label{fig:solidpidftS}
\end{figure}

\subsubsection{Melting Line}

\begin{figure}
	\includegraphics[scale=0.66]{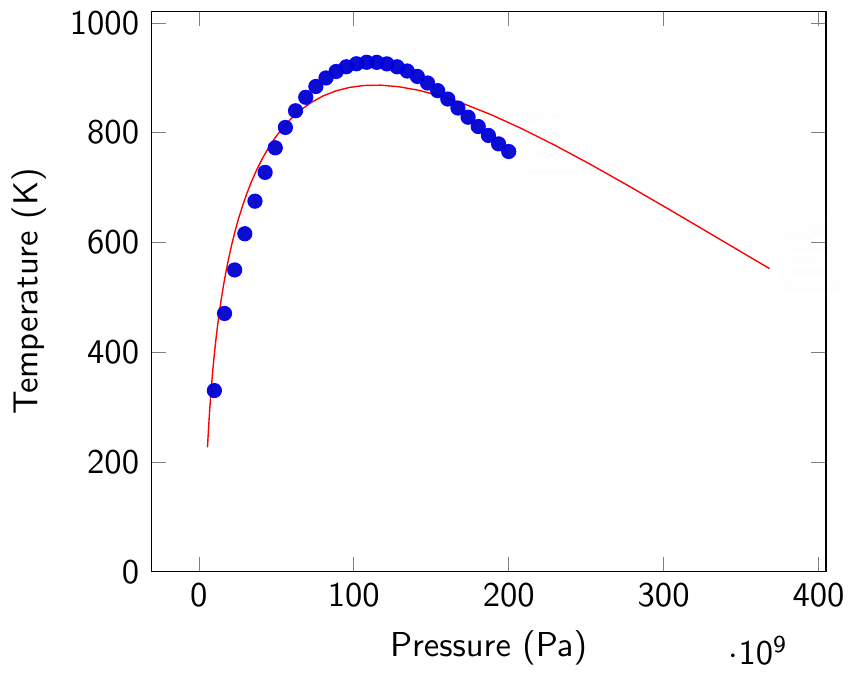}
	\includegraphics[scale=0.66]{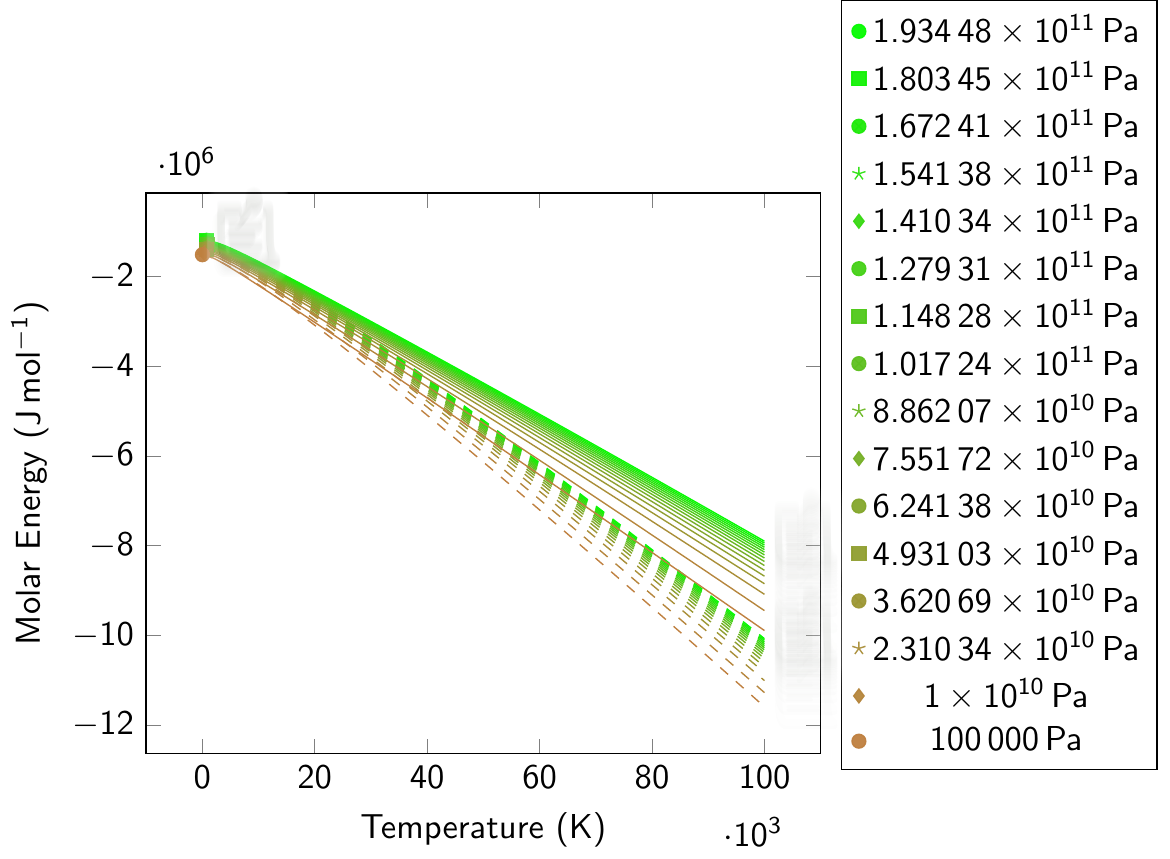}
	\caption{(left) Melting line separating the molecular solid (low-$T$) and the molecular liquid (high-$T$) at constant pressure. Symbols represent data (see text) which constrain the melt line of our EOS model (continuous line). 
		(right) Gibbs free energy isobars for the molecular solid (solid line) and the molecular liquid (dashed line).
		Symbols represent the value of the free energy at the established melting point,
		and the precise free energy values that the molecular liquid should have in order to reproduce the target melting points. 
	}
	\label{fig:melt}
\end{figure}

The melting points as obtained from ab initio simulations
are also used to adjust the parameters of the molecular liquid free energy model. 
As a result, the data used to fit this model for the \emph{liquid} depends indirectly on the already specified details of the 
molecular \emph{solid} model. The parameters are adjusted to reproduce the melt 
temperature as a function of compression by equating the liquid model and solid (fixed) model Gibbs free energies. 
Therefore a term as in Eq.~\ref{eq:minparamg} is explicitly added to the dimensionless residual, 
along with the terms associated with single-phase thermodynamic data ($P$, $U$, $S$). 
The $T$ vs. $P$ melt data (Fig.\ref{fig:melt}) result from thermodynamic integration and free energy matching, where solid and liquid free energies are computed with DFT-MD assuming {\it classical} ions, and using the PBE \pdfcomment{cite PBE} exchange-correlation functional. As discussed in Ref.\cite{Morales13a}, this particular combination (classical ions + PBE) has produced, to date, the best agreement with the experimental $T_{\rm melt}$ vs. $P$, even though {\it both the PBE and classical-ion approximations are known to be severe for this system}. In this sense, we view our choice of melt curve data to be merely a practical one; another nearly equivalent choice would have been to use the experimental data itself to constrain our free energy fits. 

\subsection{Atomic Fluid}

The basis of the atomic fluid is that the cold curve and electron thermal are subsummeed in the \textsc{Purgatorio} result, plus an ion thermal part,
\begin{subequations}\label{eq:liquid2model}
	\begin{equation}
		F(V, T) = \phi_0 + F^\text{Purga}(V, T) + \cdots
	\end{equation}
	\begin{equation}
		\cdots + \frac{k_\mathrm{B}}{\text{atom}} \mathopen{}\left(\frac{9}{8} \bar{\theta}+T 3 \log\mathopen{}\left(1-\exp\mathopen{}\left(\frac{-\bar{\theta}}{T}\right)\mathclose{}\right)\mathclose{}-T \mathcal{D}_{3}\mathopen{}\left(\frac{\bar{\theta}}{T}\right)\mathclose{}\right)\mathclose{}- \frac{T k_\mathrm{B}}{\text{atom}} \log\mathopen{}\left(w\right)\mathclose{}+\cdots
	\end{equation}
	(where $\bar{\theta}=\bar{\theta}^0 \mathopen{}\left(\frac{V}{V^0}\right)\mathclose{}^{-\gamma}$, $\log\mathopen{}\left(w\right)\mathclose{}=0.8$)
	\begin{equation}
		\cdots-\frac{1}{2} \frac{k_\mathrm{B}}{\text{atom}} \log\mathopen{}\left(\mathrm{erf}\mathopen{}\left(\sqrt{\frac{T^{*}}{T}}\right)\mathclose{}-\frac{2}{\sqrt{\pi}} \sqrt{\frac{T^{*}}{T}} \exp\mathopen{}\left(-\frac{T^{*}}{T}\right)\mathclose{}\right)\mathclose{}
	\end{equation}
	(where $T^{*}=\frac{m_\mathrm{H} k_\mathrm{B} \tilde{\theta}^{2} R^{2}}{2 \hbar}$, $R=\sqrt[3]{\frac{\frac{3}{4} 2 V}{N_\mathrm{A} \pi}}$, $\tilde{\theta}=\frac{\bar{\theta}}{w^{\frac{1}{3}}}$)
\end{subequations}

\begin{figure}
	\includegraphics[scale=0.66]{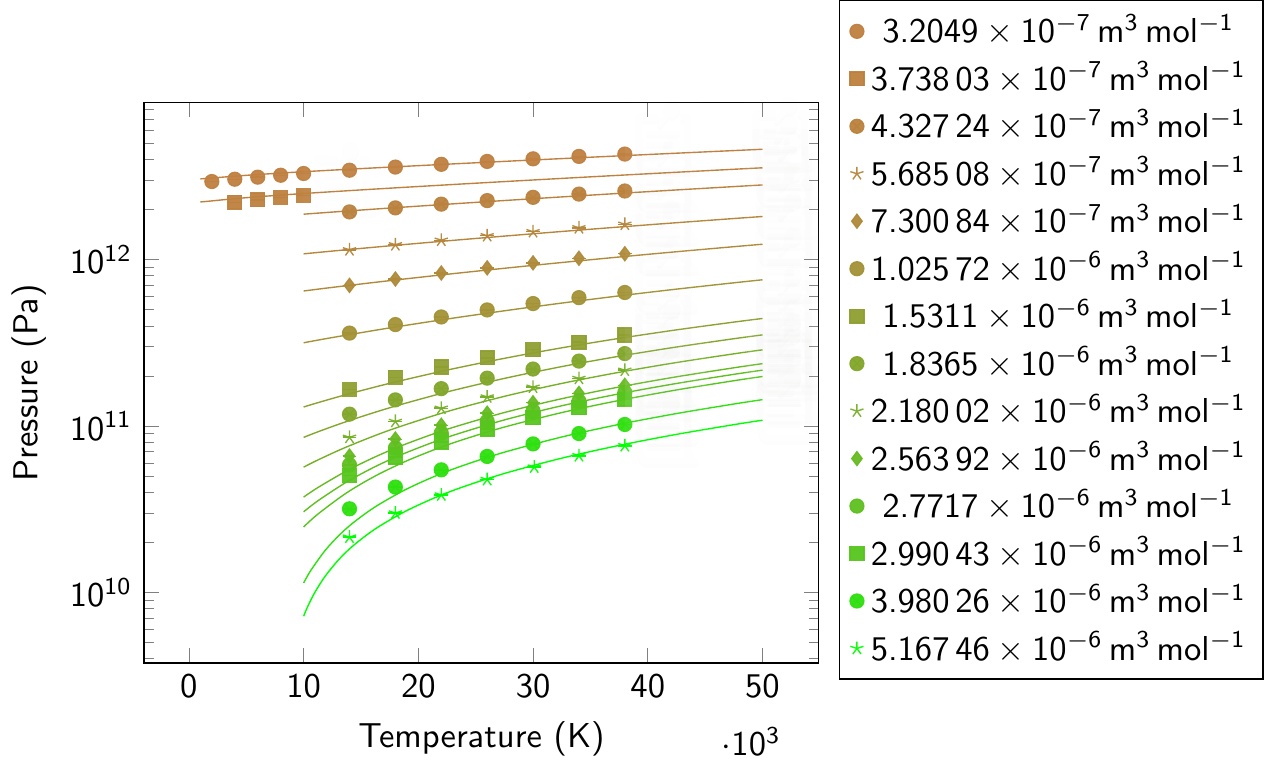}
	\caption{Atomic liquid: Pressure isotherms (left) and isochores (right). 
		Symbols represent data obtained with MD-DFT simulations.
		Continuous lines represent the results of our EOS model (derived from Eq.~\ref{eq:liquid2model}).
		Typical error bars (due to simulation statistics; not shown) are of the order of $1\times10^9~\mathrm{Pa}$.
	}
	\label{fig:liquid2mddftP}
\end{figure}

The parameters of the atomic model are fitted based in two sets of data, 
MD-DFT (classical ions) simulations resulting in internal energy and pressure 
as a function of volume and temperature (Fig.~\ref{fig:liquid2mddftP}) and the 
constant pressure transition to the modelular liquid reported in Ref.~\cite{Morales13a}, which imposes 
an equality of Gibbs free energies.

\begin{table}
	\begin{tabular}{l|l|l|l}
		\hline\hline
		Parameter         & Value (SI units)                          & (cgs units)                           & (mixed units)                        \\ \hline
		$\phi_0$          & $ -1.52996 \times 10^6~\mathrm{J/mol}$    & $\pdfcomment{complete} \times 10^{13}~\mathrm{erg/mol}$ & $\pdfcomment{complete}~\mathrm{eV/atom}$               \\
		$\theta^0$        & $1302.28~\mathrm{K}$                      & $\to$                                 & $0.112198~\mathrm{eV}/k_\mathrm{B}$       \\ 
		$(V_\theta)$      & $(2.\times 10^{-6}~\mathrm{m^3/mol})$     & ($2.~\mathrm{cm^3/mol}$)              & ($3.321~\mathrm{\AA^3/\text{atom}}$) \\
		$\ell_\text{max}$ & $40$                                      & $\to$                                 & $\to$                                \\
		\hline\hline
	\end{tabular}
	\caption{Atomic Liquid: THIS WILL BE UPDATED WITH ATOMIC MODEL PARAMETERS!!! Optimal choice of parameters obtained from our fitting procedure. 
		These parameters can be directly plugged into Eq.~\ref{eq:liquid2model} 
		$V_\theta$ is shown in parenthesis because it is {\it not} a fitting parameter, but simply the volume at which $\theta_{A}$ and $\theta_{B}$ take on the values $\theta_{A}^{0}$ and $\theta_{B}^{0}$. Note that $B_0$ does not represent the actual bulk modulus of the molecular solid at low-$T$; rather, it is the bulk modulus of the cold curve alone, which is quite different from the physical low-$T$ modulus due to the sizable zero-point energy. This applies to other quantities as well.}
	\label{tab:aliq}
\end{table}

\begin{figure}
	\includegraphics[scale=0.66]{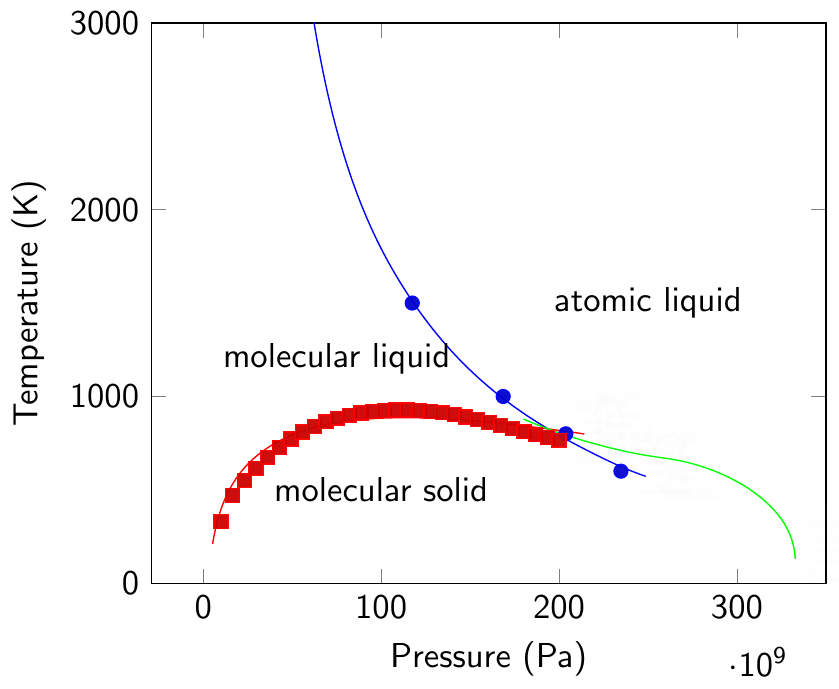}
	\caption{Hydrogen Phase diagram. Melting (at constant pressure) separating the molecular solid and the molecular liquid (red), and the atomic liquid (green).
		Transition line between molecular and atomic liquid (blue) and
		extended transition line (dashed) using the free energies of Eq.~\ref{eq:mliq_model} and Eq.~\ref{eq:liquid2model}.
		Symbols represent simulation data used to constrain the EOS model, lines are results  of the model.
	}
	\label{fig:crit}
\end{figure}

Having, at this point, defined precisely three individual phases of hydrogen, namely a molecular solid (Section~\ref{sec:modelsmsolid}) a molecular liquid and an atomic liquid, 
we can report a simplified phase diagram involving these three phases (Fig.~\ref{fig:crit}), defined by 
i) melting of the molecular solid into a molecular liquid (at low pressure)
ii) melting of the molecular solid into an atomic liquid (at high pressure)
iii) a triple point of these three phases 
iv) a molecular to atomic liquid transition that reproduces simulations at low temperature, 
but artificially continues to high temperature. 
This artificial continuation is due to the fact that, in the discussion presented so far
the atomic and molecular liquid models have \emph{independent} equations of state, this produces
necessarely a \emph{continuous} manifold where the free energies are equal.

In the simulations (and probably in reality) the molecular and atomic liquids are not independent
phases but just idealizations of a single phase at different conditions. 
We decide to model that single phase as a mix of idealized phases.
In the light of the current evidence, the fact that in certain paths (at low temperature)
the thermodynamic quantities show a discontinuous transition poses the challenge of building
a model that has a critical point.

\subsection{Global Fluid}

The global fluid (including the molecular, atomic and mixed stages) is descibed by the following formula, which includes a minimization procedure:

\begin{equation}
	\begin{split}
		F(V,T) = \inf_x \Big\{& (1-x) [F_\mathrm M(V,T) + J(V) x] + x [F_\mathrm A(V,T) + J(V)(1-x)] +\\
		& + \frac{k_\mathrm B T}{\text{atom}} \left[(1-x) \log(1-x)/2 + x \log(x)\right]\Big\}\\
	\end{split}
\end{equation}
where $J(V) = J_0 \mathrm{e}^{-V/V_0}$.

Here, $F_\mathrm{M}$ (molecular fluid) and $F_\mathrm{A}$ (atomic fluid) are already defined in Eq.~\ref{eq:mliq_model} and Eq.~\ref{eq:liquid2model} respectively.
On top of the parameters defining $F_\mathrm{M}$ and $F_\mathrm{A}$, the parameters defining $J$ is adjusted to fit the critical point to its current estimation (by simulation).

\subsubsection{Temperature and Pressure Dissociation}

Each volume and temperature evaluation requires a minimization (with respect to $x$), this defines an auxiliary function $x(V, T)$ 
(or indirectly $x(V(P,T), T)$) that can be interpreted as a the molecular (mass) fraction at a certain condition.

\section{Results and Discussion}\label{sec:results}
While we have made use of many different types of data to fit our global multiphase hydrogen EOS model, there are also many existing pieces of data we have elected not to use. Indeed, the only experimental data we have used is low-$P$ and -$T$ EOS information \cite{Silvera78,Silvera80} (since the Sivera-Goldman potential is essentially a perfect fit to experiment in that range). Experimental isotherm and shock compression data exists in large numbers; we elected not to fit to them. Regarding ab initio simulation data, there is a very notable set that we also chose not to use in the fitting we described in the preceding section: The recent wide-ranging path integral quantum Monte Carlo (PIMC) results of Hu et al. \cite{HuPRL,HuPRB}. These omissions are not merely careless on our part; rather, we have chosen to use as much low and moderate compression (and temperature) theoretical results as we deem necessary to essentially {\it determine the rest of the EOS}. 
Our motivation here is three-fold: 1. We want to examine the extent to which the current set of ab initio methods,  when trained upon the lower temperatures, together with our cell and atom-in-jellium models, determine the behavior into ultra-high compressions and temperatures (covered by the PIMC data). 2. We want to examine just how predictive these methods really are, when eventually compared to the available experimental data for hydrogen and deuterium in extreme conditions. 3. In the end, the smaller the data set to which we fit, the less complex need be our EOS model and our fitting prescription.

In this section, we compare to all of these data left out of our fitting procedure. We will see that while the resulting agreement with our model and these data is not completely perfect, it is strikingly good in most respects. This validates the (mostly) ab initio theoretical methods, our EOS models, and our fitting procedure all at once. Where bona fide predictions are made, we highlight them. We also compare our EOS, both locally (in its ability to post-dict experimental results, for instance) and globally, to the other prominent EOS models in current use for ICF and astrophysical applications. In so doing, we highlight some essential weaknesses of our approach and suggest directions for further improvement.

\subsection{Molecular-Atomic Fraction}
Before we launch into the discussion of these comparisons, we begin by presenting our results for the fraction of H$_2$ molecules (as opposed to free atoms) in the liquid phase. This depends sensitively on the individual atomic and molecular liquid EOS models, as well as on the mixing model which includes the critical behavior outlined in Section III.B.4 and Appendix C. While we know of no direct experimental measurement of the molecular phase fraction, and while its precise definition in the context of simulations is somewhat nebulous \cite{TamblynPRL,TamblynPRB}, its behavior as a function of $(P,T)$ greatly determines the behavior of the principal Hugoniot and other thermodynamic 
tracks of importance in the neighborhood of the regime of maximum compression, since in this regime, \(\mathrm{H_2}\) molecules are undergoing dissociation. 

\begin{figure}
	\includegraphics[scale=0.40]{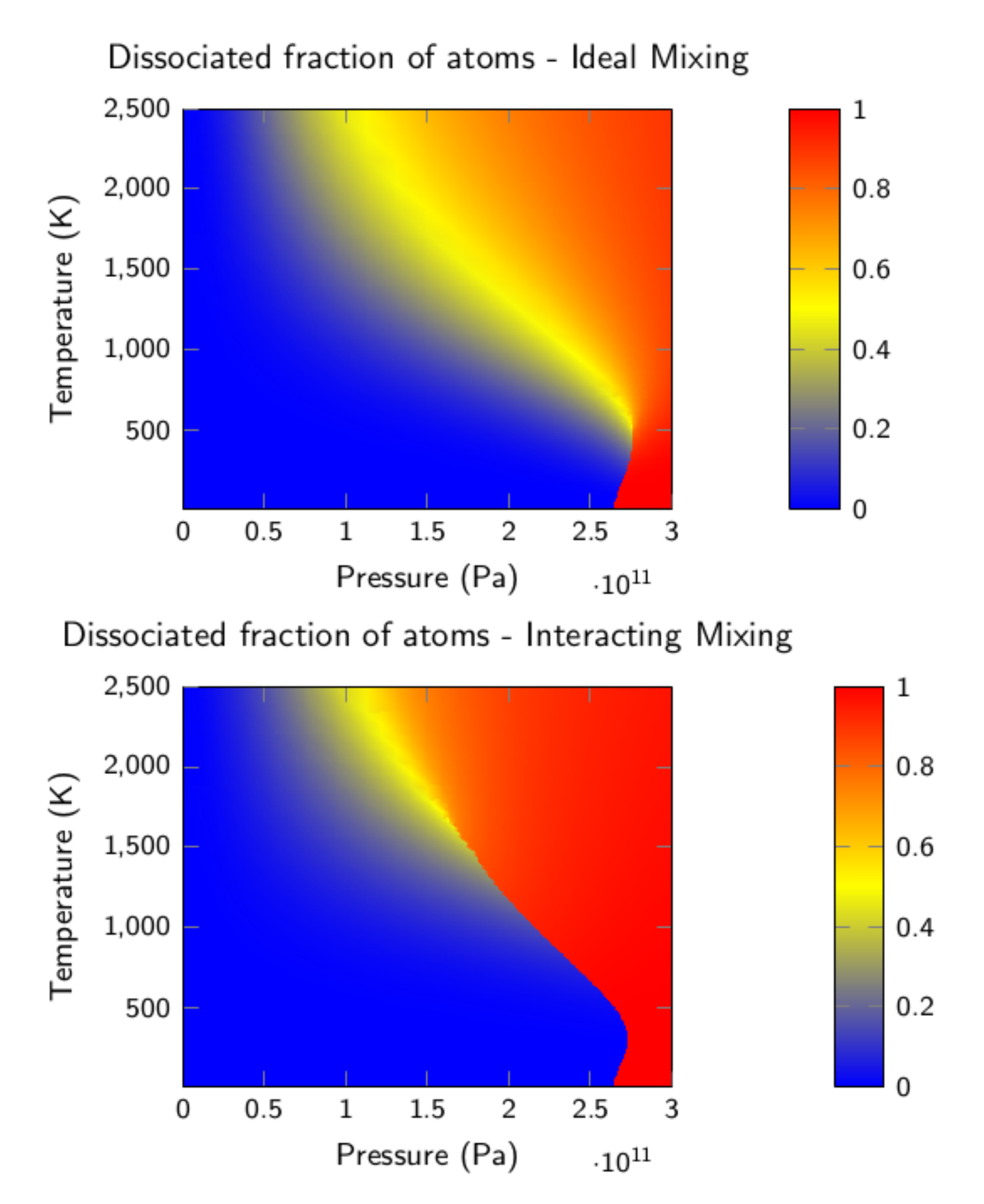}
	\caption{The dissociated fraction of H atoms, $x$, as a function of $(P,T)$ within our EOS model for two cases. Top plot: $J= 0$ (ideal atomic/molecular mixing in the liquid). Bottom plot: \pdfcomment{put value} $J > 0$ (mean-field mixing model described in detail in Appendix C). 
	}
	\label{fig:phasefraction}
\end{figure}

Fig.\ref{fig:phasefraction} shows the phase fraction, $x$, appearing in Eqs.~\ref{solx} (top plot) and \ref{solxscf} (bottom plot). 
As expected, low-$T$, low-$P$ favors molecules ($x\to 0$), while high-$T$, high-$P$ favors free atoms ($x\to 1$). 
For much of the $P,T$-range, above the critical temperature $T_\mathrm{c}$, the transition is continuous. 
However, for $T < T_\mathrm{c}$, there is an abrupt change in $x$ as $P$ is increased. 
This is the critical line. In the top plot, $T_\mathrm c$ is quite low ($\sim 600$ K);
this is because the line results merely from the fact that atomic and molecular liquid 
free energies are different and therefore coexist at different densities. 
Other than this, the atomic-molecular mixing is Saha-like and therefore continuous. 
In the bottom plot, $T_\mathrm c\sim 1500~\mathrm K$, 
which was fit to the ab initio EOS data \cite{Morales09} mainly by adjusting the parameter $J$ (see Section IV.E above).
Note that other than in a relatively narrow region of $(P,T)$, top and bottom plots are broadly similar. We have demonstrated that the principal Hugoniot computed using these two mixing fractions is completely insensitive to presence of the critical line in our EOS model. This makes sense, since the principal Hugoniot (with initial density $\rho_{0}({\rm H})\sim 0.085$ g/cc) does not intersect the critical line, as predicted in recent studies \cite{TamblynPRL,TamblynPRB,RMP,HEDP,Morales09}. However, it bears repeating that the general features exhibited in both plots are essential for a realistic representation of the Hugoniot and other thermodynamic tracks of importance.

\subsection{Comparisons to PIMC}
The recent work of Hu et al. \cite{HuPRB} presents a large $(\rho,T)$-table of internal energy and pressure values for liquid deuterium as computed with PIMC. This is the most extensive data set of its kind yet generated. The $(\rho,T)$ points were chosen to coincide with the high-$T$ end of the range relevant for simulations of ICF capsule performance. As we discussed in Section II, the PIMC approach makes no approximations other than finite simulation box size,
the Trotter decomposition of the unitary time-evolution operator, and the fixed-node approximation, all of which become ever more forgiving as $T$ as raised. Because of this, the authors were able to demonstrate perfect agreement with the ideal gas EOS at sufficiently high-$T$, and comparisons between these data and our liquid EOS model then show the extent to which our model captures this high-$T$ behavior.

\begin{figure}
	\includegraphics[scale=0.40]{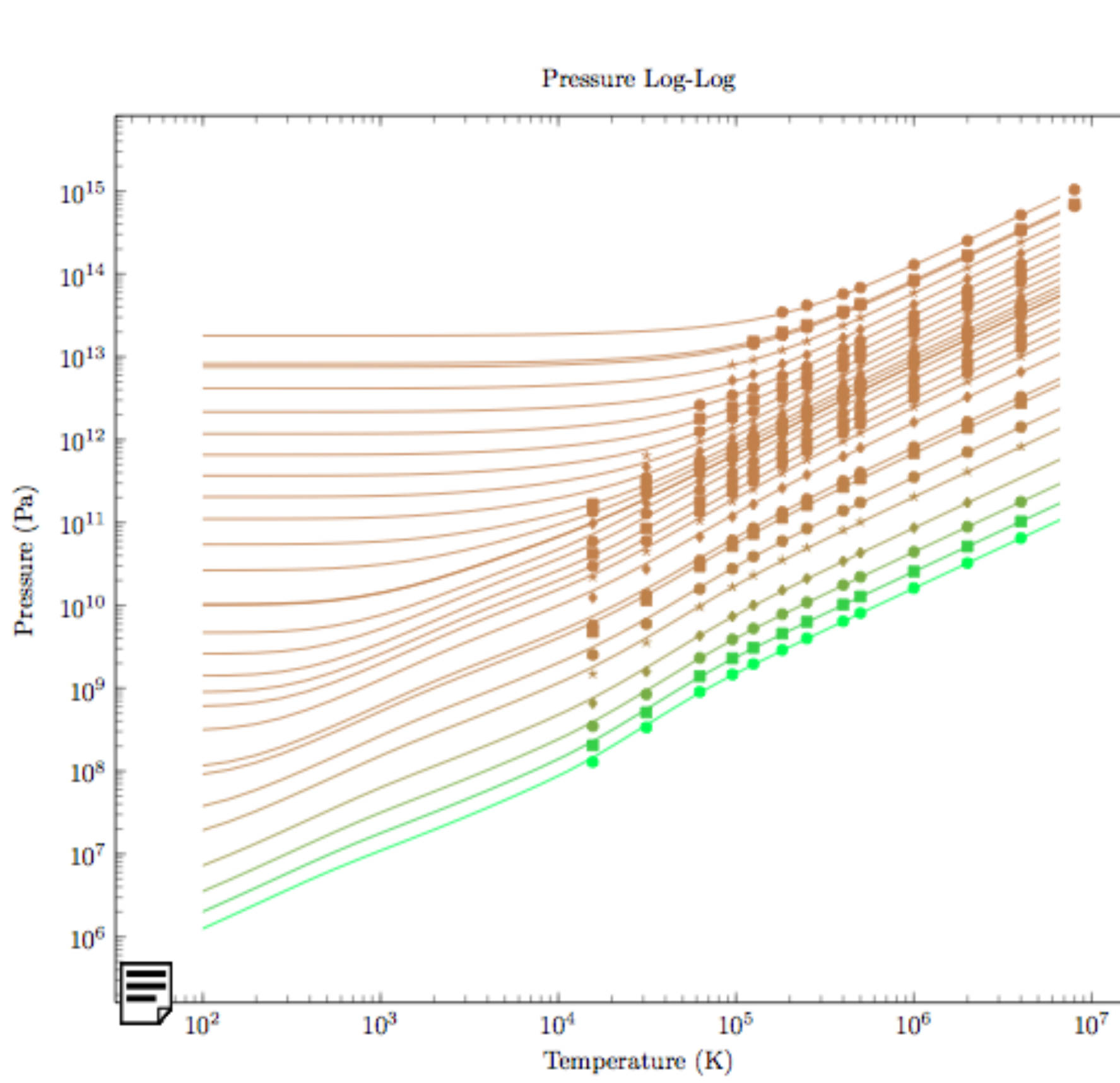}
	\caption{$P$ vs. $T$ for a large collection of isochores from our EOS model for hydrogen (solid curves) and the PIMC data of Ref.\cite{HuPRB} (density-scaled from deuterium).
	}
	\label{fig:PPIMC}
\end{figure}

Figure \ref{fig:PPIMC} shows $P$ vs. $T$ for numerous isochores. Points are the PIMC data for Deuterium and the solid curves are the results of our H EOS model; 
Note that all the PIMC data resides above $10^4~\mathrm K$, 
where the systematic and statistical uncertainties of PIMC for H are thought to be very small indeed. 
Note also that no notable deviations are seen from the corresponding isochores of our model. 
This a is significant result, for \emph{we fit to no data at such high-T in the course of constructing our EOS}. 
Our ab initio MD calculations in the lower-$T$ regime for the liquid together with 
1. The cell model for the ion-thermal contribution, and 
2. The \textsc{Purgatorio} model for the cold and electron-thermal contribution, 
correctly determine the high-$T$ behavior, and specifically the approach to the ideal gas. 
Similar favorable comparisons have been made for the other prominent H EOS models as well \cite{Kerley03,Saumon12,HuPRB}. 
Comparisons of the internal energy are also of this same high level of agreement. 
While this is very encouraging, we will see in Section~\ref{sec:weakness} that 
there are indeed some subtle aspects of the approach to the ideal gas in our EOS 
which are slightly different relative to those of these other EOS models.

\subsection{Shock Hugoniot: Comparison to experiments}\label{sec:comparison}
\begin{figure}
	\includegraphics[scale=0.70]{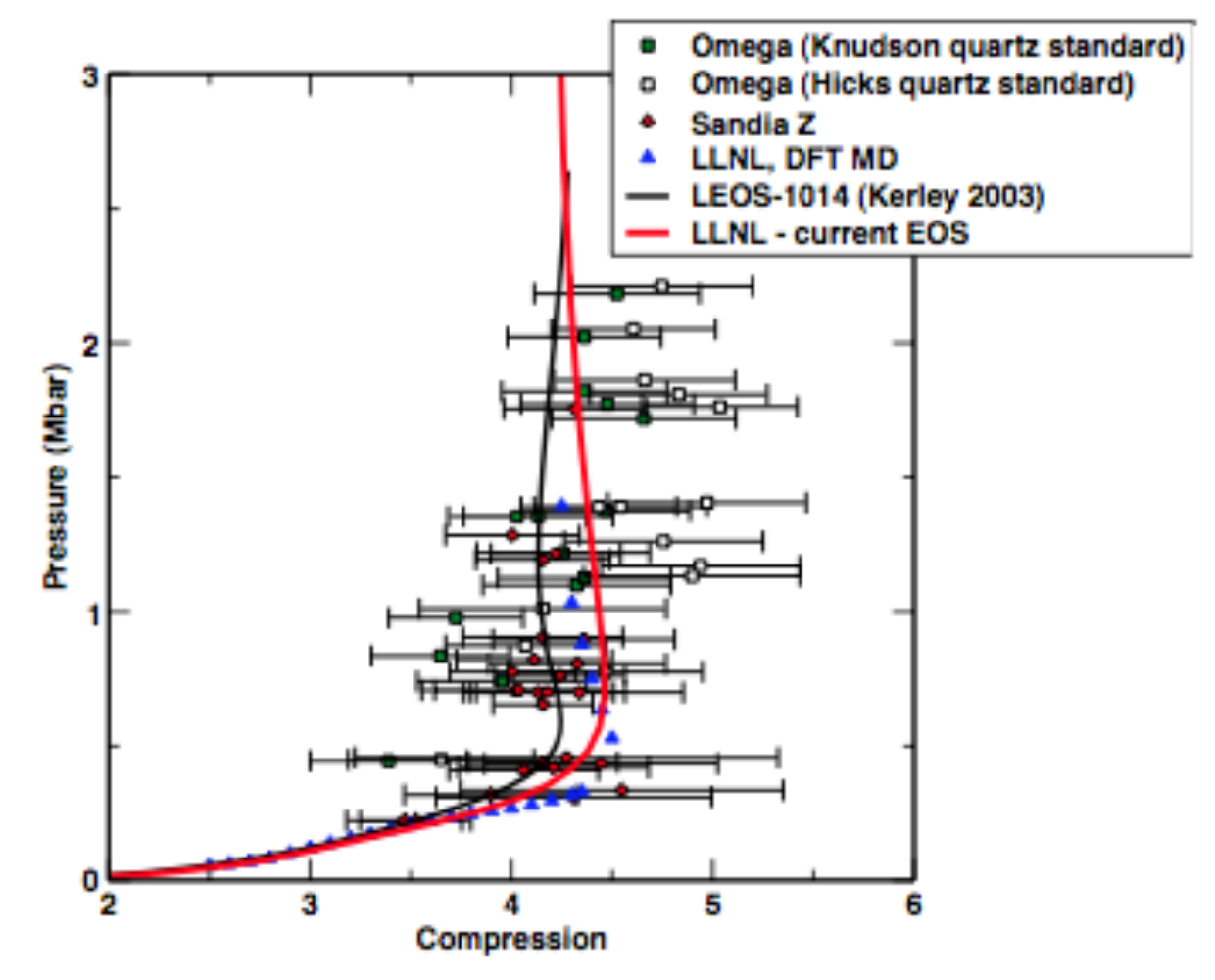}
	\caption{Principal Hugoniot for starting conditions $\rho_{0}= 0.08515$ g/cc and $T_{0}= 20$ K. EOS model results (2003-Kerley: solid black line, our model: solid red line) and experimental data from Refs.\cite{Hicks,Knudson01,Knudson04,HicksQuartz} are displayed as points with error bars. Note that the experimental data points each correspond to slightly different $(\rho_{0},T_{0})$, but these differences are negligible here, given that the abscissa is compression {\it ratio}, $\rho/\rho_{0}$. Blue points are the direct computations of the Hugoniot from our DFT-MD calculations. 
	}
	\label{fig:Hugexp}
\end{figure}

Figure \ref{fig:Hugexp} shows the principal Hugoniot of our EOS model 
(assuming initial conditions: $\rho_{0}= 0.08515$ g/cc and $T_{0}= 20$ K) together with the principal Hugoniot of the 2003-Kerley EOS \cite{Kerley03}, our DFT-MD calculations of the Hugoniot (blue points), and a host of experimental data \cite{Hicks,Knudson01,Knudson04}, some of which involve a reanalysis of older data using an updated understanding of the EOS of the quartz standard \cite{HicksQuartz}. The Hugoniot of our new EOS is a bit more compressible than that of 2003-Kerley. This is a reflection of the fact that our EOS has been constructed by fitting directly to the ab initio EOS data; the same methods predict a Hugoniot with a correspondingly larger compressibility (see Fig.\ref{fig:Hugexp}). The slight discrepancy between 2003-Kerley and the recent ab initio EOS data in this regime was also pointed out recently in a work involving some of us, in which a correction to 2003-Kerley was derived \cite{HEDP} 
to better appease agreement with the ab initio results. 

It is not a given, however, that such methods actually describe this region of maximum compressibility well enough to distinguish differences in this level. The implementations of DFT that we have used to fit the regime of interest are known to be biased towards a somewhat early onset of molecular dissociation \cite{RMP}. Correction of this deficiency would likely push the maximum compression even farther away from 2003-Kerley. Nevertheless, it is important to note two things at this stage: 1. Both EOS model curves are within all of the experimental error bars, save a couple of the more recent Sandia Z-machine points at $P\sim 0.7$ and 1.3 Mbar which our EOS misses, and a recent point from Hicks et al. at $P\sim 1.75$ Mbar which 2003-Kerley misses. 2. In our recent study \cite{HEDP}, DT EOS variations of this magnitude failed to produce noticeable differences in simulated indirect-drive ICF capsule performance. 

There is another important data set that addresses the maximum compression achievable after a single shock: The shock-reverberation experiments of Knudson et al. \cite{Knudson04}. In these measurements, liquid deuterium at cryogenic temperatures was confined between an aluminum anvil and a sapphire window. A flyer plate launched toward the anvil then sent shock waves that underwent multiple reflections from the material surfaces and reverberated through the deuterium. The measured ratios of various shock arrival times then provided tight constraints on the maximum compression after a single shock. Using a standard shock impedance matching analysis, multiple data points (corresponding to different flyer plate velocities) were analyzed using the 2003-Kerley EOS for deuterium, the Al EOS of Ref.\cite{AlEOS}, and the sapphire EOS of Ref.\cite{Al2O3EOS}. 

\begin{figure}
	\includegraphics[scale=0.30]{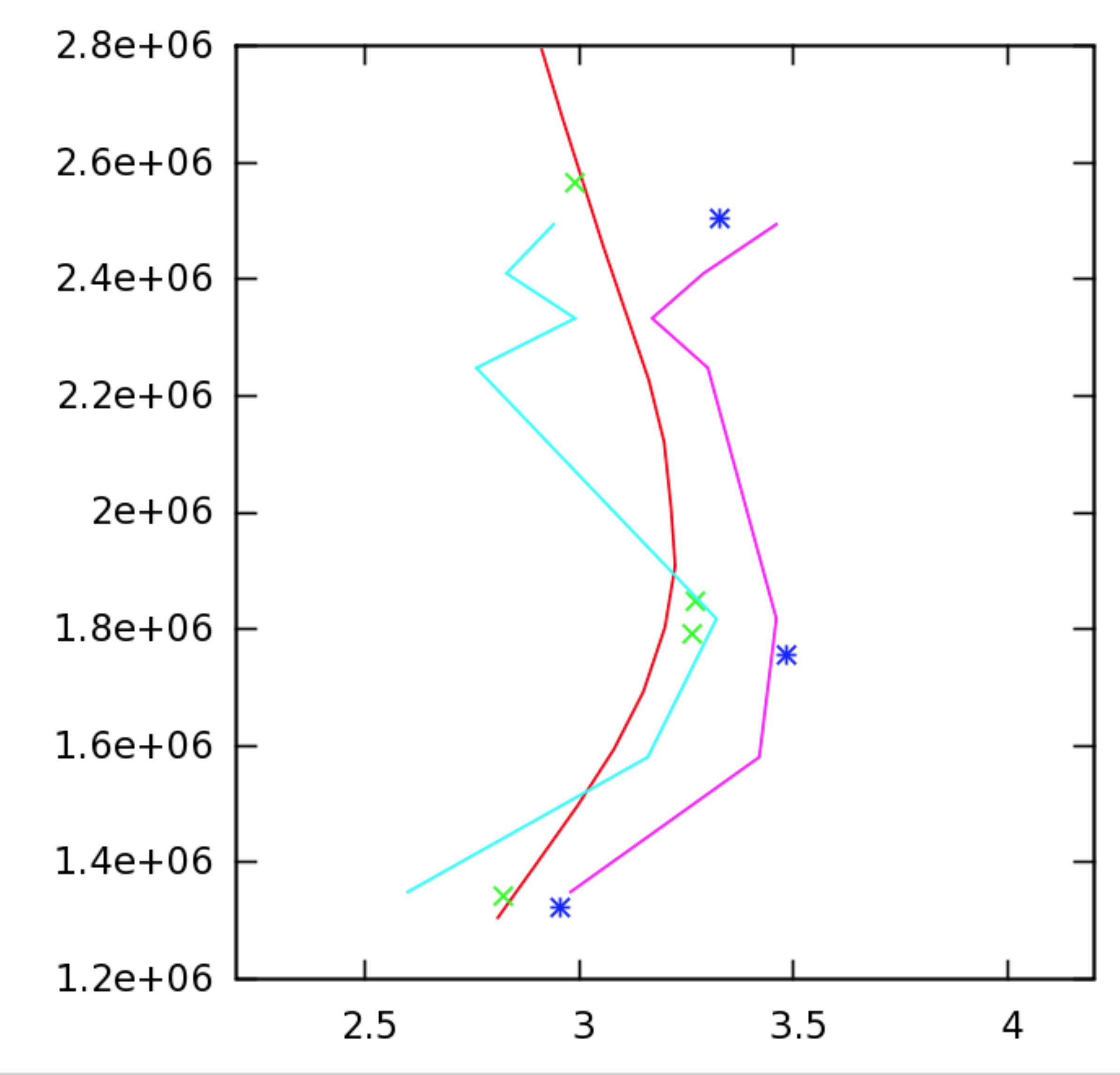}
	\caption{Initial shock velocity in deuterium vs. reverberation ratio as defined and discussed in Ref.\cite{Knudson04}. Red curve is Kerley-2003, and the blue points are our EOS for deuterium. The cyan and magenta curves indicate the range of experimental uncertainty presented in Ref.\cite{Knudson04}. These EOS predictions each rely on an identical set of Al flyer plate and sapphire window EOS models \cite{Knudson04}.
	}
	\label{fig:reverbanalysis}
\end{figure}

Fig.\ref{fig:reverbanalysis} shows the measured shock velocity in the deuterium sample vs. reverberation ratio (which is a ratio of times which is indicative of the density compression ratio \cite{Knudson04}), as plotted in Fig.8 of Ref.\cite{Knudson04}. The red solid line is the result (reported in Ref.\cite{Knudson04}) of using the 2003-Kerley EOS for deuterium. The blue points are the result of using our deuterium EOS. Note that our EOS presents a slightly larger compressibility, as already noted above. The cyan and magenta lines indicate the lower and upper bounds of the experimental error bars on the reverberation ratio reported in Table II of Ref.\cite{Knudson04}. It is apparent that our EOS is in no better or worse agreement with these data than is 2003-Kerley; our EOS is on the soft side (more compressible) while 2003-Kerley is on the stiff side (less compressible). In addition, the DFT-MD work of Desjarlais \cite{Desjarlais} shows similar features to our results for the lower shock velocities, but exhibits slightly less compressibility (smaller reverberation ratio) for the larger shock velocities. The most recent deuterium EOS of Saumon \cite{Saumon12} also makes such a comparison, which looks slightly less favorable on the whole than that presented here. We stress, however, that {\it all} of these comparisons make use of the same aforementioned EOS models for Al and sapphire. Any potential inaccuracies in those models weaken the efficacy of the experiment-model comparisons we discuss here.

\begin{figure}
	\includegraphics[scale=0.40]{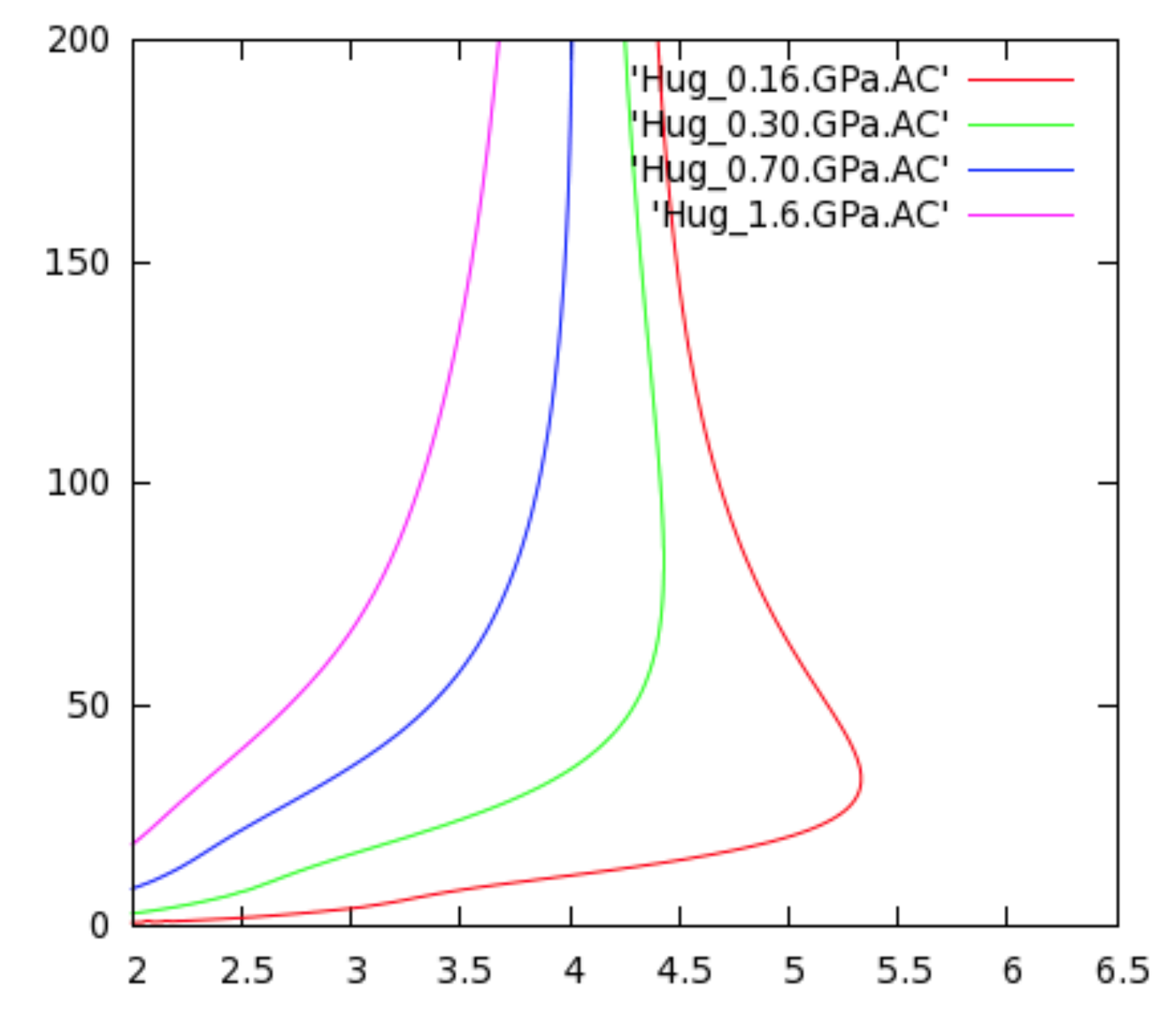}
	\caption{Hugoniot curves ($P$ vs. $\rho/\rho_{0}$) for our hydrogen EOS model, assuming the four starting pressures, $P_{0}=$ 0.16, 0.30, 0.70, and 1.0 GPa. For all curves, $T_{0}=$ 297 K. This is to be compared to Fig.5 of Ref.\cite{Loubeyre}.
	}
	\label{fig:hugprecomp}
\end{figure}

In addition to Hugoniot data in which the initial conditions are $\rho_{0}\sim 0.085$ g/cc, there are also data pertaining to shocks applied to precompressed samples. An example of this is reshock data, in which a subsequent shock is applied to a sample that has already been shocked to rather high stresses, such as reported in Ref.\cite{Knudson04}. These reshock data have fairly large error bars and are therefore quite a bit less constraining than the shock reverberation data we just discussed, though they do address a regime of higher compression. Another more recent set of precompressed shock data is the work of Loubeyre et al. \cite{Loubeyre}, in which laser-driven shocks were applied to both hydrogen and deuterium samples in diamond anvil cells. The initial pressures prior to the shock ranged from 0.16 to 1.6 GPa. This larger initial stress made possible the shock of hydrogen/deuterium to over 5-fold compression. Fig.\ref{fig:hugprecomp} shows four Hugoniot curves computed with our EOS for each of the four initial pressures, $P_{0}$=  0.16, 0.30, 0.70, and 1.6 GPa ($T_{0}= $ 297 K for all four); initial densities are computed from $(P_{0},T_{0})$ using our EOS. Comparing to Fig.5 of Ref.\cite{Loubeyre}, we see that our $P_{0}= 16$ GPa curve has a maximum compression which is very similar to the experimental data, though: 1. The error bars are quite large, and  2. The pressure at which the compression is maximum for this curve is a bit lower than that seen in the data. For the $P_{0}$= 0.16 GPa curve, the experimental data seems to show nearly constant compression for $P > 80$ GPa, while our EOS model shows compression continuing to increase with $P$. Similar features are found with the ab initio-based EOS of Ref.\cite{Caillabet}, also presented in Fig.5 of Ref.\cite{Loubeyre}. Again, the 2003-Kerley EOS exhibits somewhat lower compressibility for these same precompressed Hugoniots \cite{Loubeyre}.

\subsection{Liquid-vapor dome}
I don't know if we want to say anything here... we could just delete this subsection.

\subsection{EOS model comparisons}
We have already alluded to comparisons between our EOS and some of the other recent hydrogen/deuterium EOS models \cite{Kerley03,Caillabet,Saumon12}. Differences were mentioned above in the context of comparisons to various shock data, and the broad similarity between the models was discussed both there and in the comparison to PIMC. We now address these EOS model differences more directly, outside of the framework of comparisons to data. 

\begin{figure}
	\includegraphics[scale=0.80]{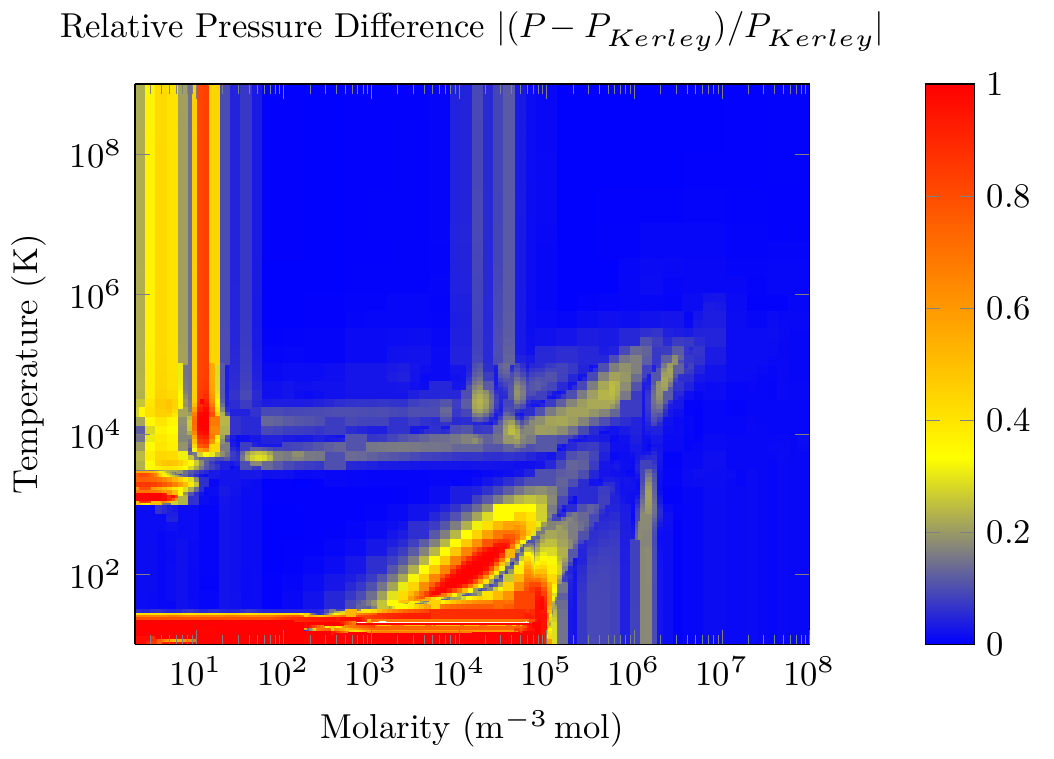}
	\caption{Percent differences between the pressures of our EOS model and the (density-scaled deuterium) EOS of Ref.\cite{Saumon12}. Blue indicates zero difference while red indicates a difference of 100\%.
	}
	\label{fig:vsKerley}
\end{figure}

First we examine global differences between our EOS and that of the most recent wide-ranging model of which we are aware: 2012-Saumon \cite{Saumon12}. Figure \ref{fig:vsKerley} shows a contour plot of percent differences in the pressure between this EOS  and ours, over the entire range of $\rho$ and $T$ which is common to both tables. ALFREDO: YOU WRITE THE REST OF THIS PARAGRAPH AFTER YOU HAVE MADE THE PLOT.

\begin{figure}
	\includegraphics[scale=0.40]{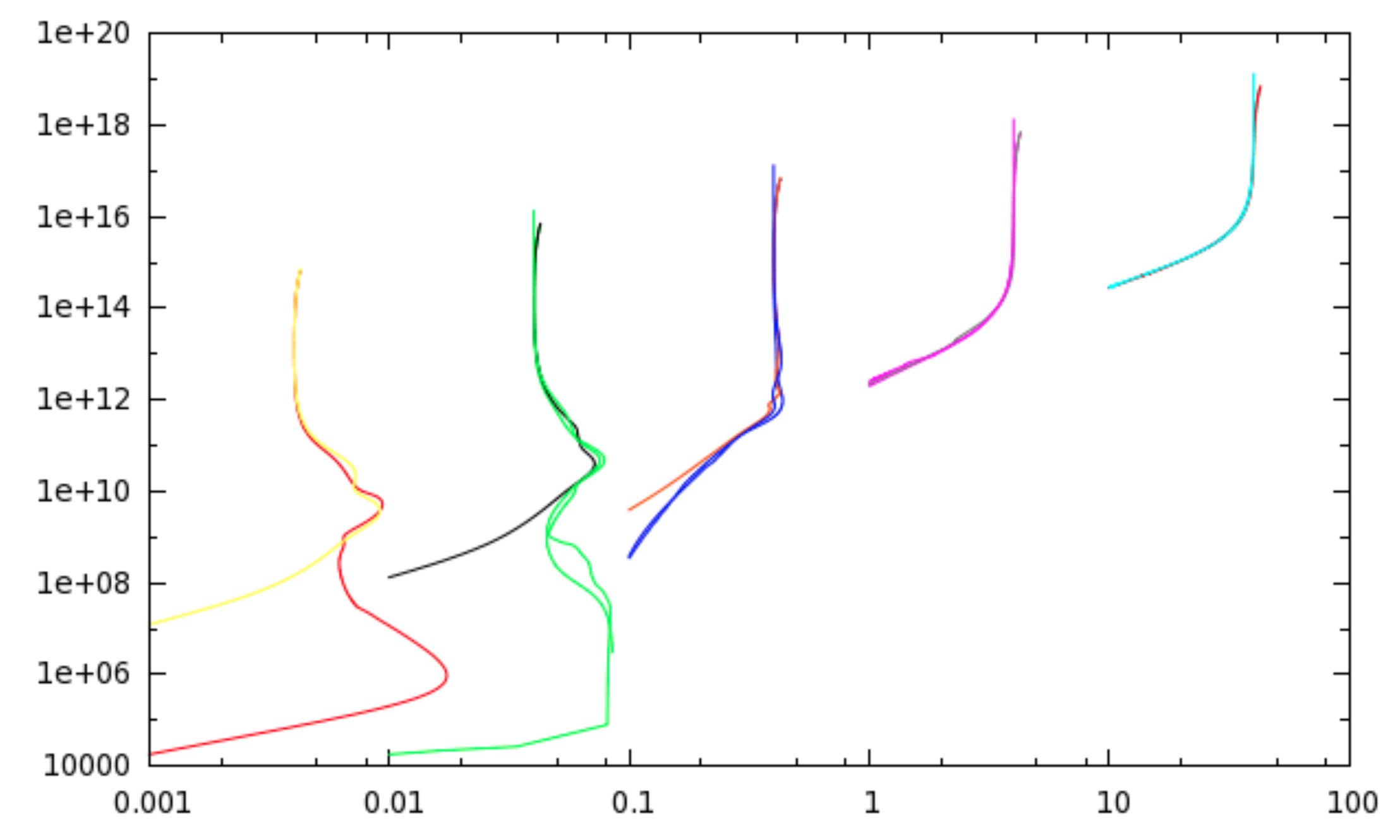}
	\caption{Hugoniot curves ($P$ vs. $\rho$) for four different starting densities: $\rho_{0}$= 0.001, 0.01, 0.1, 1.0, and 10.0 g/cc. Three different EOS models are represented: Ours, the EOS model of Ref.\cite{Kerley03}, and the (density-scaled from deuterium) EOS model of Ref.\cite{Saumon12}.
	}
	\label{fig:hugACD}
\end{figure}

Next, we return to shock Hugoniots and compare them for the different EOS models over a wide range of starting densities, $\rho_{0}$. Figure \ref{fig:hugACD} shows the $P$ vs. $\rho$ Hugoniot curves with five different $\rho_{0}$: 0.001, 0.01, 0.1, 1.0, and 10.0 g/cc. For all curves, the initial temperature is chosen to be $20~\mathrm K$. 
Three different hydrogen EOS models are represented: The EOS of this work, 2003-Kerley, and 2012-Saumon (which we density-scaled from deuterium). For $\rho_{0}= 0.001$ g/cc, our version of the 2003-Kerley EOS table failed to exhibit a solution to the Rankine-Hugoniot relation, so we only compare our model and 2012-Saumon for that particular (left-most) curve. We see that the results for small $\rho_{0}$ and for low-$P$ are quite divergent, though this is partly exaggerated by the log-log plot; indeed, the pressures where they diverge are very small fractions of a GPa. These regions of disagreement reflect different treatments of the EOS near ambient conditions (recall that our approach makes use of a particular set of low-$P$ experimental data in this regime \cite{Silvera78,Silvera80}). For lager $P$ and larger $\rho_{0}$, the various models are in extremely good accord, reflecting the fact that they all approach the ideal gas limit at sufficiently high-$T$, and that they all respect the Thomas-Fermi limit at ultra-high $\rho$. One notable discrepancy at the highest pressures for all the curves can be seen: Our \textsc{Purgatorio}-based EOS includes a fully relativistic DFT description of the electrons \cite{Purgatorio} which gives rise to a larger final compression limit for the Hugoniot \cite{Hora} (${\rm max}[\rho/\rho_{0}]=$ 7 rather than 4). This is only hinted at in Fig.\ref{fig:hugACD} however, since the maximum $T$ in the various EOS tables is still well below $m_{e}c^{2}/k_{\rm B}$.

\subsubsection{High-pressure Hugoniot: a weakness of \textsc{Purgatorio} and other ion-sphere models }\label{sec:weakness}
The stronger the shock applied to a material, the hotter the shocked material becomes. A shock of sufficient strength will eventually force a material into an ideal gas state. We now examine the specific set of final states accessible in such {\it nearly}-ideal gas conditions, as they pertain to some of the assumptions inherent in the construction of our EOS. These are the points on the $P$ vs. $\rho/\rho_{0}$ Hugoniot curves which are right below the perfectly vertical portions in Fig.\ref{fig:hugACD}, for instance.

The relation that defines a Hugoniot curve for low-$P$ initial condition is:
\begin{equation}\label{Castor1}
	E-E_0 -\frac12P\left(\frac{1}{\rho_0}-\frac{1}{\rho}\right)=0\;,
\end{equation}
where $E$ is the internal energy per unit mass of the shocked
state, $E_0$ is the internal energy at the cold, normal density
reference state, for which the pressure was neglected in Eq.~\ref{Castor1}, 
$\rho$ is the mass density at the shocked state, and $\rho_0$ is the
reference density. The absolute scale of energy admits an
arbitrary shift, so we can suppose that the zero point is so
chosen that the potential energy contributions to $E$ vanish at
high $T$, when the electrons and nuclei are completely dispersed
and weakly interacting. In that case $E_0$ is a negative quantity
and its magnitude is the amount of energy required to remove the
atoms from the liquid or solid and to completely ionize them. It
is relatively easy to estimate this from tabulated dissociation
and ionization data.

Supposing that in the hot shocked state the gas is perfectly
ideal, then the pressure and energy obey the relation
\begin{equation}\label{Castor2}
	E=\frac32\frac{p}{\rho}\;,
\end{equation}
assuming for the moment that relativity, degeneracy effects, and Coulomb corrections are negligible. If the ionization is
not complete, then $E$ will also contain binding energy terms.
We can readily eliminate $E$ between Eqs.~\ref{Castor1} and \ref{Castor2}, to
obtain this form for the $P$ \emph{vs.} $\eta$ Hugoniot:
\begin{equation}\label{hyper1}
	P=\frac{2\rho_0\eta |E_0|}{\eta-4}\;,
\end{equation}
where $\eta$ is the compression ratio $\rho/\rho_0$. This is the
equation of a hyperbola in the $P$-$\eta$ diagram, with a
vertical asymptote at $\eta=4$. At lower temperatures where there
are recombined electrons, and where electron degeneracy and
Coulomb effects become significant, Eq.~\ref{hyper1} can not be
expected to apply.

In the case of  hydrogen, for which an estimate of
$|E_0|$ gives 15.88 eV, about equal to the ionization energy of
neutral H plus one half the dissociation energy of H$_2$, it is
found that our Hugoniot does not tend asymptotically to
this hyperbola, while the Hugoniot of some other hydrogen EOS models \cite{Kerley03,Saumon07} do. We now explore why this is the case.

If there are corrections to the ideal gas relation, then we can
introduce quantities $\delta E$ and $\delta P$ such that
$\rho E=(N_\mathrm i+N_\mathrm e) k_\mathrm B T(3/2+\delta E)$ and $P=(N_\mathrm i+N_\mathrm e)kT(1+\delta P)$. Then
\begin{equation}
	E=\frac{P}{\rho}\frac{3/2+\delta E}{1+\delta P}\approx \frac{3P}{2\rho}
	+\frac{(N_\mathrm i+N_\mathrm e)k_\mathrm BT}{\rho}\left(\delta E-\frac32\delta P\right)\;,
\end{equation}
where in the last form the powers and products of the small
quantities $\delta E$ and $\delta P$ have been neglected.
Using this approximation, we can express the energy term as
\begin{equation}\label{Castor3}
	E-E_0\approx \frac{3P}{2\rho}+E^*\;,
\end{equation}
where the new quantity $E^*$ is defined by
\begin{equation}\label{Estardef}
	E^*=\frac{(N_i+N_e)kT}{\rho}\left(\delta E-\frac32\delta
	P\right)-E_0\;.
\end{equation}
The rest of the algebra leading to equation (\ref{hyper1}) is
unchanged, so we now have
\begin{equation}\label{hyper2}
	P=\frac{2\rho_0\eta E^*}{\eta-4}\;.
\end{equation}
So the question about the asymptotic form of the Hugoniot is
focussed on finding the proper value of $E^*$.

If the electrons are slightly degenerate, then the electron
pressure departs from $N_\mathrm ek_\mathrm BT$ by a factor $(2/3)F_{3/2}/F_{1/2}$,
in terms of the Fermi integrals. As it happens, for the ideal
Fermi gas the internal energy is corrected by exactly the same
factor, and therefore we still have $\delta E=(3/2)\delta P$ to all orders of
degeneracy, and degeneracy does not contribute to $E^*$.

For dense plasmas there are substantial corrections to the
equation of state arising from the electron-electron,
electron-ion and ion-ion Coulomb interactions, $\pm e^2/r$.  The
magnitude of these terms is measured by $\Gamma=e^2/(r_\mathrm sk_\mathrm BT)$, in
which $r_\mathrm s=(4\pi N_\mathrm i/3)^{-1/3}$, the ion-sphere radius. The nature
of the Coulomb corrections depends on whether $\Gamma\gg 1$, the
ion-sphere limit, or $\Gamma\ll 1$, the Debye-H\" uckel limit.
These limits are well studied, \emph{cf.}, in Ref.~\cite{HansMcD}. 
The brief summary of the result is that in
the ion-sphere limit $\delta E$ and $\delta P$ are $O(\Gamma)$,
while in the Debye-H\" uckel limit they are $O(g)$ where $g$ is
the \emph{plasma parameter}, $e^2/(r_\mathrm Dk_\mathrm BT)$ in terms of the Debye
length $r_\mathrm D$. For a $Z=1$ plasma, $g$ is related to $\Gamma$ by
$g=\sqrt{6}\Gamma^{3/2}$. This means that Coulomb corrections to the EOS are
smaller in the Debye-H\" uckel theory than in an ion-sphere theory,
for weakly-coupled plasmas ($\Gamma<1$).

A consequence of the scaling of $\delta P$ and $\delta E$ with
$\Gamma$ or $g$, in view of equation (\ref{Estardef}), is that
the Coulomb terms in $E^*$ tends to a constant at
$T\rightarrow\infty$ with ion-sphere scaling \pdfcomment{(WHAT IS THIS??)}, but tend to zero as
$1/\sqrt{T}$ with Debye-H\" uckel scaling. The constant value of
the Coulomb energy per atom is
$\mathcal{O}(e^2/r_\mathrm s)$ in the ion-sphere case, since we recall that $r_s$
itself tends to a constant as $T\to\infty$. In fact, for
the hydrogen Hugoniot, $r_\mathrm s$ is about $2a_0$, which means that
the Coulomb corrections to $E^*$ are of the order of half a
hartree per atom, and they can nearly cancel $E_0$. In the Debye-H\" uckel
case the Coulomb corrections are significantly less, so $E^*$ is
closer to $|E_0|$.

\begin{figure}
	\includegraphics[scale=0.40]{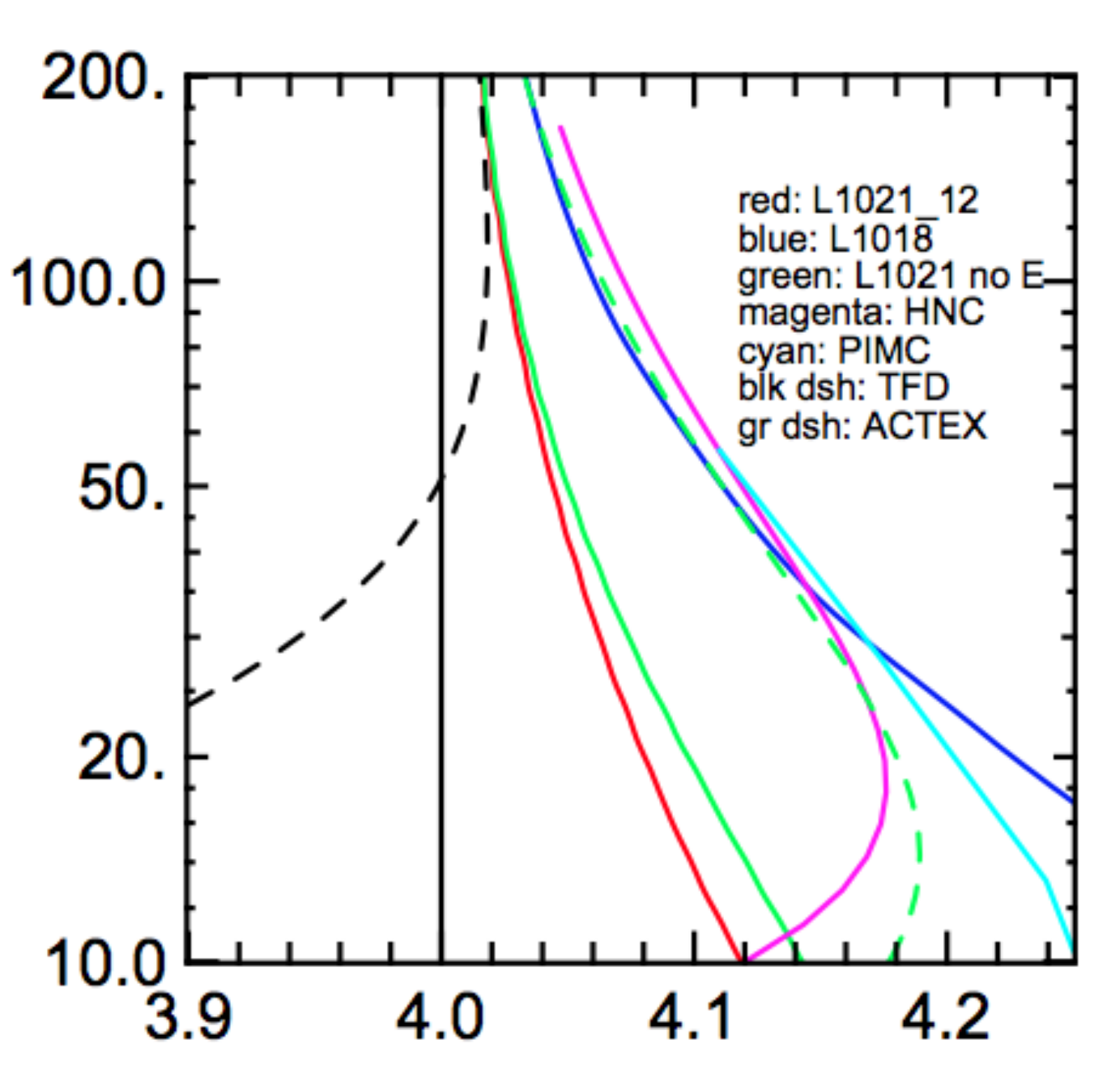}
	\caption{Principal Hugoniot curves ($P$ vs. $\rho/\rho_{0}$) for hydrogen in the regime approaching ideal gas behavior. The black vertical line indicates the (non-relativistic) ideal gas limit of $\rho/\rho_{0}= 4$. Results of various EOS models possessing different treatments of the electron-thermal component are shown (see text).
	}
	\label{fig:Castor}
\end{figure}

Our EOS model is based on \textsc{Purgatorio}, which is an 
ion-sphere model in which a single representative atom is described with single-electron orbitals from LDA-DFT embedded in a jellium background. Thomas-Fermi and Thomas-Fermi-Dirac (TFD) models are also ion-sphere models. All of 
these models assume a strictly uniform charge distribution outside the ion sphere, and
therefore do not have the expected Debye screening tail at
$r_\mathrm s< r\lesssim r_\mathrm D$. Since the electron density is not uniform
within the ion sphere, the high-$T$ behavior of the EOS is not
quite the same as that described above for the ion sphere
model. 
The comparison shown in Figure~\ref{fig:Castor} includes a 
TFD model, and two variants of our EOS which is \textsc{Purgatorio}-based (L1021\_12, and L1021\_12\_noEH2). The TFD
model shown in the figure has not been adjusted at all to match
the reference density and bulk modulus, and was made using $T=0$
Kohn-Sham exchange.

What is most evident in Fig.\ref{fig:Castor} is that in the
intermediate range 20--50 Mbar of pressure, the ion-sphere-like
models have noticeably less compression than the Debye-H\" %
uckel-like models. Notwithstanding the superior description of
the electronic states in \textsc{Purgatorio} (\emph{i.e.,} L1021), it
shares the same $T\to\infty$ limit as the Kohn-Sham TFD
model. This is not too surprising, since exchange is treated with
the same LDA in \textsc{Purgatorio}. But we see that PIMC \cite{Militzer},
as well as Kerley's 2003 model \cite{Kerley03}, have what is probably the
more accurate Debye-H\" uckel limit.

We should keep in mind that, while the ion-sphere models make a
substantial relative error in the Coulomb correction for
$\Gamma\ll 1$, this is actually a large error in a small quantity
in this case. The maximum difference in compression, around 20 Mbar,
is only of order two percent. 
Improving on PURGATORIO in this respect will require a model that
treats $g_{ii}(r)$ and $g_{ei}(r)$ at $r_s<r\lesssim
r_D$. Possibilities are the models of Blenski and
Cichocki~\cite{Blenski} (VAAQP) and Starrett and
Saumon~\cite{Saumon}. Such models should, like HNC, capture the
Debye-H\" uckel limit.

\section{Conclusions}
We have constructed a multiphase equation of state for elemental hydrogen which is fit almost entirely to ab initio electronic structure calculations of various sorts. The quantum ion physics is included using path integral methods for the protons. For all but the lowest densities, electrons are treated within DFT. The only experimental data used to fit the EOS is pressure cell data taken at cryogenic temperatures \cite{Silvera78,Silvera80}. The EOS is validated by comparing both to other calculations not used for fitting \cite{HuPRB}, and to a variety of shock data \cite{Hicks,Knudson01,Knudson04,HicksQuartz,Loubeyre}. 

The molecular solid free energy is constructed from cold and Debye model pieces, while the liquid free energy involves, in addition, vibrational and rotational terms and an electronic excitation contribution culled from DFT atom-in-jellium calculations \cite{Purgatorio}. To take into account the chemical equilibrium mixing of atomic H and molecular H$_{2}$ in the liquid, we employ an explicit free energy mixing procedure, like in the work of Refs.\cite{Kerley72,Kerley72pub,Kerley03}. However unlike in that approach, we include a mean-field term in our mixing model which gives rise to a liquid-liquid transition below a critical temperature, exactly as recently found in ab initio simulations \cite{MoralesPNAS,Scandolo2003}.

In making comparisons to other prominent EOS models, such as the most recent models of Kerley \cite{Kerley03}, Saumon \cite{Saumon12}, and Caillabet et al. \cite{Caillabet}, we see that the gross features are very similar, as expected. However, there are important differences as well: 1. As with the most recent of these EOS models \cite{Caillabet,Saumon12}, our EOS respects the current view of the hydrogen melt curve (maximum melt temperature $\sim$ 1000 K at $P\sim$ 1 Mbar), while the 2003-Kerley model does not \cite{Kerley03}, simply because its construction predated this view. 2. Though the Caillabet et al. model was also based on ab initio calculations \cite{Caillabet}, it did not address densities below $\sim$ 0.2 g/cc; as such, less care was taken to model the H-H$_{2}$ mixing physics. 3. There are important differences between our H-H$_{2}$ liquid treatment and Kerley's, though they are similar in spirit. First, Kerley performed constant-pressure mixing, while ours is constant-density (see our Appendix B). Second, Kerley modeled the H$_{2}$ units in a more sophisticated way relative to us, including both vibration-rotation coupling and a density-dependent molecular dissociation energy. 4. Our gradual transition to the ideal gas limit uses the atom-in-jellium model, \textsc{Purgatorio}\cite{Purgatorio}, 
to account for electronic excitation, and our newly-developed cell model (see our Appendix A) to provide an approximate description of strong anharmonicity leading to freely-moving ions at high-$T$. These are radically different approaches to those taken in the construction of the other EOS models. In particular, our use of the ion-sphere model, \textsc{Purgatorio}, gives us what we assume to be a more nuanced description of pressure-dependent ionization in regimes beyond where we performed the more taxing ab initio MD simulations, but its use also gives rise to subtle inaccuracies relative to the other models (see Section V.E.1) in the weakly-coupled plasma regime. 

Possibly the most important distinguishing feature of our hydrogen EOS is that it is fit to what we believe to be the most accurate simulation data currently available for this problem. Fitting to experimental data was intentionally {\it deemphasized}, partly as an exercise to see the extent to which the resulting theoretically-derived EOS model can correctly post-dict experimental results (such as shock data). In this way, our model construction philosophy is distinctly different from Kerley and Saumon et al., but is similar to that of Caillabet et al. It is therefore noteworthy that even quite detailed metrics of these different EOS models yield strikingly similar results in general (see Section V). And indeed, though we save the discussion of this for a later work, preliminary investigations have shown that ICF simulations conducted with our EOS, 2003-Kerley, and 2012-Saumon are nearly indistinguishable \cite{ICFsimulations}. Does this mean that the hydrogen EOS is a now a solved problem? Undoubtedly, this depends on the intended application- for instance, planetary physics applications tend to stress much lower temperatures than ICF, where the EOS models are more different. Furthermore, it is entirely possible that the differences between these EOS models are significantly smaller than the difference between any one of them and nature's "true" EOS for hydrogen. Further research, particularly in the realm of high energy density physics experiment, will be needed to settle this matter. 

\begin{acknowledgments}
	We thank S.A. Bonev, S. Hamel, D.S. Clark, D. Saumon, M.D. Knudson, M.P. Desjarlais, and E.D. Chisolm for helpful discussions.
	This work was performed under DOE Contract No. DE-AC52-07NA27344. 
	Computing support for this work came from the LLNL Institutional Computing Grand Challenge program.
\end{acknowledgments}

\appendix
\section{Cell model}

The objective of this Appendix is to derive a simple expression that
interpolates between the Mie-Gr\"uneisen equation of state (appropriate for a
solid or low-$T$ elemental liquid) and the ideal gas limit for the ion-thermal
part of the free energy. The functional form we adopt for this piece is
justified with a simple model. Different aspects of the ideal gas limit, $C_{\rm
	V} = \frac{3}{2} k_{\rm B}$, and the ideal gas law, $PV= k_{\rm B} T$, are
recovered independently.

\subsection{Mie-Gr\"uneisen Solid}

First, we consider an ensemble of independent nuclei (or ions), each of which is
subjected to a confining potential, $\phi(r)$. The ion-thermal part of the free
energy can be deduced from the partition function of a particle in this
potential. Assuming the nuclei motion to be classical, the partition function is

\begin{equation}\label{Zclass} 
	Z = \frac{1}{h^3}\int\mathrm d\mathbf p\int\mathrm d\mathbf r
	\mathrm e^{-\frac{[\frac{p^2}{2m} + \phi(\mathbf r)]}{k_\mathrm B T}}.
\end{equation} 

The integral in momentum space can be done directly (which is a property of
classical systems):

\begin{equation}\label{Zr}
	Z = \frac{(2\pi m k_{\rm B}T)^\frac{3}{2}}{h^\frac{3}{2}} \frac{1}{h^\frac{3}{2}} 
	\int \mathrm d\mathbf r\mathrm e^{-\frac{\phi({\bf r})}{k_{\rm B}T}}. 
\end{equation} 

For a quadratic potential well, we have $\phi(r) = m \omega^2 r^2 /2$ and

\begin{equation}
	Z = \frac{(2\pi
		m k_{\rm B} T)^\frac{3}{2}}{h^\frac{3}{2}} \frac{1}{\hbar^\frac{3}{2}}
\int_0^\infty 4\pi \mathrm dr r^2 \mathrm e^{- \frac{m \omega^2 r^2}{2 k_{\rm B}T} },
\end{equation} 

(though obvious from elementary physics, these detailed steps will be useful in
the following development). The Gaussian integral above can be done
analytically, yielding this simple form:

\begin{equation} 
	Z 
	= \frac{(2\pi m k_\mathrm B T)^\frac{3}{2}}{h^\frac{3}{2}} \frac{1}{h^\frac{3}{2}} 4\pi
	\sqrt{\frac{\pi}{2}} \left[ \frac{k_{\rm B}T}{m \omega^2} \right]^\frac{3}{2}
	=\left[\frac{k_{\rm B}T}{\hbar\omega}\right]^3 
\end{equation}

Note that the the partition function is independent of the mass $m$, and the
only relevant energy scale is $k_B\theta = \hbar\omega$ (comming from the
quantum mechanical normalization of the partition function). The associated
ion-thermal Helmholtz free energy, $F = -k_{\rm B} T \log(Z)$, is

\begin{equation}
	F = - 3 k_{\rm B} T \log\left[ \frac{T}{\theta} \right],
\end{equation} 

The characteristic temperature scale $\theta$, can be interpreted as the Debye
or Einstein temperature. For a real solid, $\theta$ is volume-dependent, because
the curvature of the confining potential depends on the density. In a more
nuanced description, $\theta$ is not given by a single frequency but is derived
from the normalized phonon density of states at a given volume $D_V(\omega)$
\cite{Wallace98,Wallace2002}, \begin{equation} k_{\rm B}\theta(V) = \hbar
\mathrm e^{1/3}\exp\left(\int \log(\omega) D_V(\omega) \,\mathrm d\omega\right);
\end{equation} this logarithmic moment of $D_{V}(\omega)$ is used if an accurate
description of the high-$T$ classical behavior ($T > \theta$) of the free energy
is desired.

The Mie-Gr\"uneisen ion-thermal free energy has an associated ion-thermal
energy, $E = F + TS = F - T \left(\frac{\partial F}{\partial T}\right)_{V}$,

\begin{equation} 
	E = 3 k_{\rm B} T.
\end{equation} 

The resulting specific heat (per particle), $C_{\rm V} = \left(\frac{\partial
	E}{\partial T}\right)_{V} = - T \left(\frac{\partial^2 F}{\partial
	T^2}\right)_{V}$, is the Dulong-Petit constant 

\begin{equation} 
	C_{\rm V} = 3
	k_{\rm B}.
\end{equation} 

The ion-thermal pressure is given by,

\begin{equation}
	P = -\left(\frac{\partial F}{\partial V}\right)_{T}= \frac{3 k_{\rm B} T \left[
		- \frac{\mathrm d\log\theta}{\mathrm d\log V} \right]}{V}. 
\end{equation}
$\theta(V)$ is, for example, the Gr\"uneisen model. 
Mie-Gr\"uneisen formulation can be used to model a dense fluid.


\subsection{The Chisolm-Wallace Fluid}

Chisolm and Wallace\cite{Chisolm-Wallace,ChisolmAl} hypothesized that a dense
fluid behaves essentially like a solid that is allowed to oscillate in a
potential landscape consisting of a large number of potential wells in
multi-particle configuration space. Each potential well has a characteristic
vibration energy (i.e. a Debye temperature) and the time that the system spends
traveling from one potential well to another is negligible. On average, the
individual potential wells have a weighted average Debye temperature,
$\bar\theta$. The fact that the system can explore a certain number of
configurations, $W$, around which to oscillate adds to an extensive additional
configurational entropy given by $k_\mathrm B\log(W)$. An extra hypothesis made
by Chisolm and Wallace is that the total number of such wells increases
exponentially with the number of particles $N$ in the macroscopic limit (which
make entropy an extensive quantity), and the stronger hypothesis that $W$ is
independent of temperature (at least within the range of validity of the model),
so  $W = w^N$. The resulting ion- thermal free energy has the form:

\begin{equation}
	F = 
	3 k_{\rm B} T \log
	\left[
		\frac{\bar \theta}{T} \right] - k_{\rm B} T \log(w) = 3 k_{\rm B} T \log\left[\frac{\bar \theta/ w^{1/3} }{T}
	\right] 
\end{equation} 

Since the first and second terms have compatible forms, the ion-thermal free
energy is indistinguishable from that of a classical solid with an
\emph{effective Debye
	temperature} given by $\tilde\theta = \bar \theta / w^{1/3}$. Moreover they find
that $w$ is a rather universal number of around $0.8$ for a wide range of
elements\cite{Wallace2002}.

This model describes certain monoatomic dense fluids quite well in the
neighborhood of melting, as long as they have a heat capacity close to the
Dulong-Petit value \cite{Correa,LXBBe}. This model can not hold at all
temperatures, in particular because $C_V^\mathrm{ion}$ is fixed at $3 k_\mathrm
B/\text{atom}$ even at high $T$. The following is an attempt to restore the right high
temperature limit for the heat capacity of the Chisolm-Wallace fluid by adding a
correction to the free energy based on a modified partition function.

\subsection{The monoatomic ideal gas limit for the specific heat}

The ideal gas limit can be recovered from the Mie-Gr\"uneisen free energy after
a simple modification to Eq.\ref{Zr}; we impose a radial cutoff, $R$:

\begin{equation}\label{Zcell}
	Z = \frac{(2\pi m k_{\rm B} T)^\frac{3}{2}}{h^\frac{3}{2}} \frac{1}{h^\frac{3}{2}}
	\int_0^R 4\pi\mathrm dr r^2\mathrm e^{-\frac{m \omega^2 r^2}{2 k_{\rm B}T}}
\end{equation}

A posteriori, the parameter $R$ has at least three possible physical
interpretations no mutually exclusive:

\begin{description}
		
	\item[Anharmonicity:] $R$ can represent a distance beyond which the potential
	becomes  anharmonic due to greatly increased stiffness. Of course, this is a
	simplified picture; in real space the displacements are given by combinations
	of normal modes.
		
	\item[Phase space constraint:] Even if we decide that the potential remains
	harmonic at \emph{every amplitude of motion}, we have to impose a volume (per
	particle) constraint. Irrespective of other considerations, such a spatial
	cutoff must be imposed to constrain the total volume available, given a fixed
	number of particles.
	
	\item[Hard sphere repulsion/exclusion volume:] As the displacement of a
	particle is increased, it must eventually feel the repulsive force produced by
	neighboring \emph{effective} atoms. This is related to the concept of excluded
	volume, invoked for instance in the Van der Waals EOS model \cite{Hill}.
	
\end{description}
							
The partition function of Eq.\ref{Zcell} can be obtained in an analytic form,
although the expression involves the `error function' (${\rm erf}$):

\begin{equation}
	Z = \frac{(2\pi m k_{\rm B} T)^\frac{3}{2}}{h^\frac{3}{2}}
	\frac{1}{h^\frac{3}{2}} 4\pi\sqrt{\frac{\pi} {2}} \left[ \frac{k_{\rm
		B}T}{m\omega^2} \right]^\frac{3}{2} \left\{{\rm erf}\left(\sqrt{\frac{m
		\omega^2}{2 k_{\rm B} T}} R\right) - \frac{2}{\sqrt{\pi}} \mathrm e^{-\frac{m
		\omega^2 R^2}{2 k_{\rm B} T}} \frac{ \sqrt{m \omega^2} R}{\sqrt{2 k_{\rm B} T}}
\right\}.
\end{equation}
This expression is simplified in the same way as for the harmonic partition
function, by replacing $\hbar \omega$ by $k_{\rm B}\theta$, 

\begin{equation}
	Z = \left[ \frac{T}{\theta} \right]^3 \left\{{\rm erf}\left(\sqrt{\frac{T^*}{T}}\right) - \frac{2}{\sqrt\pi} \sqrt{\frac{T^*}{T} } \mathrm e^{-\frac{T^*}{T}} \right\},
\end{equation}

where we have introduced a second scale of temperature

\begin{equation}\label{TstarR}
	k_{\rm B} T^* = \frac{m \omega^2 R^2}{2} = \frac{m k_{\rm B}^2 \theta^2 R^2}{2 \hbar^2}.
\end{equation}

This \emph{saturation} temperature $T^*$ is the thermal energy at which the
oscillator starts being affected by the radial cutoff, $R$, imposed in the
calculation of the partition function. The associated free energy is

\begin{equation}
	F = -3 k_{\rm B} T \log\left[ \frac{T}{\theta} \right] - k_{\rm B} T\log\left\{{\rm erf}\left(\sqrt{\frac{T^*}{T}}\right) - \frac{2}{\sqrt\pi} \sqrt{\frac{T^*}{T} } \mathrm e^{-\frac{T^*}{T}} \right\}.
\end{equation}

We recognize the first term as the classical high-temperature limit of harmonic
oscillators with Debye temperature $\theta$. By construction, we must have $T >
\theta$ to ensure that the classical limit assumed at the outset (see
Eq.\ref{Zclass}) is valid. (Eventually this term can be replaced by its quantum
mechanical counterpart, which includes the classical Mie-Gr\"uneisen form as a
limit; the analysis is unaffected as long as $T^*>\theta$.) There is, however,
no fundamental restriction on $T$ in relation to $T^*$.
							
To obtain limiting behaviors, we note that the error function exhibits the limiting behaviors,

\begin{equation}
	{\rm erf}(x\to 0) = \frac{1}{\sqrt\pi} \left(2 x - \frac{2}{3} x^3 +
	\frac{1}{5} x^5  - \frac{1}{21} x^7 \right)+ \mathcal{O}(x)^9,
\end{equation}

and 

\begin{equation}
	{\rm erf}(x\to \infty) = 
		1 - \frac{\mathrm e^{-x^2}}{\sqrt{\pi}} 
			\left( 
				  x^{-1} 
				- \frac{x^{-3}}{2} 
				+ \frac{3 x^{-5}}{4} 
				+ \mathcal{O}(x)^{-7} 
			\right).
\end{equation}

For $T\ll T^*$ we have 
\begin{equation}
	F_{T\ll T^*} = -3 k_\mathrm B T \log\left[ \frac{T}{\theta} \right]  -
	k_\mathrm B T \log\left\{ 1 - \frac{2}{\sqrt{\pi}} \sqrt{\frac{T^*}{T}} \mathrm
	e^{-\frac{T^*}{T}} + \cdots \right\}.
\end{equation}
The heat capacity in this limit is seen to asymptote to the Dulong-Petit value: 
\begin{equation}
	{C_{\rm V}}_{T \ll T^*} = 3 k_{\rm B} - \frac 2{\sqrt\pi} k_\mathrm B \left(\frac{T^*}{T} \right)^{5/2} \mathrm e^{-\frac{T^*}{T}} + \cdots
\end{equation}

(Note that the function $\mathrm e^{-1/x}$ is a very flat --non-analytic smooth
function-- at $x\to 0^+$, so the correction of the second term is exponentially
small at low temperature and not expressable as a power series).

For $T\gg T^*$, the free energy tends to 
\begin{equation}
	F_{T/T^* \gg 1} = -3 k_\mathrm B T \log\left[ \frac{T}{\theta} \right] +
	\frac32 k_\mathrm B T \log\left[\frac{T}{T^*}\right] - k_\mathrm B T
	\log\left[ \frac4{3\sqrt\pi} \right] + \frac35 k_\mathrm B T^* -
	\frac6{175} k_\mathrm B\frac{(T^*)^2}{T} + \cdots
\end{equation}

By differentiating the last expression with respect to $T$,  we obtain

\begin{equation}
	{C_{\rm V}}_{T \gg T^*} = \frac{3}{2} k_{\rm B} + \frac{12}{175} k_{\rm B}
	\left(\frac{T^*}{T}\right)^2 + \cdots,
\end{equation}
which tends to the ideal gas heat capacity as $T\to\infty$.
							
It turns out that (assuming $T^*$ to be independent of $T$) the roughly
quadratic decay in our model of $C_V$ from $3k_\mathrm B$ to $\frac32 k_\mathrm
B$ as $T \to \infty$ is faster than that predicted by legacy
	classical MD conducted
with fixed ($T$-independent) interparticle soft-sphere potentials\cite{Grover}.
Such simulations suggest a power-law decay with a smaller exponent instead \pdfcomment{(WHAT VALUE???)}. 
Our $C_\mathrm V\to\frac32 k_\mathrm B+\mathcal O(T)^{-2}$ dependence
results from the assumption of the radial cutoff in Eq.\ref{Zcell}, together
with the simple single-frequency Einstein oscillator treatment. Other more
sophisticated models can certainly be derived which exhibit different
$T$-dependent decays; we choose this variant because it has an analytic free
energy, and as the main body of our manuscript shows, it is largely sufficient
for our purposes of fitting our H EOS model to the results of our ab initio
calculations. The generality of this assumption and practical usage is yet to be
investigated in future work.
							
It is important for our discussion to confirm that the entropy has a functional
form that is \emph{compatible} with the ideal gas entropy (the Sackur-Tetrode
equation \cite{Hill}). In our model, we have

\begin{equation}
	S_{T/T^*\gg 1} = k_{\rm B} \log\left[ \left(\frac{\sqrt{T^*}}{\theta}\right)^3
	T^\frac{3}{2} \right] + k_{\rm B}\left(\frac{3}{2} + \log\left[
	\frac4{3\sqrt\pi} \right]\right) - \frac{6}{175} k_{\rm B} \left(\frac{T^*}{T}
	\right)^2 + \cdots
\end{equation}s

Whether or not this entropy corresponds to the ideal gas law still depends on
the (yet unspecified) volume dependence of the parameters $\theta$ and $T^{*}$.
To this end, we now move on to a discussion of this $V$-dependence after a short
digression in which we consider 1D and constraints in momentum space.
							
							
\subsubsection{1D case}\label{sec:cell1d}
In the 1D case the equations simplify to:
\begin{equation}
	Z = 2\pi \frac{k_\mathrm B T}{\hbar\omega}{\rm erf}(\sqrt{T^*/T}) 
\end{equation}
\begin{equation}
	F = -k_\mathrm B T \log\left(\frac{k_\mathrm B T}{\hbar\omega}\right) - k_\mathrm B T \log\left[ {\rm erf}(\sqrt{T^*/T}) \right] 
\end{equation}
We use this 1D result to model the changes to the diatomic vibrational partition function (and free energy) resulting from the effects of dissociation.
							
\subsubsection{Momentum space constraint}\label{sec:cellmoment}
In the previous section we imposed a spatial constraint on the system in order to
reproduce the specific heat of the free particle (ideal) gas. We can just as easily impose a  constraint in momentum space. We think of this as a constraint which limits the particle momenta, as is natural for a system with long-range order. Indeed, in a solid, both the momenta and the positions of particles are constrained in a symmetric way. The resulting equations are:
\begin{equation}
	F = -3 k_{\rm B} T \log\left[ \frac{T}{\theta} \right] - 2 k_{\rm B} T\log\left\{{\rm erf}\left(\sqrt{\frac{T^*}{T}}\right) - \frac{2}{\sqrt\pi} \sqrt{\frac{T^*}{T} } {\rm e}^{-\frac{T^*}{T}} \right\},
\end{equation}
obtained by imposing cutoffs to the classical oscillator partition function in both real space and momentum space simultaneously.
Note the factor of 2 multiplying the second term, 
which causes the resulting $C_{V} \to 0$ for $T \gg T^{*}$. The one dimensional counterpart (useful to model bounded vibrations of a diatomic molecule)
is:
\begin{equation}
	F = -k_\mathrm B T \log\left(\frac{k_\mathrm B T}{\hbar\omega}\right) - 2 k_\mathrm B T \log\left[ {\rm erf}(\sqrt{T^*/T}) \right].
\end{equation}
							
\subsection{Pressure}
							
The heat capacity is only one aspect of the ideal gas limit, the other being the
mechanical equation of state, $P(T,V)$. To obtain the associated pressure we
need an explicit volume dependence of the parameters (in the same way that we need
the volume dependence of $\theta$ in the Mie-Gr\"uneisen model). Although the
partial derivatives of the free energy are, in general, complicated and are not
presented fully here, the expression for the pressure is very important and
deserves careful attention. At low temperature, the pressure is equal to the
Mie-Gr\"uneisen thermal pressure with an exponentially small correction (the
correction being proportional to ${\rm e}^{-T^*/T}$). At high temperature, the
pressure is: 
\begin{equation}\label{P} P_{T/T^*\gg 1} = \frac{3 k_{\rm B} T
		\left[ - \frac{V}{\theta} \frac{{\rm d}\theta}{{\rm d}V} +
		\frac{1}{2}\frac{V}{T^*}\frac{{\rm d}T^*}{{\rm d}V} \right] }{V} + \frac{3}{5} k_{\rm B}
\frac{{\rm d}T^*}{{\rm d}V} - \frac{12}{175} k_{\rm B} \frac{T^*}{T} \frac{{\rm d}T^*}{{\rm d}V} +
\cdots \end{equation} 
We have yet to specify the $V$-dependence of $\theta$ and
$T^{*}$, yet the heat capacity of the monoatomic ideal gas is recovered at each
volume, \emph{irrespective of the volume dependence of these parameters}. On the
other hand, the pressure and its limiting values depend explicitly on the
assumed $V$-dependence of $\theta$ and $T^*$.

The thermodynamic equation of state, $P(V,T)$, is given once $T^*(V)$ or $R(V)$ is specified, assuming the relation in Eq.\ref{TstarR},
\begin{equation}
	k_{\rm B} T^*(V) = \frac{m k_{\rm B}^2 \theta(V)^2 R(V)^2 }{2 \hbar^2}.
\end{equation}
The ideal gas equation of state, $\label{ideal}P(V, T) = k_{\rm B} T/V$,
is recovered for a \emph{particular} choice of this $T^{*}(V)$ \cite{caveat5}. The volume dependence of $T^{*}$ (or $R$) that recovers the ideal gas law can be constructed by appealing to Eqs.\ref{P} and \ref{ideal}, and requiring that the 
linear-in-$T$ term of Eq.\ref{P} is equal to the right hand side of Eq.\ref{ideal}. This means that the expression in brackets in Eq.\ref{P} becomes
\begin{equation}
	-\frac{\mathrm d\log\theta}{\mathrm d\log V} + \frac{1}{2} \frac{\mathrm d\log T^*}{\mathrm d\log V} = \frac{1}{3},
\end{equation}
or
\begin{equation}
	\frac{\mathrm d\log[\frac{\sqrt{T^*}}{\theta}]}{\mathrm d\log V} = \frac{1}{3},
\end{equation}
which integrates to 
\begin{equation}
	\frac{\sqrt{T^*}}{\theta} \propto V^{1/3},
\end{equation}
and finally
\begin{equation}\label{Tprop}
	T^*(V) \propto \theta(V)^2 V^{2/3}.
\end{equation}
We obtain this only by requiring the ideal gas law at high $T$. 
The proportionality constant in Eq.~\ref{Tprop} can be obtained by requiring that the entropy of the ideal gas asymptotes to the ideal gas entropy as $T \to \infty$.

\begin{equation}
	k_\mathrm B T^*(V) = \frac{m k_\mathrm B^2 \theta(V)^2 \left( \frac{3\mathrm e}{4\pi} V
	\right)^{2/3} }{2 \hbar^2}.
\end{equation}

In other words, $R(V) = (\frac{3\mathrm e}{4\pi} V)^{1/3}$, where the factor
`$\mathrm{e}$' originates from the fact that the system transitions from $N$
\emph{distinguishable} oscillators at low temperature to $N$
\emph{indistinguishable} particles at high temperature.

\section{Mixing model}

The partition function of a collection of $M$ independent classical molecules and $A$ independent atoms mixed in a volume $V$ is:
\begin{equation}\label{ZmixA}
	Z_\mathrm{mix}(A,M,V) = \frac{z_\mathrm{M}(V)^M}{M!} \frac{z_\mathrm{A}(V)^A}{A!}.
\end{equation}
where $z_{\{\mathrm{M, A}\}}$ are defined in terms of the partition functions of a system composed by $M_0$ molecules only or $A_0$ atoms only:
\begin{equation}
	Z_\mathrm{M}(M_0,V) = \frac{z_\mathrm{M}(V)^{M_0}}{M_0!},
\end{equation}
and
\begin{equation}\label{ZA}
	Z_\mathrm{A}(A_0,V) = \frac{z_\mathrm{A}(V)^{A_0}}{A_0!}. 
\end{equation} 
The last two equations are the partition functions in systems where either only
molecules or only atoms are allowed; Eq.~\ref{ZmixA} allows for arbitrary
ratios. When applied to ideal gases (i.e., gases of noninteracting particles)
these expressions are exact, their use is the basis of the Saha equation for
molecular-atomic equilibrium \cite{Hora}.

The idea described here is an extension of the theory to the case of nonideal
gases, in which the free energies of the pure cases (A and M, individually) are
assumed to be known \emph{a priori}. Note that for nonideal gases the
molecules/atoms cannot be thought to be independent of each other, however the approximation
we employ is that the partition functions can be factored in this way, for the
purpose of mixing. In Appendix C below, we attempt to remove this simplifying assumption by introducing a coupling which allows the free energies of molecules and atoms to depend on their relative concentrations.

The partition functions labeled A and M describe different possible states of
the \emph{same} system; conservation of the total number of atoms relates the
parameters of the first three equations, assuming the molecules (M) to be
diatomic: \begin{equation} 2 M  + A = 2 M_0 = A_0. \end{equation}
If we assume that the partition functions $Z_\mathrm{M}$ and $Z_\mathrm{A}$ (or their free energies) are known (and for example are given by the sum of idealized `cold', `IT' and 'ET' pieces), we can express $Z_\mathrm{mix}$ in terms of them, where $A$ and $M$ are considered variational parameters. 
Solving for the effective one-particle partition functions $z_\mathrm M$ and $z_\mathrm A$, one obtains:
\begin{equation}
	z_\mathrm{M} = (M_0! Z_\mathrm{M}(M_0))^{1/M_0},
\end{equation}
and
\begin{equation}
	z_\mathrm{A} = (A_0! Z_\mathrm{A}(A_0))^{1/A_0}.
\end{equation}
Substituting these into Eq.\ref{ZmixA}, we get
\begin{equation}
	Z_\mathrm{mix}(A,M)= \frac{(M_0! Z_\mathrm{M}(M_0))^{M/M_0}}{M!} \frac{(A_0! Z_\mathrm{A}(A_0))^{A/A_0}}{A!}
\end{equation}
The variational parameters $A$ and $M$ are obtained by maximizing (`sup') the value of $Z_\mathrm{mix}(A, M)$ (or equivalently minimizing the free energy) subject to constraints \cite{Nickel1980},
\begin{equation}
	Z_\mathrm{mix}(A_0, V) = \sup_{A, M, 2 M  + A = A_0} Z_\mathrm{mix}(A,M, V).
\end{equation}
The maximizing values can be obtained by making a small change in $A$ and $M$, consistent  with the constraint: $M \to M - 1$, $A \to A+2$. At the maximum, the value of $Z_\mathrm{mix}$ is unchanged by this substitution, i.e. $Z_\mathrm{mix}(A, M) = Z_\mathrm{mix}(A + 2, M-1)$,
\begin{equation}
	\frac{z_\mathrm{M}(V)^M}{M!} \frac{z_\mathrm{A}(V)^A}{A!} = \frac{z_\mathrm{M}(V)^{M-1}}{M-1!} \frac{z_\mathrm{A}(V)^{A+2}}{A+2!},
\end{equation}
or
\begin{equation}
	(A + 2)(A + 1)/ M = z_\mathrm{A}(V)^2/z_\mathrm{M}(V).
\end{equation}
For large numbers, $A$ and $M$, we have,
\begin{equation}\label{large}
	A^2/ M = z_\mathrm{A}(V)^2/z_\mathrm{M}(V) \equiv \xi(V).
\end{equation}
Together with the constraint, 
\begin{equation}
	2 M  + A = 2 M_0 = A_0,
\end{equation}
Eq.\ref{large} can be solved for $A$ and $M$,
\begin{equation}\label{4A}
	4 A = \sqrt{\xi(V)} \sqrt{\xi(V)+ 8 A_0} - \xi(V)
\end{equation}
\begin{equation}\label{8M}
	8 M = 4 A_0 + \xi(V) -  \sqrt{\xi(V)} \sqrt{\xi(V)  + 8 A_0}
\end{equation}

At this point we work instead with the free energy of the system. In particular, we will reconstruct the partition function from the free energy.
A crucial point is that we choose to reconstructruct the partition function of single molecules/atoms from the partition function of the pure molecular or pure atomic state \emph{at the same density}. For a single atom, we have
\begin{equation}
	z_{\rm A}(V) = A_0!^{1/A_0} Z_\mathrm{A}(A_0, V)^{1/A_0} = A_0!^{1/A_0} \mathrm e^{-\frac{F_\mathrm{A}(A_0, V)/A_0}{k_\mathrm BT}}, 
\end{equation}
while for a single molecule,
\begin{equation}
	z_{\rm M}(V) = M_0!^{1/M_0} Z_\mathrm{A}(M_0, V)^{1/M_0} = M_0!^{1/M_0} \mathrm e^{-\frac{F_\mathrm M(M_0, V)/M_0}{k_\mathrm BT}},
\end{equation}
where $M_0 = A_0/2$.  
This construction assumes that the effective molecule or atom will behave the same irrespective of the ratios of molecules vs. atoms in the mixture, as long as the density is the same. This is only correct for ideal gases, i.e., dilute mixtures of atoms and molecules; hence, the construct is questionable at high densities.

We then define the free energies of the atomic and molecular systems at a given voume \emph{per atom}, 
$f_\mathrm{A}(V) = F_\mathrm A(V)/A_0$ and $f_\mathrm{M}(V) = F_\mathrm M(V)/(2 M_0)$:
\begin{equation}\label{Bmol}
	z_{\rm A}(V) = \mathrm e^{-\frac{f_\mathrm A(V)}{k_\mathrm B T} - 1}  A_0,
\end{equation}
and
\begin{equation}\label{Bat}
	z_{\rm M}(V) = \mathrm e^{-\frac{2 f_\mathrm{M}(V)}{k_\mathrm B T} - 1} A_0/2,
\end{equation}
where we have used $N!^{1/N} \simeq N/\mathrm e$. Appealing to Eq.\ref{large} gives
\begin{equation}
	z_{\rm A}(V)^2/z_{\rm M}(V) = \xi(V) = \mathrm e^{-2 \frac{f_\mathrm{A}(V) - f_\mathrm{M}(V)}{k_\mathrm B T} } A_0 2/\mathrm e =  \mathrm e^{-2 \frac{\Delta f(V)}{k_\mathrm B T}} 4 A_0/\mathrm e/2.
\end{equation}
Here we have introduced the free energy difference, per atom, $\Delta f(V)$.
Note that the factor in the exponent looks like a Boltzman factor, however the variable appearing in it is the intensive {\it free} energy and not the mechanical energy. Moreover, the factor of 2 gives the free energy \emph{per pair of atoms}.
This normalization to 2 atoms stems from the fact that we assumed that molecules and atoms are the independent entities for the purposes of mixing.
Plugging into Eqs.\ref{4A} and \ref{8M} yields the atom and molecule fractions: 

\begin{equation}
	x = A / A_0 = \left( \sqrt{\mathrm e^{-2 \frac{\Delta f(V)}{k_\mathrm BT}}/\mathrm e/2}
	\sqrt{\mathrm e^{-2 \frac{\Delta f(V)}{k_\mathrm B T}}/\mathrm e/2 + 2} - e^{-2
		\frac{\Delta f(V)}{k_\mathrm BT}}/\mathrm e/2 \right),
\end{equation}
and 
\begin{equation}
	1 - x = 2 M / A_0 = M / M_0 = 1 -  \left(\sqrt{\mathrm e^{-2 \frac{\Delta f(V)}{k_\mathrm BT}}/\mathrm e/2} \sqrt{e^{-2 \frac{\Delta f(V)}{k_\mathrm B T}}/\mathrm e/2 + 2} - \mathrm e^{-2 \frac{\Delta f(V)}{\mathrm k_B T}}/\mathrm e/2\right).
\end{equation}

Note these important limits: 
\begin{itemize}
	\item If at a certain density and temperature $\Delta f = 0$, then $A/A_0 = ({\sqrt{1 + 4 \mathrm e}-1})/2/\mathrm e \sim 0.45$ and $M/M_0 = 1- ({\sqrt{1 + 4\mathrm e}-1})/2/\mathrm e \sim 0.55$. 
	\item If the temperature is low (relative to $\Delta f = f_\mathrm A - f_\mathrm M$) and $f_\mathrm M < f_\mathrm A$, then $A/A_0 = 0$ and $M/M_0 = 1$. Otherwise, if $f_M > f_A$, then $A/A_0 = 1$ and $M/M_0 = 0$. 
	\item For the high temperature limit to be well defined, 
	one has to make assumptions on the asymptotic behavior of $f_\mathrm A$ and $f_\mathrm M$. 
	Physically, the difference between them must increase superlinearly with temperature, for the limit to be expected one. 
	This is indeed the case, since the limiting specific heat of the atomic free energy is larger than that of the molecular one. This gives $f_\mathrm A < f_\mathrm M$, with the difference increasing faster than $T$, resulting in the limit: $A/A_0 = 1$ and $M/M_0 = 0$.
\end{itemize}

From a statistical standpoint $M/M_0$ and $A/A_0$ are just variational parameters tuned to maximize $Z_{\rm mix}$, but as we have implied above, we are compelled to interpret them as the fraction of dissociated atoms and molecules. What ultimately matters most, however, is the way in which these parameters participate in the final expression for the free energy of the mixture:
\begin{equation}
	F_\mathrm{mix}(A_0, V) = -k_\mathrm B T \log Z_\mathrm{mix}(A_0, V)
\end{equation}
\begin{equation}
	F_\mathrm{mix}(A_0, V) = -k_\mathrm B T(M \log z_M(V) - \log M! +  A\log z_A(V) - \log A!)
\end{equation}
\begin{equation}
	F_\mathrm{mix}(A_0, V) = 2 M f_\mathrm{M}(V) + A f_\mathrm{A}(V) + k_\mathrm B T\left[M\log(M/M_0) + A\log(A/A_0)\right];
\end{equation}
or per atom:
\begin{equation}\label{solfmix1}
	f_\mathrm{mix}(V) = (1-x) f_\mathrm{M}(V) + x f_\mathrm{A}(V) + k_\mathrm B T \left[(1-x) \log(1-x)/2 + x \log(x)\right]
\end{equation}
where
\begin{equation}\label{solx}
	x = \mathrm e^{-2 \frac{f_\mathrm{A}(V) - f_\mathrm{M}(V)}{k_\mathrm BT} -1} \left( \sqrt{1 + 4 \mathrm e^{2 \frac{f_{\rm A}(V) - f_{\rm M}(V)]}{k_\mathrm B T} + 1}} - 1\right)/2.
\end{equation}
Note that $x$ depends on the temperature and volume as well, therefore the partial derivatives of $f_\mathrm{mix}$ with respect to $T$ and $V$ will be affected by this dependence. Therefore, the last term in Eq.\ref{solfmix1} is not the full entropy but only a contribution to it.

\section{Non-ideal mixing model for the critical fluid}

By constructing a theory of the mixed phase from two ideal phases (see Appendix B above) we find an explicit free energy for the former in terms of the latter. However the
construction holds only under the assumption that the two species are statistically independent, and the theory can be only justified in this case.
In order to construct a more general theory that departs from this simplification we
introduce a coupling between the species. The idea is that constituents of one species can be thermodynamically favored
or penalized when surrounded by constituents of the same or different type. This
effect will be important in the neighborhood of conditions where the free energy difference between the species, 
($\frac{\Delta f}{k_\mathrm B T} \sim 0$) and when the temperature is similar to or smaller than this coupling energy.

If we assume that this favoring or penalizing is independent of the temperature
(or at least is fairly constant in the regime where the effect matters- i.e.,
where $\Delta f = 0$), we can construct the mean field (MF) relations that
describe this situation. The partition functions of the two species will now 
depend on the (still undetermined) average composition $x$ (compare to the composition-independent expressions of Eqs.\ref{Bmol} and \ref{Bat}):
\begin{equation}
	z_\mathrm A(V, x) = \mathrm e^{-  \beta [f_\mathrm A(V) +   g(x)] - 1} A_0
\end{equation}
\begin{equation}
	z_\mathrm M(V, x) = \mathrm e^{-2 \beta [f_\mathrm M(V) - 2 h(x)] - 1} A_0/2
\end{equation}
The terms $g(x)$ and $h(x)$ are the mean field representation of a local molecular
field due to the environment. Since this molecular field is symmetric with
respect to molecules and atoms, we posit $g(1-x) = h(x)$.
Because $z_\mathrm{A,M}$ and the resulting $f_\mathrm{A, M}$ describe 
pure phases, the only restriction on $h(x)$ is that $h(0) = 0$. 
Other than this, $h(x)$ can have any behavior. A simple positive $h(x)$ will promote the tendency for free atoms to favor nearby free atoms, and molecules to favor nearby molecules. The model interaction $h$ can in principle depend implicitly on density and temperature, but we do not expect this to be crucial since its importance is only felt in a narrow range of conditions in which $\Delta f$ is nearly zero. 

If the interaction between species is {\it pairwise} (as in a lattice-gas model for instance), $h(x)$  will be a linear function of composition:
\begin{equation} h(x) =
	g(1-x) = J x, 
\end{equation} 
where $J$ can be interpreted as the energy cost of
having a molecule surrounded by all neighboring atoms (or an atom surrounded by all
neighboring molecules), relative to the pure molecular (atomic) configuration.
A many-body molecular field can be modeled by choosing a non-linear $h(x)$.
Given this assumption of linearity, the equation for $x$ obtained by requiring that the total mixed free energy is minimum is (compare to Eq.\ref{solx}):
\begin{equation}\label{solxscf}
	x = \mathrm e^{-2 \beta[\Delta f(V) + 2J(1-2x)] - 1} \left(\sqrt{1 + 4
		e^{2\beta [\Delta f(V) + 2J(1-2x)]+1}} - 1\right).
\end{equation} 
Note that this requires a {\it self-consistent} solution for the species fraction $x$. This equation will always have at least one stable solution and can be solved numerically. The resulting mean field mixed free energy is:
\begin{equation}\label{solfmix}
	f^\mathrm{MF}_\mathrm{mix}(V) = (1-x) [f_\mathrm M(V) + J x] + x [f_\mathrm A(V) + J(1-x)] + kT \left[(1-x) \log(1-x)/2 + x \log(x)\right].
\end{equation}


\begin{thebibliography}{100}
\bibitem{caveat1}The ionic excitations need not be harmonic, though our treatment typically assumes so.
\bibitem{caveat2}We choose to include the ionic zero-point motion energy in $f_{\rm IT}(T)$ rather than $\phi_{\rm cold}(V)$, even though it is $T$-independent, because it arises from the phonon density of states also used to construct the $T$-dependent part of $f_{\rm IT}$. This choice was also made, for instance, in Ref.\cite{Correa}.
\bibitem{caveat3}This lowering of $C_{V}^{\rm ion}$ for the solid at high-$T$ can also be viewed as arising from anharmonicity, though we prefer to view it as a fundamental requirement needed to produce a sensible multiphase EOS in cases where we choose to define the free energy of each phase throughout all $(\rho,T)$.
\bibitem{caveat4}The highest temperatures we consider in this work (corresponding to those of interest for inertial confinement fusion) are such that $k_{B}T$ is well below $m_{p}c^{2}$, where $m_{p}$ is the mass of the proton.
\bibitem{caveat7}We have also conducted preliminary work in which an electron-thermal term, also PURGATORIO-based but for a system containing two electrons, is included in the model for the free energy of the molecular liquid. After mixing with the atomic liquid to produce the final liquid-H free energy, we established that the H EOS produced this way is strikingly similar to the H EOS produced by taking $f_{\rm ET}$ for the molecular liquid to be zero. This is because, at the temperatures high enough for the molecular ET term to be significant, the chemical equilibrium mixing mandates that the {\it atomic} liquid is, by far, the larger component.
\end{thebibliography}
\end{document}